\documentclass[acmsmall,screen,review=false]{acmart}

\AtBeginDocument{%
  }

\setcopyright{acmcopyright}
\usepackage{soul} % For highlight text
\usepackage{multirow}
\usepackage{mathtools}
\DeclarePairedDelimiter\norm{\lVert}{\rVert}%
\usepackage{bm}
\usepackage[capitalise]{cleveref}
\usepackage{caption}
\usepackage{subcaption}

\usepackage{tabularx}
\usepackage{pseudo}
\pseudodefinestyle{fullwidth}{
  begin-tabular = \tabularx{\linewidth}{@{}
    r % Labels
    >{\pseudosetup} % Indent, font, ...
    X % Code (flexible)
    >{\leavevmode\small\color{black!60}} % Comment styling
    p{,\linewidth} % Comments (fixed)
    @{}},
  end-tabular = \endtabularx,
  setup-append = \pseudoeq
}
\pseudoset{
  ctfont=\color{black!55}\textit,
  ct-left=//\ ,
  ct-right=,
}
\usepackage{float}
\floatstyle{plaintop}
\newfloat{algorithm}{tbp}{alg}[section]
\floatname{algorithm}{Algorithm}
\newcommand{\ceil}[1]{\left\lceil #1 \right\rceil}

\usepackage{makecell}

\begin{document}

\title[Inference Latency Prediction at the Edge]{Inference Latency Prediction at the Edge}
\author{Zhuojin Li}
\affiliation{%
  \institution{University of Southern California}
  \city{Los Angeles}
  \country{USA}}
\email{zhuojinl@usc.edu}

\author{Marco Paolieri}
\affiliation{%
  \institution{University of Southern California}
  \city{Los Angeles}
  \country{USA}}
\email{paolieri@usc.edu}

\author{Leana Golubchik}
\affiliation{%
  \institution{University of Southern California}
  \city{Los Angeles}
  \country{USA}}
\email{leana@usc.edu}

\renewcommand{\shortauthors}{Li et al.}

% *CRITICAL: Do Not Use Symbols, Special Characters, Footnotes, 
% or Math in Paper Title or Abstract.

\begin{abstract}
% Abstract:
With the growing workload of inference tasks on mobile devices, state-of-the-art neural architectures (NAs) are typically designed through Neural Architecture Search (NAS) to identify NAs with good tradeoffs between accuracy and efficiency (e.g., latency). 
Since measuring latency of a huge set of candidate architectures during NAS is not scalable, approaches are needed for predicting end-to-end inference latency on mobile devices. Such predictions are challenging due to hardware heterogeneity, optimizations applied by ML frameworks, and diversity of neural architectures.
Motivated by these challenges, in this paper, we first quantitatively assess characteristics of neural architectures and mobile devices that have significant effects on inference latency. Based on this assessment, we propose a latency prediction framework which addresses these challenges by developing operation-wise latency predictors, under a variety of settings and a number hardware devices, with multi-core CPUs and GPUs, achieving high accuracy in end-to-end latency prediction, as shown by our comprehensive evaluations.
%%%This is due to a number of factors, including our framework's ability to accurately
%%%deduce the operations and kernels running on mobile CPUs and GPUs - without deploying the target NA on the actual device - by exposing the principles of ML framework-specific optimizations.
%%%%
To illustrate that our approach does not require expensive data collection, we also show that accurate predictions can be achieved on real-world NAs using only small amounts of profiling data.

\end{abstract}

\begin{CCSXML}
<ccs2012>
   <concept>
       <concept_id>10010147.10010257.10010293.10010294</concept_id>
       <concept_desc>Computing methodologies~Neural networks</concept_desc>
       <concept_significance>500</concept_significance>
       </concept>
   <concept>
       <concept_id>10003120.10003138.10003141.10010898</concept_id>
       <concept_desc>Human-centered computing~Mobile devices</concept_desc>
       <concept_significance>500</concept_significance>
       </concept>
   <concept>
       <concept_id>10002944.10011123.10011674</concept_id>
       <concept_desc>General and reference~Performance</concept_desc>
       <concept_significance>500</concept_significance>
       </concept>
   <concept>
       <concept_id>10002944.10011123.10010912</concept_id>
       <concept_desc>General and reference~Empirical studies</concept_desc>
       <concept_significance>500</concept_significance>
       </concept>
 </ccs2012>
\end{CCSXML}

\ccsdesc[500]{General and reference~Performance}
\ccsdesc[500]{General and reference~Empirical studies}
\ccsdesc[500]{Computing methodologies~Neural networks}
\ccsdesc[500]{Human-centered computing~Mobile devices}

\keywords{neural networks, NAS, latency, prediction, mobile, GPU, CPU}

\maketitle

\section{Introduction} \label{introduction}

%\hl{\textbf{(Background: ML on mobile devices - Limited computing resources)}}
Due to significant breakthroughs in machine learning (ML), inference tasks using neural networks are being deployed to an increasing number of edge devices (e.g., smartphones, smartwatches, tablets), largely for computer vision and natural language tasks. In comparison with powerful cloud servers, edge devices have limited resources, which restricts the choice of deployed neural architectures (NAs).

%\hl{\textbf{(Background: NAS - Good trade-off between accuracy and efficiency; direct measure is unscalable)}}
In such a limited resource setting, state-of-the-art neural architectures \cite{howard2019searching, tan2019mnasnet, tan2019efficientnet} are typically designed through Neural Architecture Search (NAS) \cite{zophL17nas} by searching for an architecture with a good trade-off between accuracy and efficiency. 
%
% Recent works focus on architecture design satisfying target latency and energy constraints \cite{dai2019chamnet, ma2018shufflenet, zhang2018shufflenet}, which take into account the limited computation resources of edge platforms.
%
For example, recent works \cite{tan2019mnasnet, yang2018netadapt} propose to optimize accuracy under constraints on efficiency metrics (e.g., latency) that are measured directly on a target platform. However, neural architectures exhibit distinct performance characteristics across platforms \cite{tang2021bridge}, and it is impractical to measure end-to-end latency of every architecture on all possible platforms during model search.
% I removed this part, which didn't make sense to me: due to the cost of porting the NAS architectures trained from cloud to edge devices
%
%\hl{\textbf{(Background: Alternative to direct measure - Prediction model)}}
As an alternative to direct measurements, existing approaches for evaluating the efficiency of a neural architecture can be categorized as those using: (1)~Proxy metrics \cite{tan2019efficientnet, zoph2018nasnet} (e.g., FLOPs), which are usually platform-independent and cannot accurately reflect the actual performance due to the diversity of platforms \cite{ma2018shufflenet, tang2021bridge}. (2)~Look-up tables \cite{cai2019once, dai2019chamnet, wu2019fbnet}, which are collected for pre-defined building blocks in the search space, but cannot cover every possible configuration in a potentially huge search space and require comprehensive measurements on each platform. (3)~Prediction models \cite{abbasi2021maple, cai2019proxylessnas, dudziak2020brp}, which broadly rely on machine learning techniques (e.g., MLPs) and have the potential to predict the performance of any configuration in the search space. 
%
%\hl{\textbf{(Difficulties of prediction model - Why existing works do not solve them)}}
However, it is difficult to build accurate prediction models for efficiency metrics on \emph{mobile devices} due to the following challenges. 

\emph{(1) Hardware heterogeneity}: %
% Compared to cloud servers equipped with powerful GPUs, the limited computing and energy capacities restrict the capability of edge devices to process ML workloads. 
% Distinct from the cloud servers where modern GPUs achieve times of speedup over CPUs, edge CPUs can achieve comparable throughput to edge GPUs based on multi-threading and quantization techniques;
Existing prediction models mainly focus on cloud servers \cite{abbasi2021maple,gao2021runtime,geoffrey2021habitat,justus2018predicting} where Nvidia GPUs dominate the market for ML workloads; instead, the heterogeneity of mobile CPUs and GPUs makes performance prediction more difficult.
In particular, inference tasks are frequently performed on mobile devices using CPUs \cite{wu2019machine}, due to the support of a broader set of available operations (e.g., Channel Shuffle \cite{zhang2018shufflenet} is currently unavailable on the TensorFlow-Lite GPU Delegate \cite{lee2019device}). Modern mobile CPUs typically use the ARM big.LITTLE architecture, which consists of heterogeneous core clusters, e.g., high-performance cores and high-efficiency cores \cite{wang2019high}; when an inference task takes advantage of this multi-core architecture, the schedule of threads on different cores has a significant impact on performance (\cref{sec:multithreading}). 
In addition, multi-core speedups on a given device can vary for different neural architectures; for instance, MobileNet (with width multiplier of 0.75) and ResNet18 (with width scale of 0.25) achieve comparable inference latency (28.4~ms and 28.1~ms, respectively) on Pixel~4 with one medium core, but differ by 24.6\% with three medium cores (11.8~ms and 14.7~ms, respectively).
Therefore, it is necessary to evaluate prediction approaches using heterogeneous hardware resources, in particular on multi-core CPUs; this is \emph{not taken into consideration by existing works} on latency prediction for mobile CPUs \cite{lu2021one,zhang2021nn}.

\emph{(2) ML framework optimizations}: %
Modern ML frameworks introduce optimizations that can significantly accelerate inference tasks. For example, operator fusion \cite{niu2021dnnfusion} reduces overhead in the invocation of OpenCL kernels on GPUs; 
our tests show that disabling OpenCL kernel fusion in TensorFlow Lite (TFLite) \cite{tensorflowlite} can lead to an average 22\% performance degradation over 102 real-world NAs on PowerVR GE8320 (\cref{sec:kernel_fusion}).
Similarly, the choice of algorithms used to implement each operation can considerably affect inference performance; e.g., TFLite uses the faster Winograd \cite{lavin2016fast} algorithm for some (but not all) convolution layers on GPUs. \emph{Existing works} on latency estimation for mobile GPUs \cite{bouhali2021execution,bouzidi2021performance} \emph{do not consider such optimizations} (which are specific to ML frameworks) but predict inference latency only from the architecture of the neural network.

\emph{(3) Neural architecture diversity}: %
During the exploration of the search space by NAS algorithms, the properties of neural architectures (e.g., the number of operations and their latency) can vary considerably; in addition, novel architectures are proposed by manual design \cite{howard2019searching, ma2018shufflenet, zhang2018shufflenet}, prompting the definition of new NAS search spaces. \emph{Existing ML-based performance prediction models use training and test datasets with very similar neural architectures} \cite{abbasi2021maple, bouhali2021execution, syed2021generalized}, \emph{or with a small set of popular architectures} \cite{bryzgalov2021predicting, geoffrey2021habitat, hafeez2020empirical}; in contrast, practical applicability of performance prediction to NAS requires accuracy on a large set of \emph{diverse} neural architectures.

% %\hl{\textbf{(Highlights on our work - How to conquer the aforementioned obstacles)}}
% \ul{
% In this work, we propose an approach to predict latency of ML inference tasks on mobile CPUs and GPUs using machine learning models. We conquer the aforementioned obstacles as follows.
% %
% (1) For hardware heterogeneity, we study the performance of inference tasks on mobile CPUs by analyzing the effects of different choices of heterogeneous cores for the multi-thread scheduling. We profile multi-threading execution time under different combinations of cores, and we construct separate models to predict performance under each setting.
% % not sure what is meant by separate models?
% (2) For ML framework optimizations, we reveal the principles adopted by TensorFlow lite for kernel fusion and kernel selection. Without the need for compiling the model on actual hardware devices, we are able to deduce the OpenCL kernels that will be selected on mobile GPUs.
% (3) For neural architecture diversity, we design a NAS space covering the majority of representative operations and building blocks; we use this NAS space to generate synthetic neural architectures for our training sets and we evaluate our latency prediction models on a test set with the same data distribution as well as a test set including many state-of-the-art neural architectures.
% }

Motivated by the above stated challenges, in this paper, we first quantitatively assess characteristics of neural architectures and mobile devices that have significant effects on inference latency. Based on this assessment, we develop
a framework to predict latency of inference tasks on mobile CPUs and GPUs using machine learning models as well as carry out a comprehensive evaluation study to demonstrate the accuracy of the proposed approach. 
Our approach predicts inference latency using ML models trained to estimate latency of neural architecture components as ``building blocks'' of end-to-end latency.
%why advantageous - last sentence of the paragraph?
A \emph{per-component} approach provides learning efficiency (i.e., models can be trained quickly from small datasets), in contrast with complex learning models \cite{abbasi2021maple,gao2021runtime,dudziak2020brp} predicting latency from graphs of tensor operations (including parameters of all operations).

Our prediction framework allows us to address several shortcomings in existing literature.
We develop a training dataset that is representative of real-world neural architectures but provides broader coverage of possible neural architectures, leading to better generalizability. In addition, while existing works \cite{lu2021one,zhang2021nn} study the performance of mobile CPUs on a single core, our dataset includes measurements over a broader set of practical scenarios, e.g., choices of CPU cores and use of data representations (floating-point or integer quantization); our prediction models allow NAS approaches to account for realistic scenarios, leading to better estimation of inference latency on mobile devices.
Also differently from previous work \cite{zhang2021nn}, which builds black-box models to estimate effects of ML framework optimizations, we expose important principles from open-source ML framework, which enables accurate estimation of GPU kernels without deploying candidate neural architectures on actual devices.
%
% Moreover, from the CPU operations and GPU kernels of a target NA, we construct ML predictors to capture the underlying relationship between input features and latency, which provides better estimation of each NA's component's latency and consequently end-to-end latency prediction as compared to approaches that use direct end-to-end predictions \cite{abbasi2021maple,gao2021runtime,dudziak2020brp}.

%\hl{\textbf{(Contributions)}}
Thus, the main contributions of our work are as follows.
\begin{itemize}
\item By collecting measurements for 102 state-of-the-art neural architectures from 25 papers on 4 mainstream mobile platforms (or SoCs), based on quantitative evidence, we identify aspects of hardware and ML frameworks which substantially affect the latency of inference tasks on mobile devices.
For mobile CPUs, we expose performance characteristics under various settings, including multithreading over ARM heterogeneous core clusters and quantization with lower-bit representations (\cref{sec:background_cpu_performance}). On mobile GPUs, we categorize two types of optimization strategies due to ML framework compilation: kernel fusion and kernel selection (\cref{sec:background_gpu_performance}). As a representative example, we expose the principles of both strategies in TFLite, and empirically evaluate resulting speedups to highlight their impact on inference latency.

\item Based on the results of our performance study, we develop a framework for estimating end-to-end inference latency on mobile devices by composing accurate latency predictions of individual NA components (\cref{sec:prediction_model}). 
To address hardware heterogeneity, we profile execution times of neural architectures under various settings of multi-core and data representations, and train ML models to predict performance under each setting. For ML framework optimizations, we are able to deduce the OpenCL kernels that are selected on mobile GPUs, without the need for deploying and compiling the target neural architecture on the actual hardware (\cref{sec:kernel_deduction}).
By conducting one-time training data collection on each device, we are able to utilize learning-based models to accurately predict latency of inference tasks under various settings of mobile CPUs and GPUs, which can be used by existing NAS techniques to evaluate inference latency without access to actual hardware.

\item We build a synthetic dataset of 1000 neural architectures sampled from a NAS space covering a majority of configurations for common operations and building blocks (\cref{sec:synthetic_dataset}). For each neural architecture, we comprehensively measure latency under 72 scenarios across 4~mainstream mobile platforms, including the combination of multiple cores and the utilization of integer representations after quantization.
In addition to accurate latency prediction, this provides insight to (i)~NA developers for how to build efficient neural architectures and (ii)~mobile developers for how to choose suitable optimizations for running inference tasks.

\item To evaluate how our approach addresses the aforementioned challenges, in addition to the default setting of NAS (\cref{sec:result_default_setting}), we show that our predicting framework also achieves accurate estimations under hardware heterogeneity (\cref{sec:result_hardware_heterogeneity}), neural architecture diversity (\cref{sec:result_neural_architecture_heterogeneity}), and ML framework optimizations (\cref{sec:result_framework_heterogeneity}).
To address the concerns of the cost of training data collection \cite{lu2021one}, we evaluate accuracy of predictions with limited amounts of training data, using multiple ML methods (\cref{sec:result_limited_training_data}).
Our results highlight that, when trained with sufficient data of 1000 synthetic neural architectures, more powerful ML methods can achieve accurate predictions for neural architectures with similar characteristics to the training data (e.g., by using GBDT \cite{friedman2001greedy}, 2.4\% average error when using one large CPU core and 6.3\% when using GPUs); with limited training data of only 30 neural architectures, the linear approach of Lasso \cite{tibshirani1996regression} can generalize well to real-world neural architectures, even when their characteristics differ from training data (e.g., 6.9\% average error for CPUs with one large core, and 9.1\% for GPUs).

\end{itemize}

\section{Background: Neural Architectures on Mobile Devices} \label{sec:background}

\begin{figure}[t]
    \centering
	\includegraphics[width=.65\linewidth]{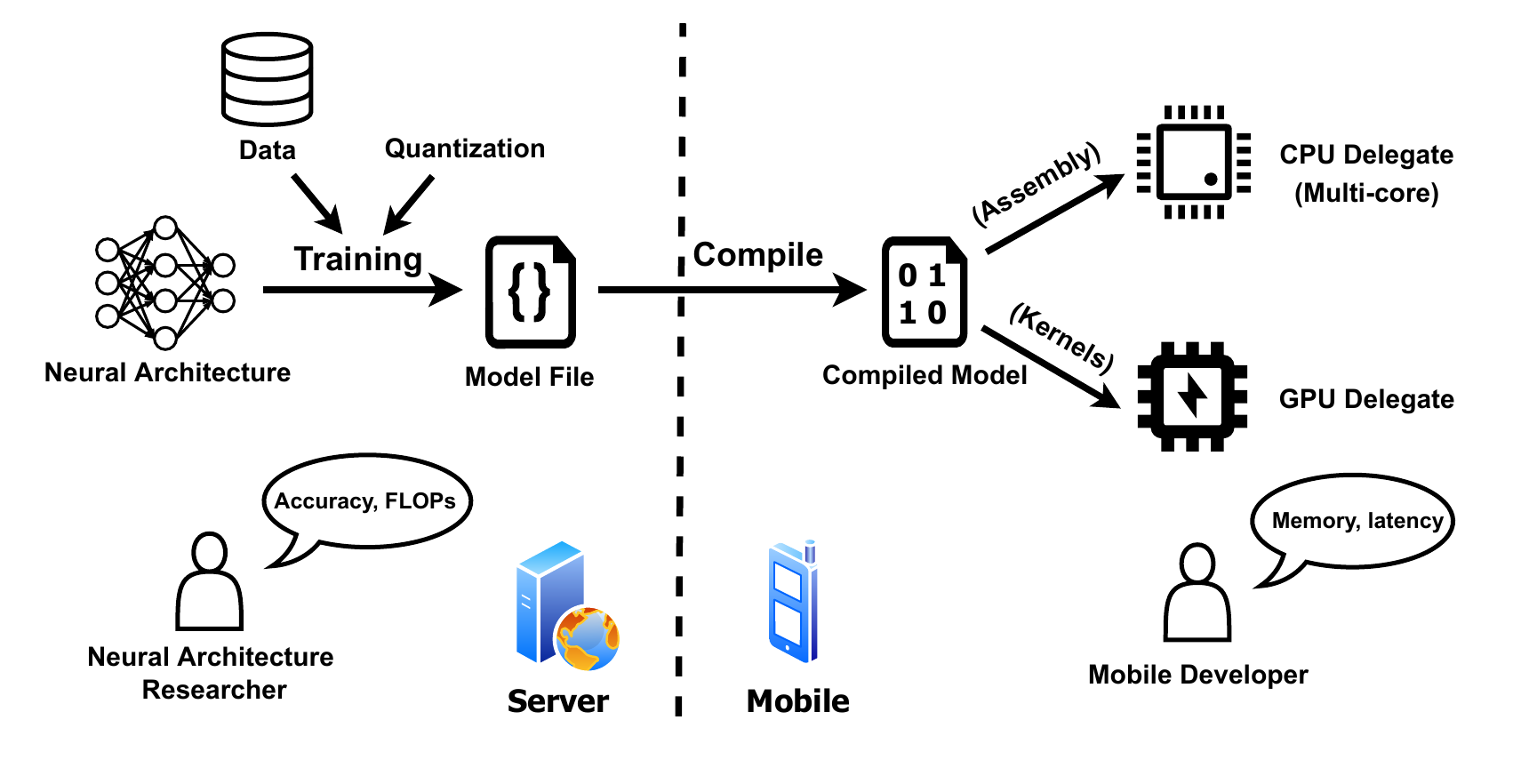}
	\vspace{-1em}\caption{\vspace{-1em}Lifecycle of Neural Architecture Development and Deployment on Mobile Devices}
	\label{fig:lifecycle}
\end{figure}

As illustrated in \cref{fig:lifecycle}, the lifecycle of neural architecture development and deployment on mobile devices consists of (1)~designing and training a neural network model on cloud servers, and (2)~deploying the model on a target mobile device where it is executed, i.e., where inference tasks are performed on CPU cores or GPU.

State-of-the-art neural architectures are developed by both manual design~\cite{he2016deep,howard2017mobilenets,ma2018shufflenet} and NAS~\cite{sandler2018mobilenetv2,howard2019searching, tan2019mnasnet, tan2019efficientnet}. Due to scarce computing and memory resources, neural architectures intended for inference tasks on mobile devices are designed not only to maximize prediction accuracy, but also to satisfy performance constraints such as end-to-end latency and memory consumption. To achieve these goals, \emph{model quantization} \cite{jacob2018quantization,nagel2021white} is frequently applied: fixed-width integers are used to represent the model parameters and to perform computations with low precision, reducing memory requirements and computation times (as shown in \cref{sec:quantization}).

After the identified neural architecture is trained on cloud servers, it is stored as a model file, which can be distributed to heterogeneous mobile platforms for inference tasks. For instance, in TFLite, a neural architecture is described as a computational graph, where each node represents an operation and each edge represents the flow of intermediate results between operations; the complete computational graph is included in the \texttt{.tflite} model file.

A mobile device can be equipped with multiple hardware accelerators to serve inference tasks (e.g., CPU, GPU, DSP and Edge TPU are available on Pixel 4). To be executed on different hardware, the model is ``compiled'' to select an optimized CPU implementation or a platform-specific GPU kernel for each operation of the computational graph. Notably, the same operation can be executed using different algorithms on different devices; for example, the TFLite GPU Delegate can select different kernels for convolution operations on Adreno GPUs vs.\ Mali GPUs (as detailed in \cref{sec:kernel_selection}).
In addition, the computational graph can be optimized during model compilation; for instance, two consecutive operations can be ``fused'' and executed with a single GPU kernel (as detailed in \cref{sec:kernel_fusion}).
Eventually, a compiled model is executed on the target hardware: on GPUs, kernels are dispatched to a command queue for execution; on CPUs, operations are executed sequentially, while multithreading is used only to accelerate the execution of individual operations on multiple cores (as detailed in \cref{sec:multithreading}).

% In addition to prediction accuracy, mobile developers aim at improving memory utilization and \emph{end-to-end latency} of inference tasks on devices; in contrast, researchers typically adopt \emph{proxy metrics} such as FLOPs to suggest efficient neural architectures, normally conducting NAS only with cloud resources.
% %
% However, existing studies have shown that neural architectures with better proxy metrics can exhibit worse inference latency on some mobile platforms \cite{ma2018shufflenet, tang2021bridge}, due to the substantial impact of the optimizations applied by mobile developers and ML frameworks, which is not captured by proxy metrics.

\section{Performance Characteristics of Inference on Mobile Devices} \label{section:performance_characteristics}

In this section, we present the results of our empirical study on the performance of state-of-the-art neural architectures on mobile platforms; in particular, we analyze thread scheduling and model quantization in multicore mobile CPUs (\cref{sec:background_cpu_performance}), and kernel fusion and selection in mobile GPUs (\cref{sec:background_gpu_performance}), evaluating their impact on inference latency. The insight gained here will be used in \cref{sec:methodology} to develop our latency prediction framework.

\subsection{Performance Characteristics of Mobile CPUs} \label{sec:background_cpu_performance}

\subsubsection{Effects of Multithreading} \label{sec:multithreading}

\begin{table}[t]\centering\footnotesize
\renewcommand\arraystretch{1.3}
\begin{tabular}{l l p{0.35\linewidth} l}\toprule
Device & Platform & CPU Cores & GPU \\
\toprule
Google Pixel 4 & Snapdragon 855 & Large: 1x Kryo 485 Prime (2.84 GHz)\newline Medium: 3x Kryo 485 Gold (2.32 GHz)\newline Small: 4x Kryo 485 Silver (1.8 GHz) & Adreno 640 \tabularnewline
Xiaomi Mi 8 SE & Snapdragon 710 & Large: 2x Kryo 360 Gold (2.2 GHz)\newline Small: 6x Kryo 360 Silver (1.7 GHz) & Adreno 616 \tabularnewline
Samsung Galaxy S10 & Exynos 9820 & Large: 2x M4 Cheetah (2.73 GHz)\newline Medium: 2x Cortex-A75 (2.31 GHz)\newline Small: 4x Cortex-A55 (1.95 GHz) & Mali G76  \tabularnewline
Samsung Galaxy A03s & Helio P35 & Large: 4x Cortex-A53 (2.3 GHz)\newline Small: 4x Cortex-A53 (1.8 GHz) & PowerVR GE8320 \tabularnewline
\bottomrule
\end{tabular}
\caption{Mobile Platforms in Our Study}
\label{table:mobile_platforms}\vspace{-3em}
\end{table}

Modern mobile platforms typically adopt the ARM big.LITTLE architecture, which allows multiple types of CPU cores to be integrated on the same system; each group of homogeneous cores is operated as a ``core cluster'' running at the same clock speed. The ``big cores'' with higher clock speed can handle computationally intensive tasks, while the ``LITTLE cores'' with lower clock speed can reduce power consumption.
\cref{table:mobile_platforms} lists the core clusters of the 4 SoCs in our study, providing a range of mobile device hardware. For example, Snapdragon~855 uses three clock domains for prime, gold and silver core clusters, respectively: tasks with high priority are usually scheduled on prime and gold cores for higher performance, while non-urgent tasks are scheduled on silver cores to reduce energy consumption.

\begin{figure}[t]
	\centering
	\begin{subfigure}[b]{.49\linewidth}
		\centering
		\includegraphics[width=\linewidth]{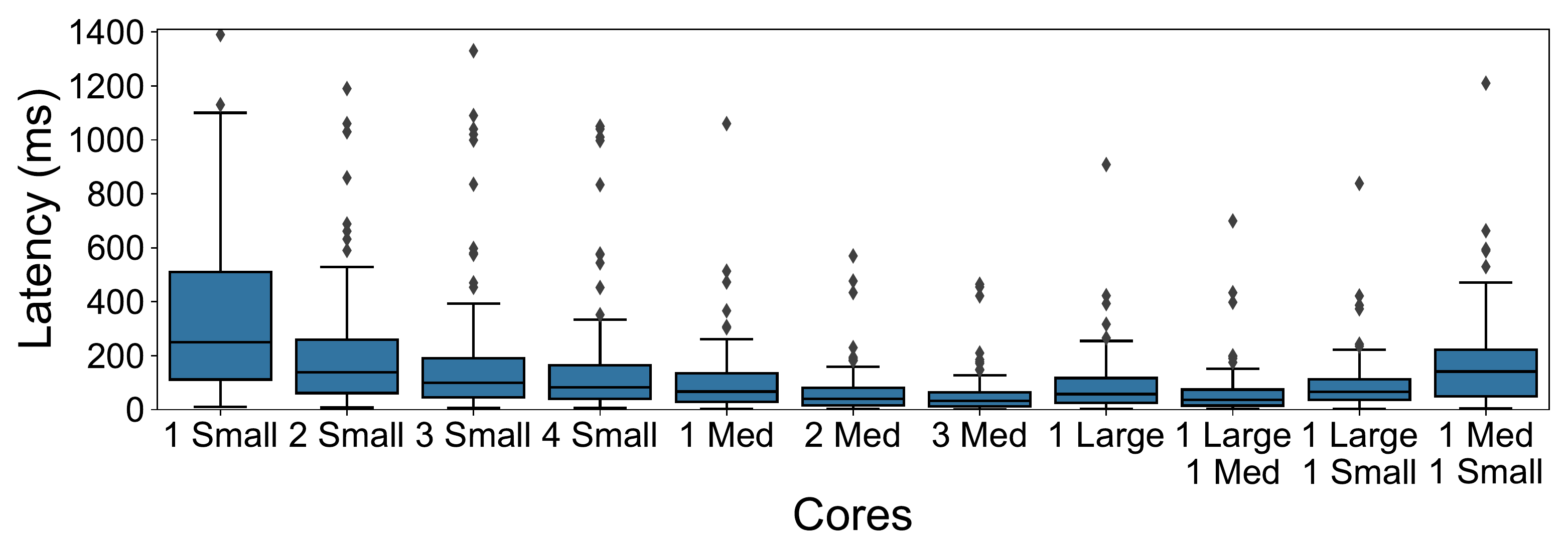}
		\caption{Snapdragon 855}\label{fig:multithread_pixel4}
	\end{subfigure}
	\begin{subfigure}[b]{.49\linewidth}
		\centering
		\includegraphics[width=\linewidth]{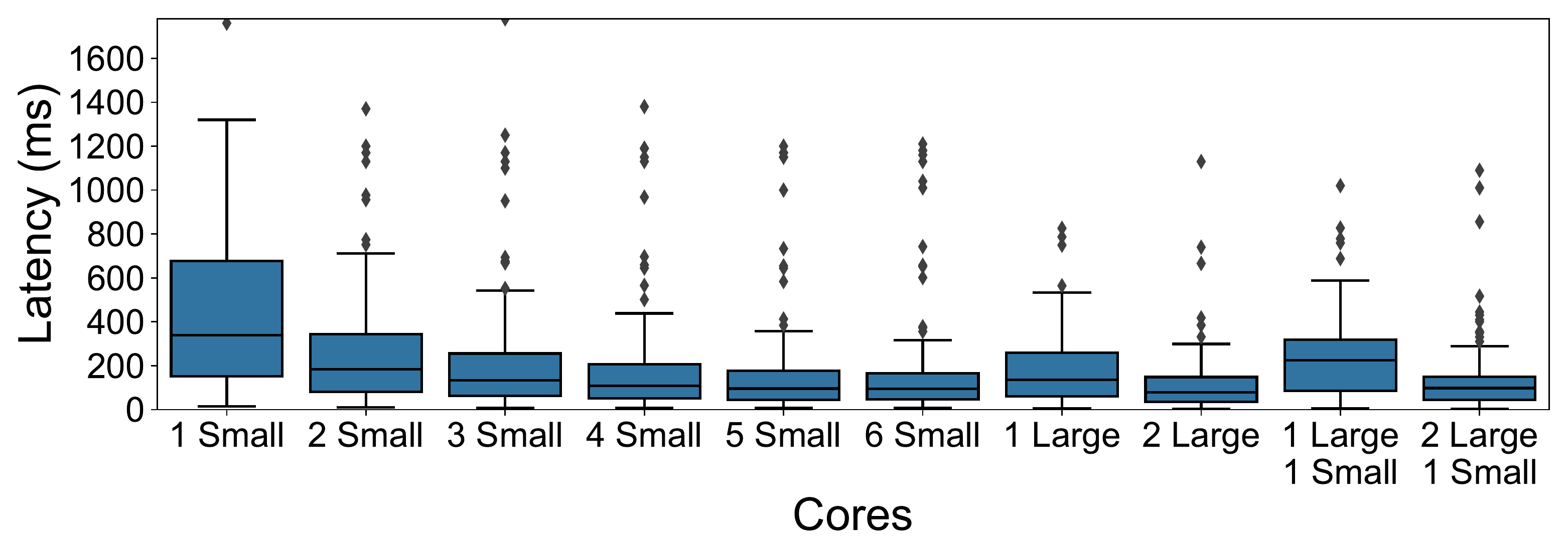}
		\caption{Snapdragon 710}\label{fig:multithread_mi8se}
	\end{subfigure}\\
	\begin{subfigure}[b]{.49\linewidth}
		\centering
		\includegraphics[width=\linewidth]{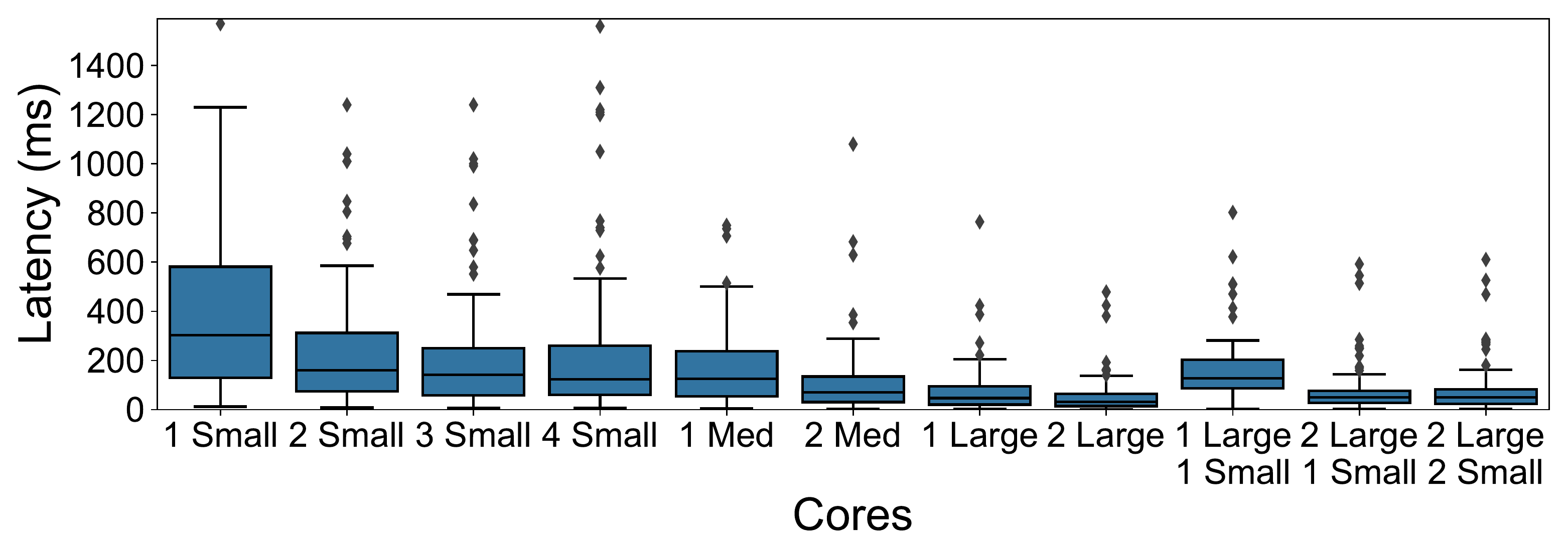}
		\caption{Exynos 9820}\label{fig:multithread_s10}
	\end{subfigure}
	\begin{subfigure}[b]{.49\linewidth}
		\centering
		\includegraphics[width=\linewidth]{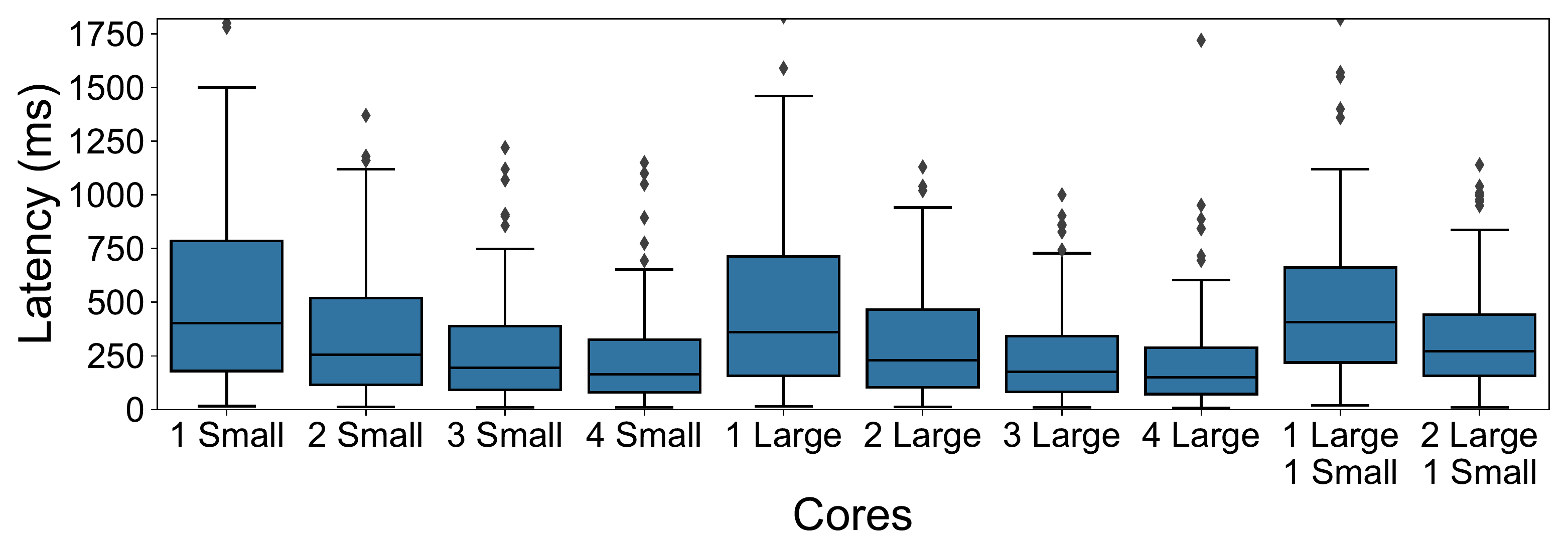}
		\caption{Helio P35}\label{fig:multithread_a03s}
	\end{subfigure}
	\caption{Effects of Multithreading / Multicore on End-to-end Latency}
	\label{fig:multithread}\vspace{-1em}
\end{figure}

An inference task can be accelerated with multithreading over multiple cores. \cref{fig:multithread} uses boxplots to depict end-to-end latency of 102 state-of-the-art neural architectures (details of these neural architectures are reported in \cref{appendix:common_models}) on Snapdragon 855, Snapdragon 710, Exynos 9820, and Helio P35 platforms for different multicore configurations; in these experiments, we use a number of threads equal to the number of cores and specify CPU affinity for each thread to schedule on a specific core. In \cref{fig:multithread} (and in the rest of the paper) boxplots indicate 1st quartile, median, and 3rd quartile of the data; whiskers extend for 1.5x the interquartile range; points outside of whiskers are denoted as outliers.
%%%with minimum identify the lowest (and highest) data points within the 1.5 times interquartile range below the 25 percentile (and above 75 percentile);
For clarity of presentation, in \cref{fig:multithread} we omit some outliers with substantially higher latency (<4\% of data points per configuration); due to lack of space, complete data are reported in \cref{fig:appendix_multithread} of the Appendix. Counterintuitively, \emph{using multiple heterogeneous cores can result in performance degradation}: for example, on Snapdragon 855 (\cref{fig:multithread_pixel4}), the combination of a medium core and a small core exhibits worse performance (on average) than a medium core; on Exynos 9820 (\cref{fig:multithread_s10}), the combination of a large core and a small core is slower than a large core.
After inspecting the source code of TFLite and of its library Ruy for CPU execution \cite{TFLiteThreads,RuyThreads}, we attribute this performance degradation to: (1)~the overhead of multi-threading across different clusters of CPU cores (e.g., large and small cores), and (2) the approach used to distribute work, which is split \emph{equally} among the number of available threads; with heterogeneous CPU cores, threads assigned to slower cores can become the stragglers.

\begin{figure}[t]
    \captionsetup{justification=centering}
	\centering

	\begin{subfigure}[b]{.49\linewidth}
		\centering
		\includegraphics[width=\linewidth]{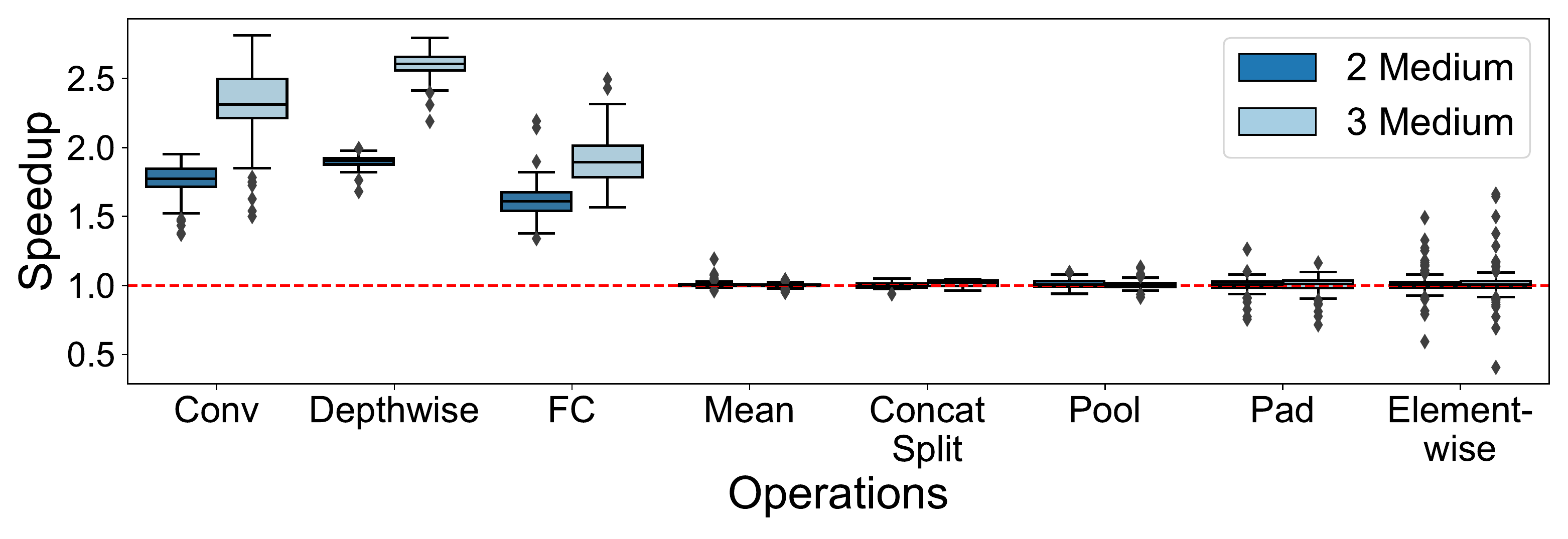}
		\caption{Snapdragon 855}\label{fig:multithread_ops_pixel4}
	\end{subfigure}
	\begin{subfigure}[b]{.49\linewidth}
		\centering
		\includegraphics[width=\linewidth]{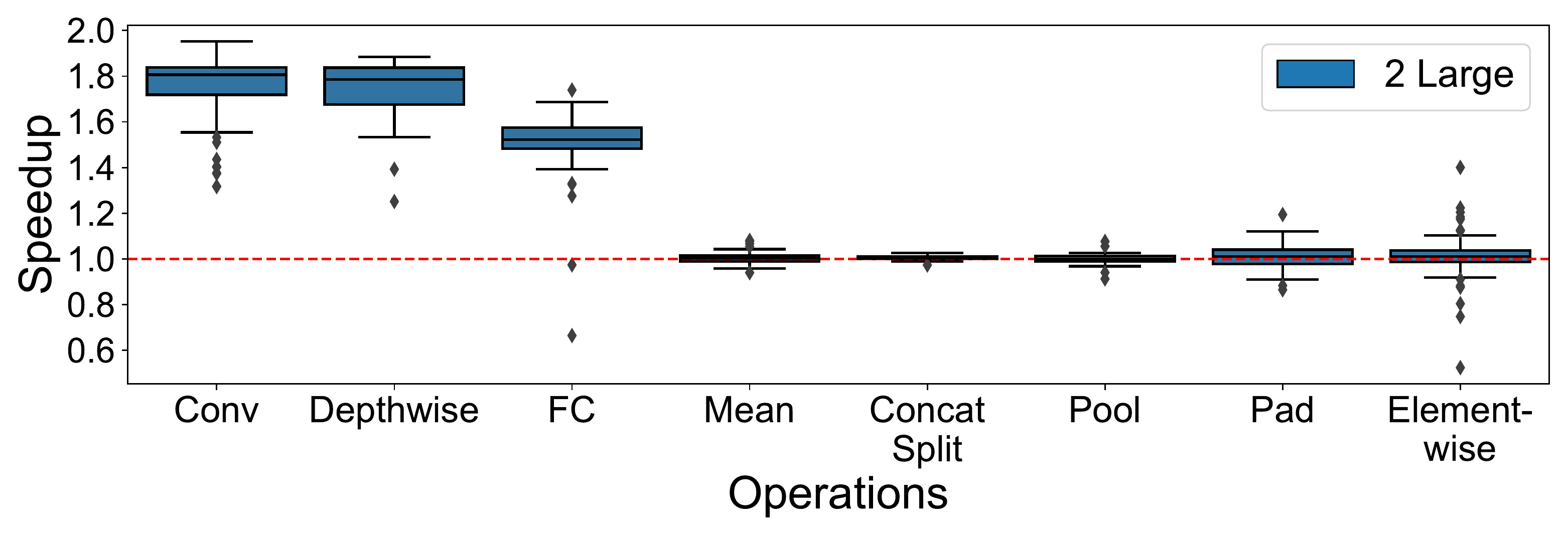}
		\caption{Snapdragon 710}\label{fig:multithread_ops_mi8se}
	\end{subfigure}
	\begin{subfigure}[b]{.49\linewidth}
		\centering
		\includegraphics[width=\linewidth]{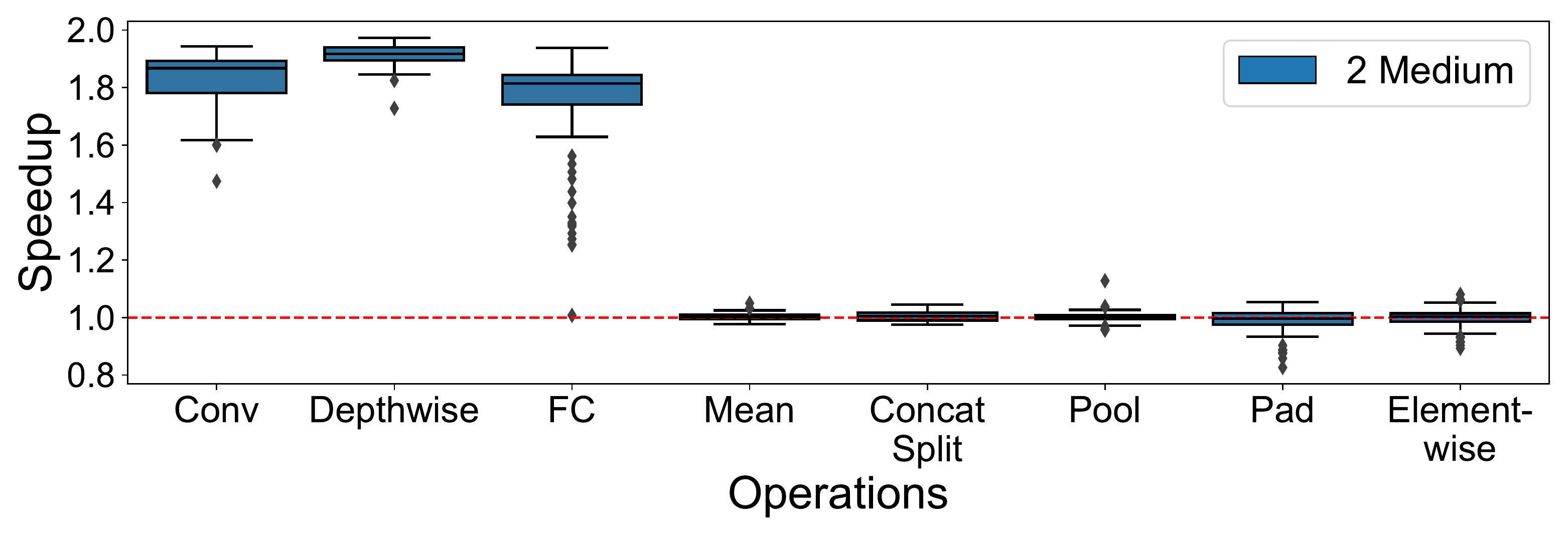}
		\caption{Exynos 9820}\label{fig:multithread_ops_s10}
	\end{subfigure}
	\begin{subfigure}[b]{.49\linewidth}
		\centering
		\includegraphics[width=\linewidth]{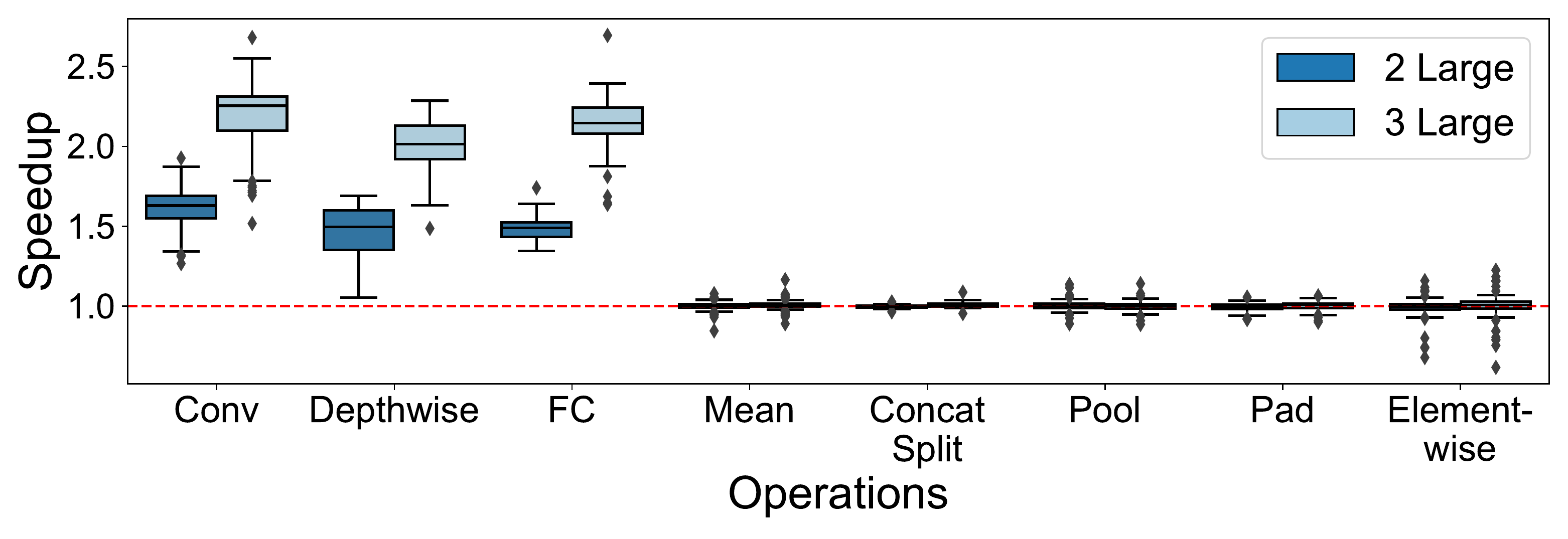}
		\caption{Helio P35}\label{fig:multithread_ops_a03s}
	\end{subfigure}

	\caption{Effects of Multithreading / Multicore on Operation-wise Latency (Speedup over One Core)}
	\label{fig:multithread_ops}
\end{figure}

For multithreading with homogeneous cores in \cref{fig:multithread}, we observe a sublinear speedup with respect to the number of cores.
\cref{fig:multithread_ops} shows the speedup of different operation types with respect to the number of homogeneous cores. We observe that convolution, depthwise convolution and fully-connected operations achieve sublinear speedups as the number of threads increases.
However, performance improvements on the remaining operations are negligible, due to the lack of support for parallel execution of these operations in the current TFLite implementation.

\vspace{.5em}
\noindent
\framebox{\parbox{\dimexpr\linewidth-2\fboxsep-2\fboxrule}{%
\textbf{Insight 1.} On mobile CPUs, multithreading has a significant impact on the performance of inference tasks. On homogeneous cores, multithreading leads to \textit{sublinear} reduction of latency for convolution, depthwise convolution and fully-connected operations in TFLite; however, on heterogeneous cores, multithreading can result in \textit{performance degradation} when small cores become stragglers of operations that support parallel execution.
}}

\subsubsection{Effects of Quantization} \label{sec:quantization}

\begin{figure}[t]
	\centering

	\begin{subfigure}[b]{.49\linewidth}
		\centering
		\includegraphics[width=\linewidth]{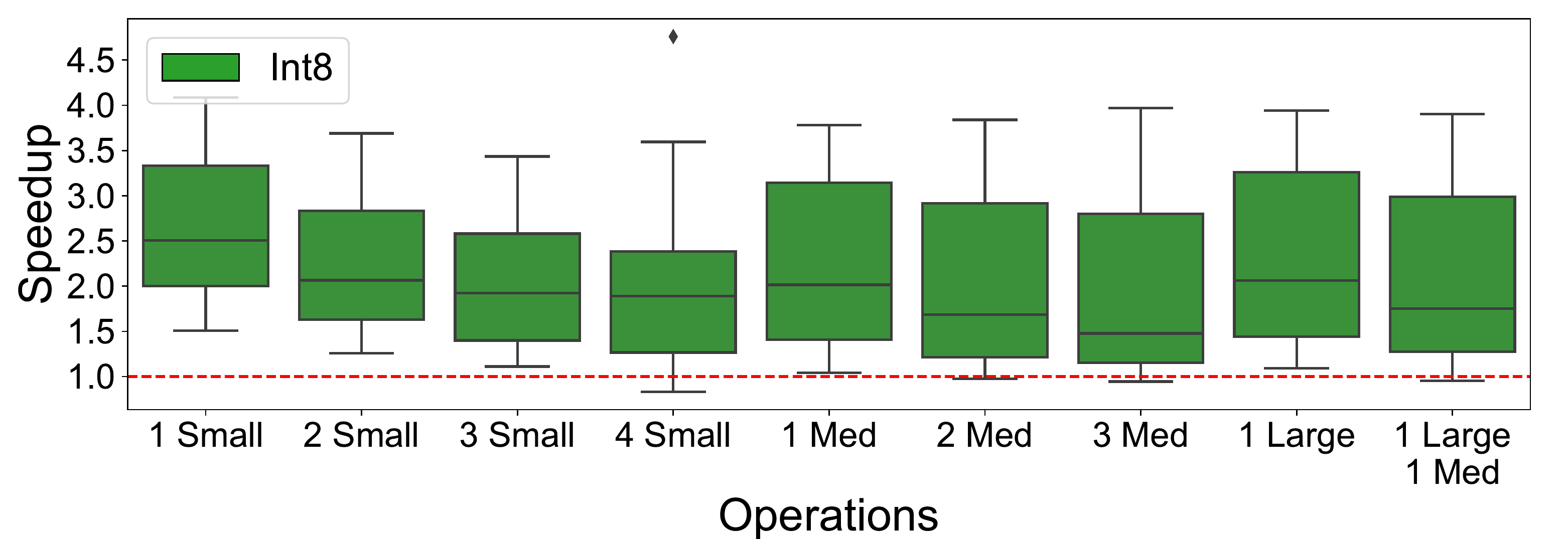}
		\caption{Snapdragon 855}\label{fig:quantization_pixel4_2}
	\end{subfigure}
	\begin{subfigure}[b]{.49\linewidth}
		\centering
		\includegraphics[width=\linewidth]{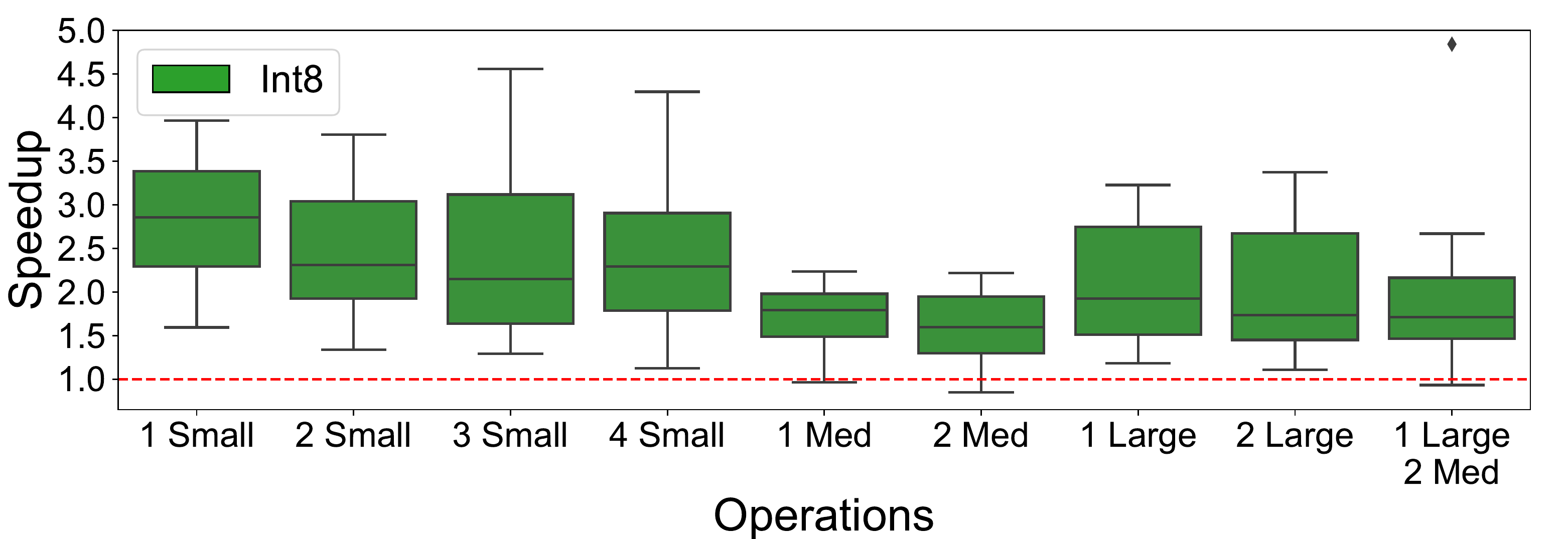}
		\caption{Exynos 9820}\label{fig:quantization_s10_2}
	\end{subfigure}
	\begin{subfigure}[b]{.49\linewidth}
		\centering
		\includegraphics[width=\linewidth]{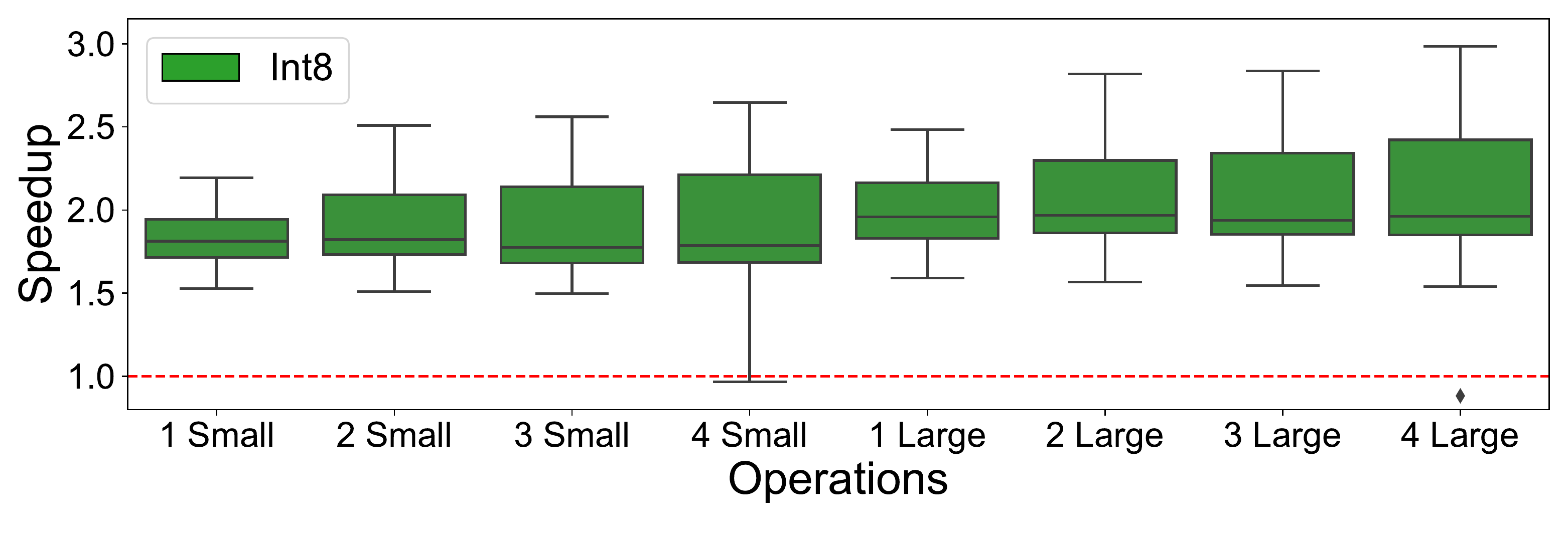}
		\caption{Helio P35}\label{fig:quantization_a03s_2}
	\end{subfigure}
	\begin{subfigure}[b]{.49\linewidth}
		\centering
		\includegraphics[width=\linewidth]{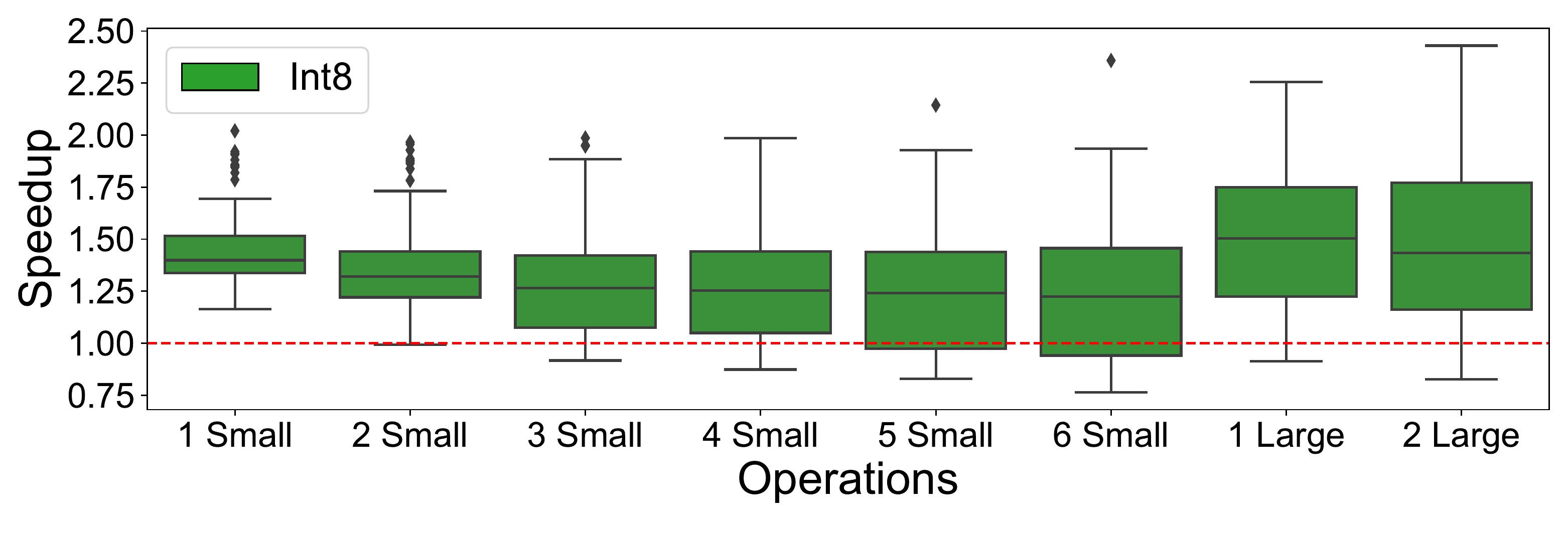}
		\caption{Snapdragon 710}\label{fig:quantization_mi8se_2}
	\end{subfigure}

    \caption{Effects of Quantization on End-to-end Latency}
	\label{fig:quantization}\vspace{-1em}
\end{figure}

On mobile devices with limited power and computing resources, neural architectures can be converted into lower-precision representations (e.g., 16-bit floating point or 8-bit integers) to reduce memory utilization and computational demand, without substantial accuracy loss. We focus on the approach of integer-arithmetic-only inference \cite{jacob2018quantization} available in TFLite, where both weights and activations are represented as 8-bit integers during inference.\footnote{We study the effects of quantization only on mobile CPUs, because using 8-int integers can cause significant overhead in the current implementation of the TFLite GPU delegate, to invoke GPU kernels of quantization and dequantization.}
\cref{fig:quantization} compares inference latency using an 8-bit integer representation and a floating point representation. Similarly to \cref{fig:multithread}, we omit outliers (only of a couple of points) for better visualization, and report complete data in \cref{fig:appendix_quantization} of the Appendix.
As can be seen, quantization shows a distinct speedup on various combinations of cores on all devices.

\begin{figure}[t]
    \captionsetup{justification=centering}
	\centering
	\begin{subfigure}[b]{.49\linewidth}
		\centering
		\includegraphics[width=\linewidth]{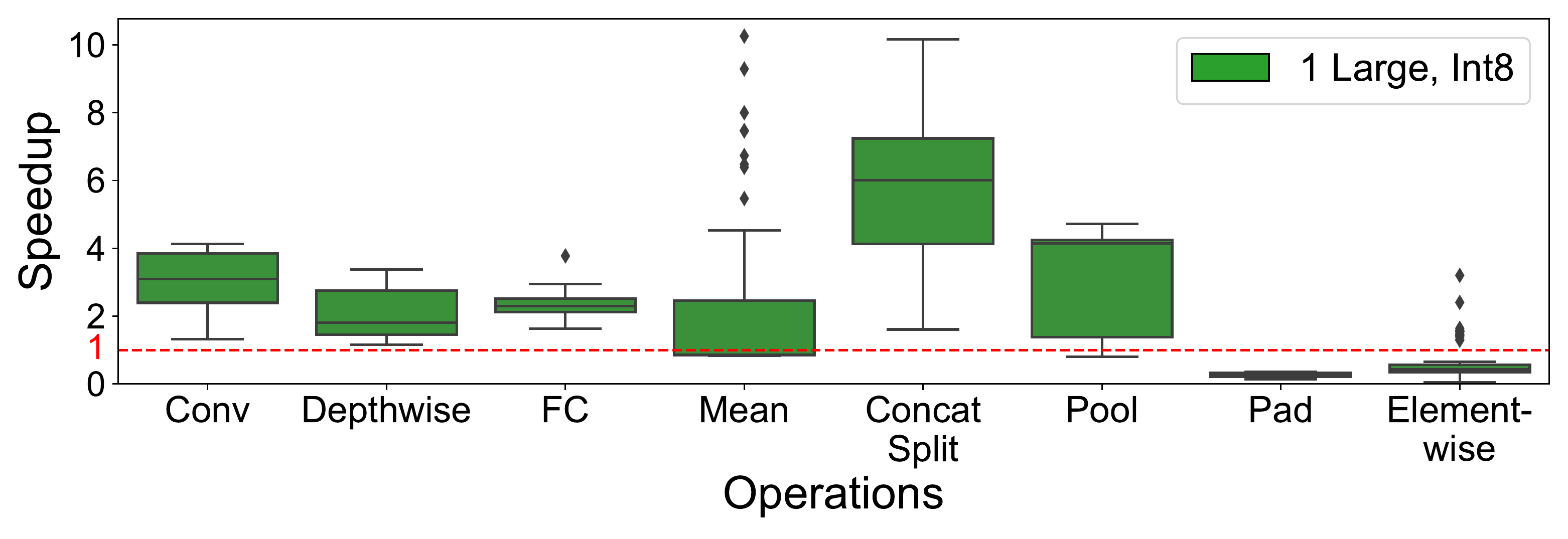}
		\caption{Snapdragon 855}\label{fig:quantization_ops_pixel4_1}
	\end{subfigure}
	\begin{subfigure}[b]{.49\linewidth}
		\centering
		\includegraphics[width=\linewidth]{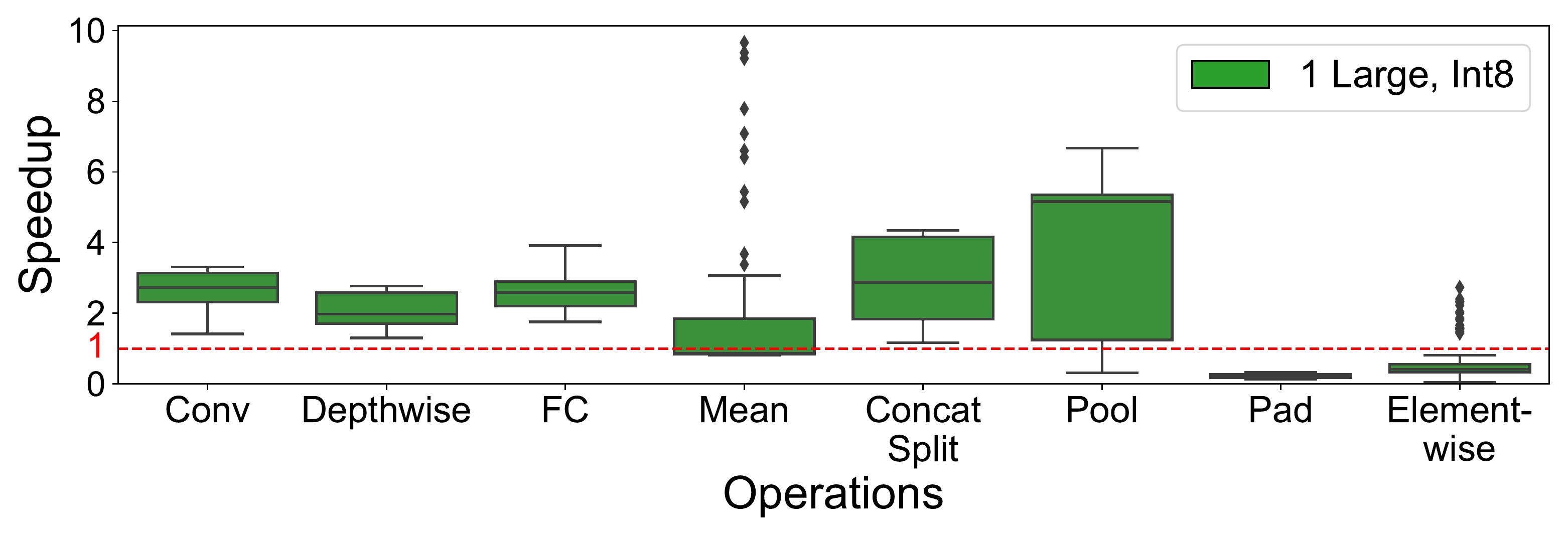}
		\caption{Exynos 9820}\label{fig:quantization_ops_s10_1}
	\end{subfigure}
	\begin{subfigure}[b]{.49\linewidth}
		\centering
		\includegraphics[width=\linewidth]{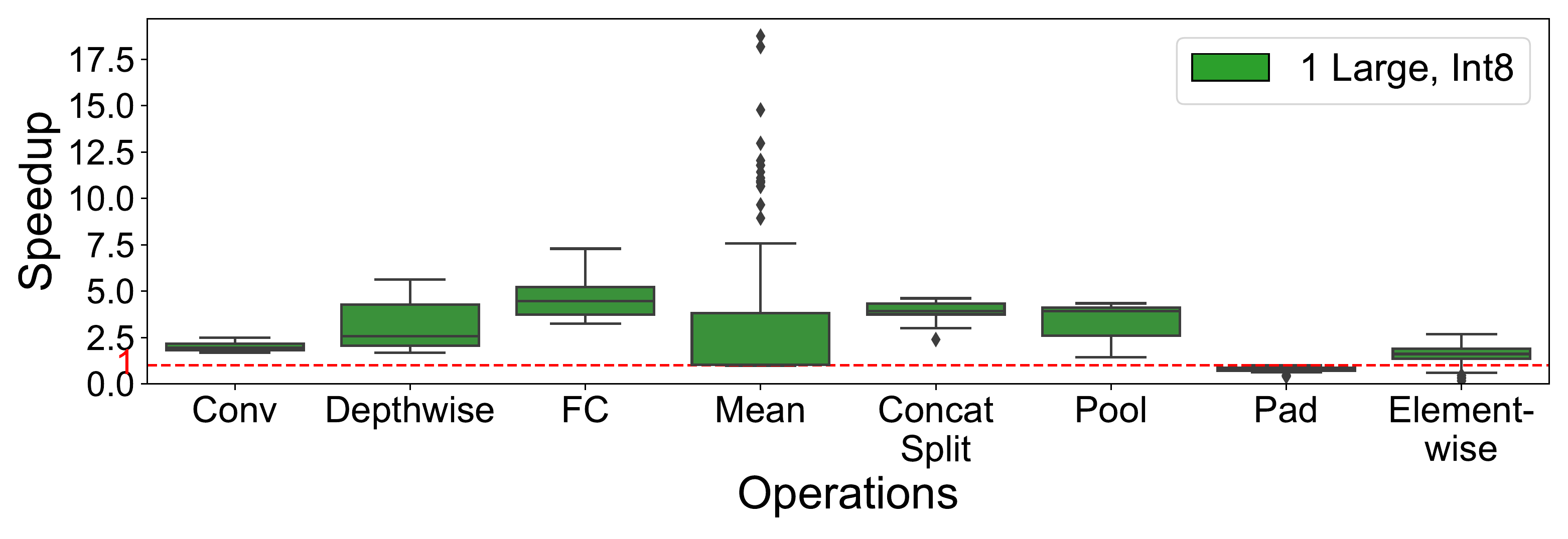}
		\caption{Helio P35}\label{fig:quantization_ops_a03s_1}
	\end{subfigure}
	\begin{subfigure}[b]{.49\linewidth}
		\centering
		\includegraphics[width=\linewidth]{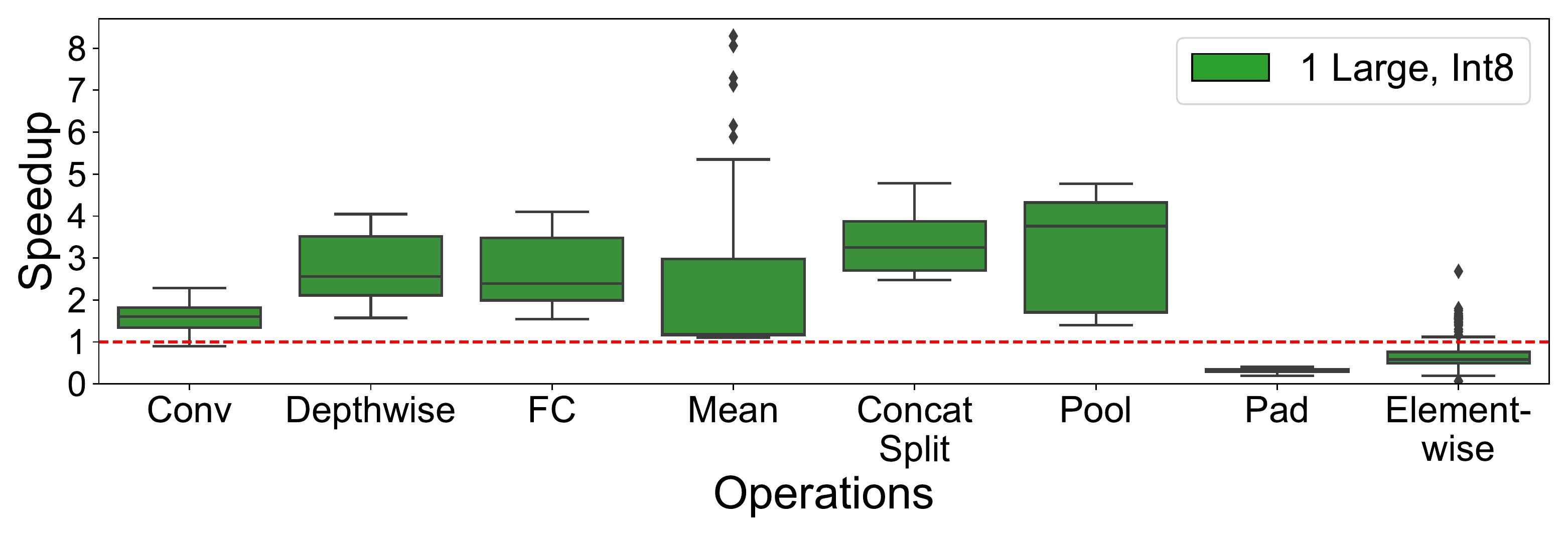}
		\caption{Snapdragon 710}\label{fig:quantization_ops_mi8se_1}
	\end{subfigure}

	\caption{Effects of Quantization on Operation-wise Latency}
	\label{fig:quantization_ops}
\end{figure}

\cref{fig:quantization_ops} depicts performance improvement of each type of operation after quantization. On all devices, most operations achieve significant speedup when using 8-bit integers; however, padding and element-wise operations show \emph{performance degradation} after quantization. 
For example, the average latency of element-wise operations is increased to 2.55x and 2.60x on Snapdragon 855 and Exynos 9820 respectively.
Previous work \cite{jacob2018quantization, nagel2021white} suggests that this degradation is due to the overhead of matching quantization ranges (i.e., the scale) of all inputs of quantized operations (e.g., element-wise addition).

\vspace{0.5em}
\noindent
\framebox{\parbox{\dimexpr\linewidth-2\fboxsep-2\fboxrule}{%
\textbf{Insight 2.}  Quantization can reduce latency and memory utilization of a model, significantly improving the performance of inference tasks on mobile CPUs. However, quantization can cause \emph{performance degradation} for some operations due to the cost of scaling its inputs.
}}

\subsection{Performance Characteristics of Mobile GPUs} \label{sec:background_gpu_performance}

% On mobile GPU, operations on the computational graph are compiled as shaders. To improve the performance, TFLite fuses neighboring kernels and selects optimized algorithm for operations.

\subsubsection{Effects of Kernel Fusion} \label{sec:kernel_fusion}

Kernel fusion has been broadly adopted to reduce the overhead of dispatching kernels \cite{zhang2021nn}. 
We analyzed the implementation of kernel fusion available in TFLite \cite{tfliteFusion}, which we report in the Appendix (\cref{alg:kernel_fusion}); two consecutive operations of the computational graph are fused when (1)~the first operation has only one output tensor, (2) the second operation is the only operation in the graph using this output tensor, (3) uses this output tensor as its first input and produces a single output, and (4) has a compatible type.

\begin{figure}[t]
	\centering
	\begin{subfigure}[b]{.4\linewidth}
		\centering
		\includegraphics[width=\linewidth]{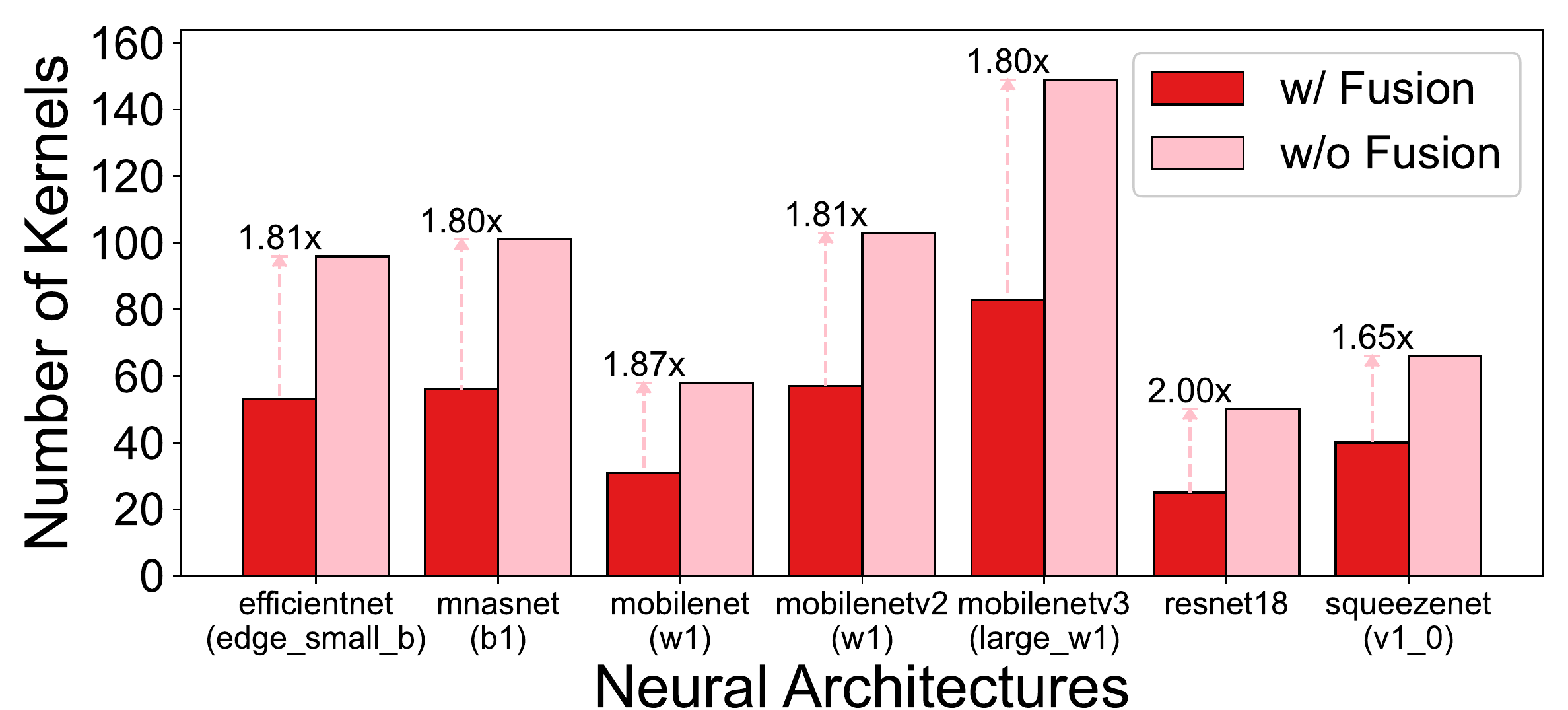}
		\caption{Number of Kernels}\label{fig:kernel_fusion_kernel_num}
	\end{subfigure}
% 	\begin{subfigure}[b]{.35\linewidth}
% 		\centering
% 		\includegraphics[width=\linewidth]{figures/performance_box_plot/kernel_fusion_performance.pdf}
% 		\caption{End-to-end Latency}\label{fig:kernel_fusion_performance}
% 	\end{subfigure}
	\begin{subfigure}[b]{.4\linewidth}
		\centering
		\includegraphics[width=\linewidth]{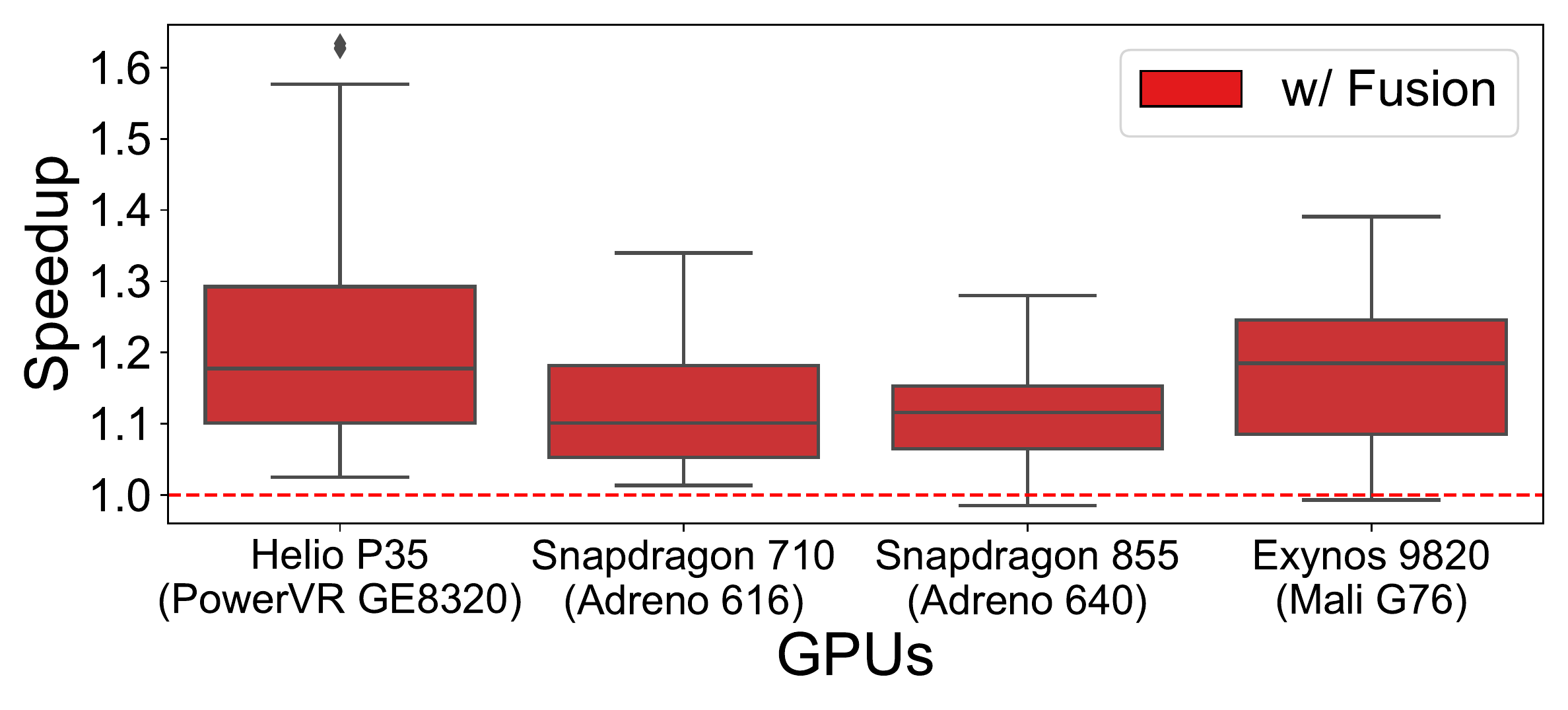}
		\caption{Latency Speedup}\label{fig:kernel_fusion_speedup}
	\end{subfigure}
	\caption{Effects of Kernel Fusion}
	\label{fig:kernel_fusion}
\end{figure}

\begin{figure}[t]
	\centering

% 	\begin{subfigure}[b]{.3\linewidth}
% 		\centering
% 		\includegraphics[width=\linewidth]{figures/performance_box_plot/kernel_fusion_ops_pixel4_outliers.pdf}
% 		\caption{Snapdragon 855 (Adreno 640)}\label{fig:kernel_fusion_ops_pixel4_1}
% 	\end{subfigure}
% 	\begin{subfigure}[b]{.3\linewidth}
% 		\centering
% 		\includegraphics[width=\linewidth]{figures/performance_box_plot/kernel_fusion_ops_a03s_outliers.pdf}
% 		\caption{Helio P35 (PowerVR GE8320)}\label{fig:kernel_fusion_ops_a03s_1}
% 	\end{subfigure}

	\begin{subfigure}[b]{.49\linewidth}
		\centering
		\includegraphics[width=\linewidth]{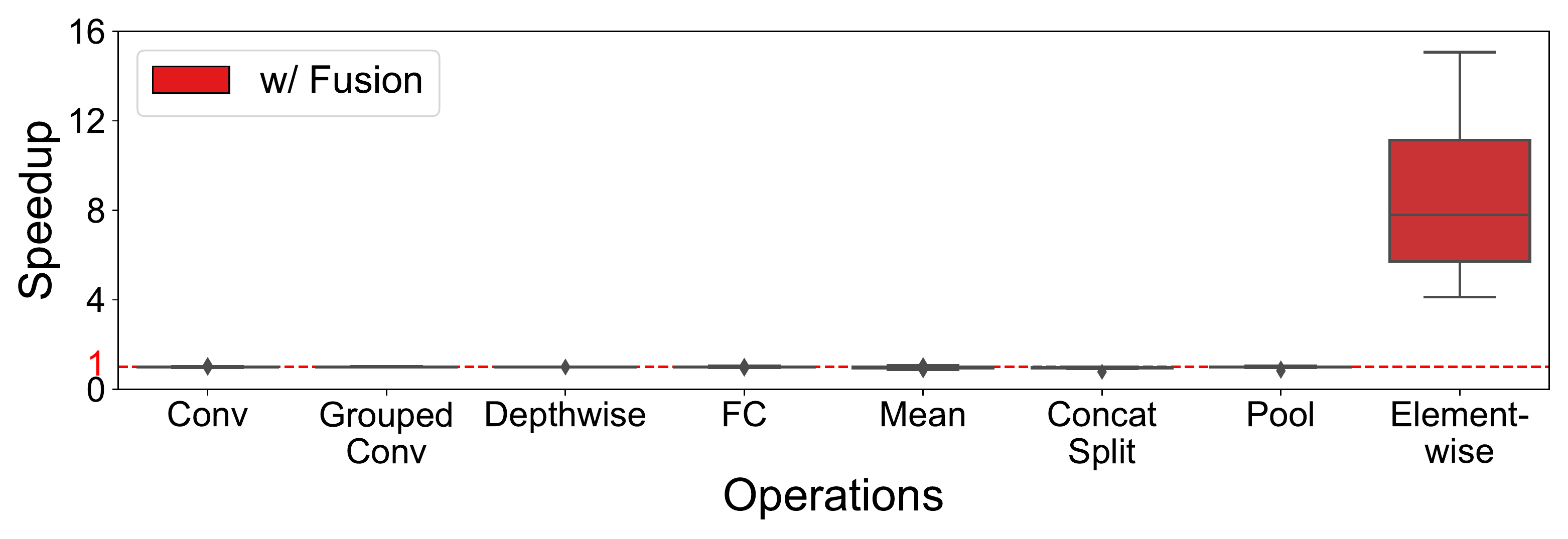}
		\caption{Snapdragon 855 (Adreno 640)}\label{fig:kernel_fusion_ops_pixel4}
	\end{subfigure}
	\begin{subfigure}[b]{.49\linewidth}
		\centering
		\includegraphics[width=\linewidth]{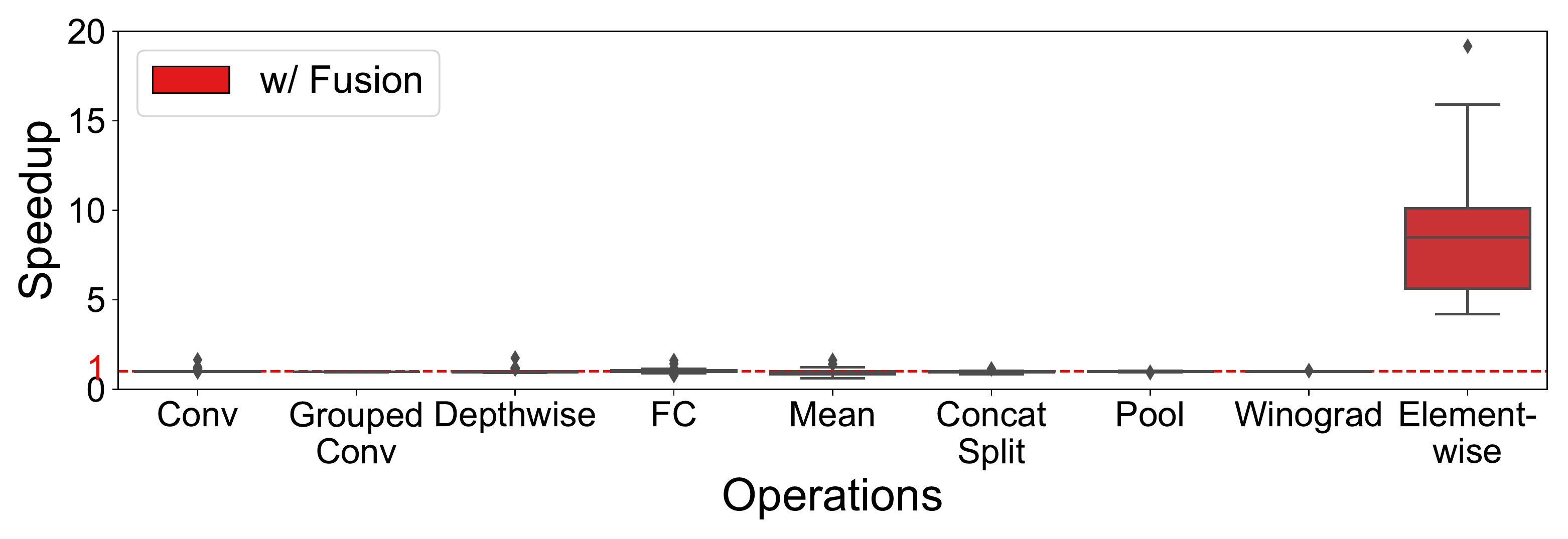}
		\caption{Helio P35 (PowerVR GE8320)}\label{fig:kernel_fusion_ops_a03s}
	\end{subfigure}

	\caption{Effects of Kernel Fusion on Operation-wise Latency}
	\label{fig:kernel_fusion_ops}
\end{figure}

\cref{fig:kernel_fusion_kernel_num} illustrates the difference in number of OpenCL kernels observed when kernel fusion is enabled: kernel fusion results in a reduction in the number of kernels of over 45\% for these state-of-the-art neural architectures. 
\cref{fig:kernel_fusion_speedup} shows the performance improvements from kernel fusion on four mobile devices; the outliers (only a couple of data points) are removed to improve visualization.
We observe 1.22x speedup of the average end-to-end latency over all the neural architectures on four mobile devices, due to a reduction in the cost of kernel dispatching.

% On the other hand, we find that kernel fusion does not significantly affect the latency of the original kernel.
%
As shown in \cref{fig:kernel_fusion_ops} (where a few outliers with large speedups on element-wise operations are reported in \cref{fig:appendix_fusion_ops} of the Appendix), kernel fusion can significantly reduce the latency of element-wise operations by merging them with other kernels; at the same time, there is no substantial latency increase for other operations after fusion. This observation is in line with Line~\ref{line:linkable_operation} in \cref{alg:kernel_fusion}: the operations fused into other operations are mainly element-wise operations.

\vspace{0.5em}
\noindent
\framebox{\parbox{\dimexpr\linewidth-2\fboxsep-2\fboxrule}{%
\textbf{Insight 3.} By substantially reducing the number of operation kernels, kernel fusion can improve the performance of inference tasks on mobile GPUs. However, only \textit{element-wise operations} provide substantial performance improvements; effect on other operations is negligible.
%%%are not affected.
%
% In our experiments, element-wise operations can achieve up to 3.26x speedup, while other operations show performance degradation less than 4\%.
}}

\subsubsection{Effects of Kernel Selection} \label{sec:kernel_selection}

\begin{figure}[t]
	\centering
% 	\begin{subfigure}[b]{.3\linewidth}
% 		\centering
% 		\includegraphics[width=\linewidth]{figures/performance/optimized_kernel_winograd_a03s.pdf}
% 		\caption{Helio P35 (PowerVR GE8320)}\label{fig:optimized_kernel_winograd_a03s}
% 	\end{subfigure}
% 	\begin{subfigure}[b]{.3\linewidth}
% 		\centering
% 		\includegraphics[width=\linewidth]{figures/performance/optimized_kernel_winograd_s10.pdf}
% 		\caption{Exynos 9820 (Mali G76)}\label{fig:optimized_kernel_winograd_s10}
% 	\end{subfigure}
	\begin{subfigure}[b]{.49\linewidth}
		\centering
		\includegraphics[width=\linewidth]{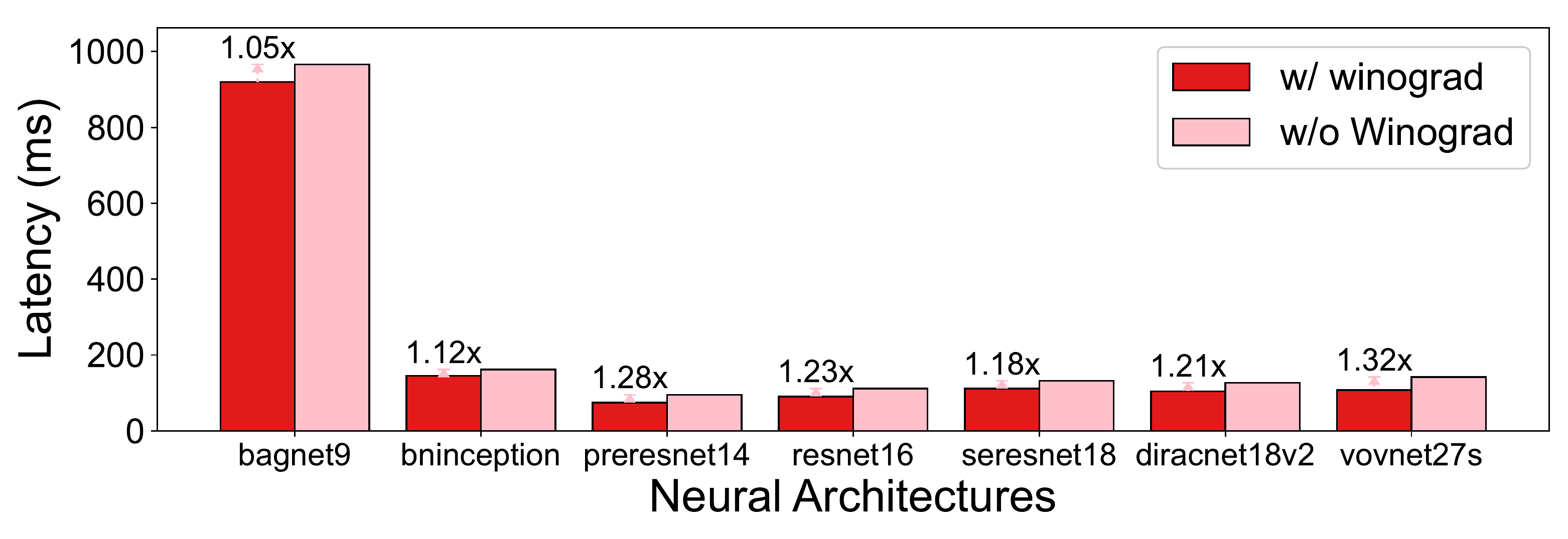}
		\caption{Helio P35 (PowerVR GE8320)}\label{fig:optimized_kernel_winograd_a03s}
	\end{subfigure}
	\begin{subfigure}[b]{.49\linewidth}
		\centering
		\includegraphics[width=\linewidth]{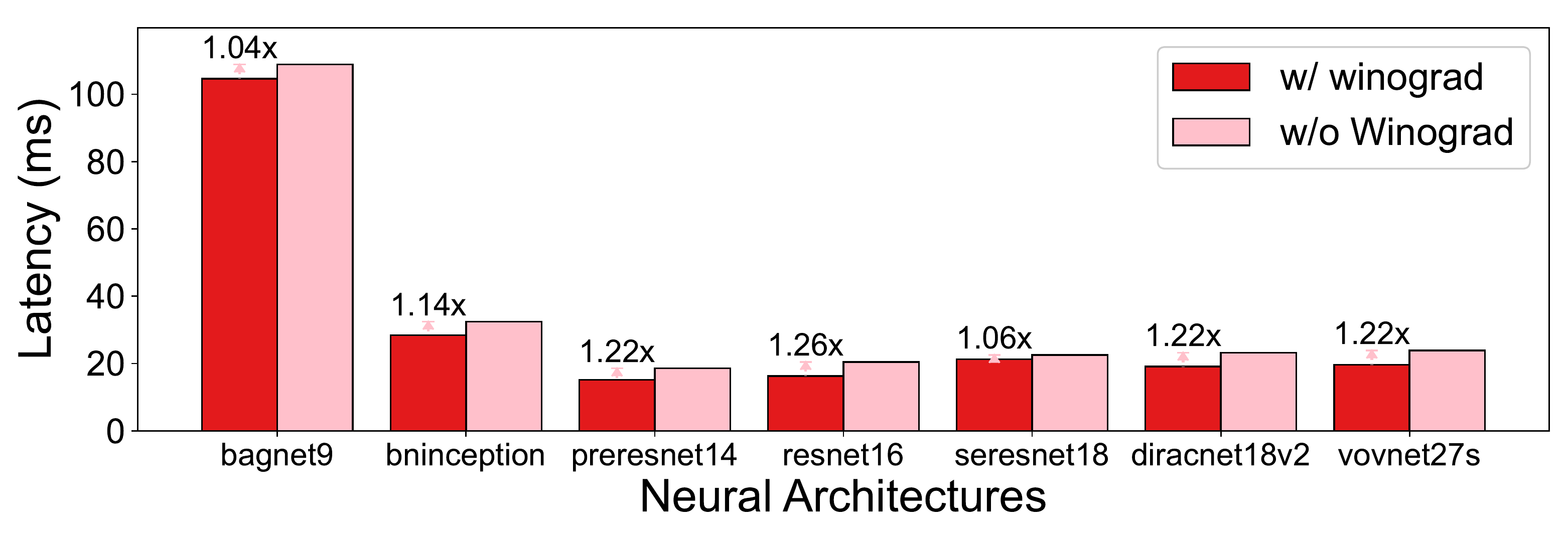}
		\caption{Exynos 9820 (Mali G76)}\label{fig:optimized_kernel_winograd_s10}
	\end{subfigure}
	\caption{Effects of Using Winograd Kernels on End-to-end Latency\vspace{-1em}}
	\label{fig:optimized_kernel_winograd}
\end{figure}

Machine learning frameworks can use different optimized algorithms to implement operations of the computational graph. For example, \cref{alg:select_conv2d_kernel} summarizes the criteria used by TFlite to enable the use of the Winograd algorithm for convolution operations: when the input tensor and kernel size of a convolution operation both satisfy certain criteria (defined by the \textsc{CheckWinograd} function), the kernel of Winograd will be selected for the operation.
\cref{fig:optimized_kernel_winograd} shows the performance improvement from using Winograd kernels in state-of-the-art neural architectures; the application of Winograd kernels results in performance improvements up to 1.32x for PowerVR GE8320 and 1.26x for Mali G76.

Notably, kernel selection is hardware-dependent. We observe that none of these neural architectures obtains performance improvements on Adreno 640 or Adreno 616, because the requirements for applying Winograd algorithm on Adreno GPUs are stricter than Mali and PowerVR GPUs in current TFLite implementation.
For example, \cref{table:winograd_condition_check} presents three convolution operations in ResNet16, which all have only one convolution group, kernel size 3x3 and stride 1. For convolution (1), src\_depth and dst\_depth fail to satisfy the conditions for Adreno GPUs (Line~\ref{line:adreno_depth_check}), but meet the requirements for Mali and PowerVR GPUs (Line~\ref{line:others_depth_check}). For convolution (2), total\_tiles is too small for Adreno 600-level GPUs (Line~\ref{line:adreno600_tiles_check}), but large enough for Mali and PowerVR GPUs (Line~\ref{line:others_tiles_check}). Convolution (3) cannot be implemented using the Winograd algorithm in either GPU because of the small total\_tiles (Line~\ref{line:others_tiles_check}).

\begin{table}[t]\centering\footnotesize
\renewcommand\arraystretch{1.1}
\begin{tabular}{c c c c c c c c c}\toprule
\multirow{2}{*}{Index} & \multicolumn{3}{c}{Configurations} & \multicolumn{3}{c}{Conditions in \cref{alg:select_conv2d_kernel}} & \multicolumn{2}{c}{If use Winograd} \\
\cmidrule(lr){2-4} \cmidrule(lr){5-7} \cmidrule(lr){8-9}
 & \makecell{Input \\ channels} & \makecell{Output \\ channels} & \makecell{Output \\ height} & src\_depth & dst\_depth & total\_tiles & Adreno & Mali \\
\toprule
(1) & 64 & 64 & 56 & 16 & 16 & 196 & No & Yes \tabularnewline
(2) & 128 & 128 & 28 & 32 & 32 & 49 & No & Yes \tabularnewline
(3) & 256 & 256 & 14 & 64 & 64 & 16 & No & No \tabularnewline
\bottomrule
\end{tabular}
\caption{Applicability of Winograd Kernels to Convolutions in ResNet16 (1 group, 3x3 kernel, stride 1)}
\label{table:winograd_condition_check}\vspace{-2em}
\end{table}

\begin{figure}[t]
	\centering
	\begin{subfigure}[b]{.49\linewidth}
		\centering
		\includegraphics[width=\linewidth]{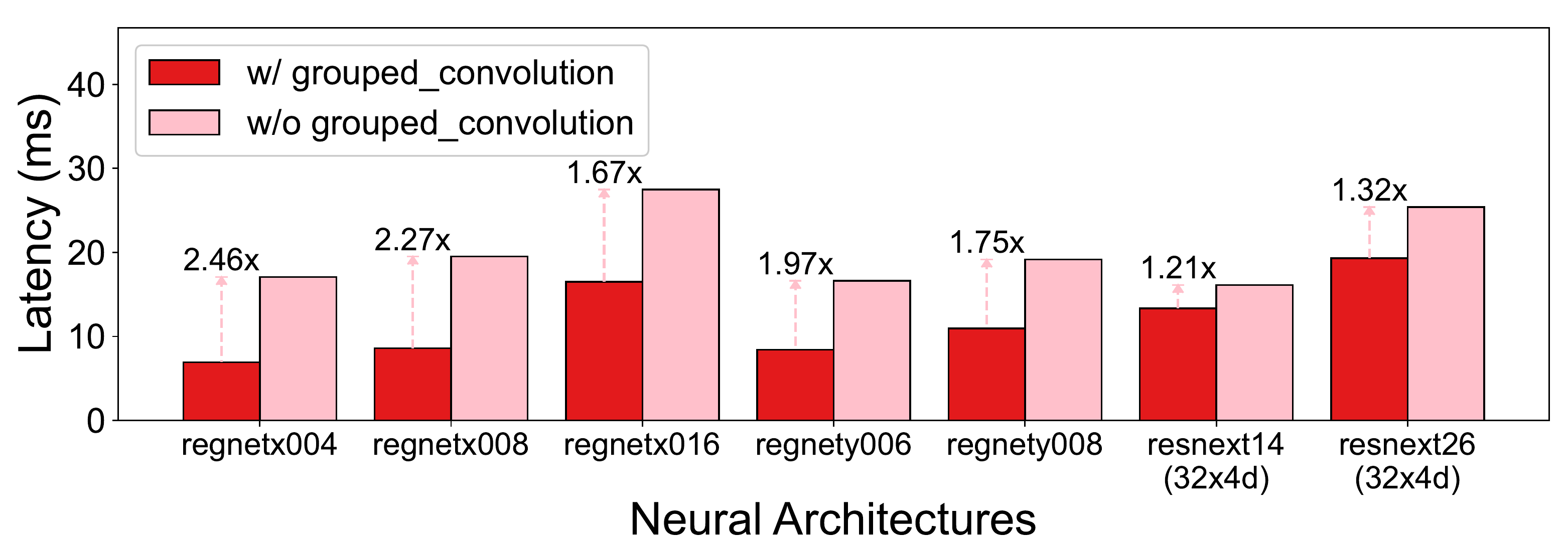}
		\caption{Snapdragon 855 (Adreno 640)}\label{fig:optimized_kernel_grouped_conv_pixel4}
	\end{subfigure}
	\begin{subfigure}[b]{.49\linewidth}
		\centering
		\includegraphics[width=\linewidth]{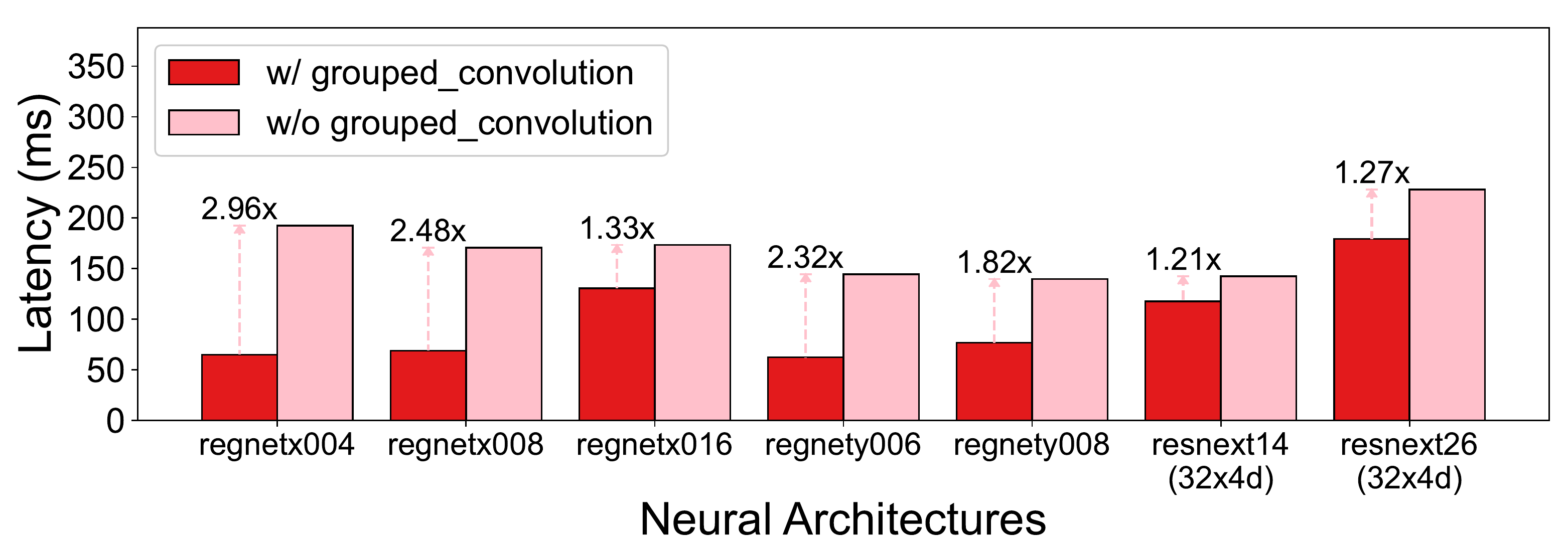}
		\caption{Helio P35 (PowerVR GE8320)}\label{fig:optimized_kernel_grouped_conv_a03s}
	\end{subfigure}
	\caption{Effects of Using \textit{grouped\_convolution\_2d} Kernels on End-to-end Latency}
	\label{fig:optimized_kernel_grouped_conv}
\end{figure}

Another operation allowing optimized implementations is grouped convolution, which consists of three stages: (1)~splitting the input tensor over channel size, (2)~performing a convolution on each resulting tensor (i.e., on each group), and (3) concatenating all output tensors. A naive implementation of grouped convolution uses an independent convolution kernel for each group, and two kernels for the split and concatenation operations. TFLite supports an optimized implementation of  \textit{grouped\_convolution\_2d} using only one kernel.
\cref{fig:optimized_kernel_grouped_conv} illustrates the performance improvement of the optimized \textit{grouped\_convolution\_2d} kernel over a naive implementation; we observe substantial improvements, e.g., 2.96x speedup for RegNetX004 on PowerVR GE8320.

\vspace{.5em}
\noindent
\framebox{\parbox{\dimexpr\linewidth-2\fboxsep-2\fboxrule}{%
\textbf{Insight 4.} Framework-dependent optimizations have significant impact on the performance of inference tasks.
In TFLite, convolution operations with certain shapes of input tensors and kernel sizes can use the Winograd algorithm to accelerate the execution; grouped convolution can make use of an optimized implementation to achieve considerable performance improvement.
Therefore, for an accurate performance prediction model, it is important to understand which kernels are executed during inference.
}}

\section{Methodology} \label{sec:methodology}

% \begin{figure}[t]
%     \centering
% 	\includegraphics[width=.6\linewidth]{figures/methodology/framework_overview.pdf}
% 	\caption{Predicting Framework Overview}
% 	\label{fig:framework_overview}\vspace{-1em}
% \end{figure}

Given a model file (e.g., a .tflite file of TFLite) generated on a cloud server (e.g., during NAS), we aim at accurately predicting the end-to-end latency over different mobile CPUs and GPUs, without deploying the neural architecture on the actual mobile devices; this framework includes the following steps:
(1) from an input model file, we first extract the information of the operations on the computational graph, which are the execution units on mobile CPUs; (2) for mobile GPUs, we deduce (without using the mobile device) the actual kernels executed after kernel fusion and kernel selection (\cref{sec:kernel_deduction}); (3) for each operation type (e.g., convolution, fully-connected), we use ML models to predict its inference latency on the target device from the operation parameters (e.g., input shape, number of channels, \cref{sec:prediction_model}); (4) end-to-end latency is estimated as the sum of predicted operation latencies plus the additional latency due to ML framework overhead.
%
% To improve the prediction model, we propose to sample training data from a NAS space (\cref{sec:data_sampling}), and implement a profiling tool to accurately measure the latency of each kernel (\cref{sec:profiling}).
To train the ML models and to evaluate our approach, we collect latency measurements on a synthetic dataset including 1000 neural architectures from a NAS space (\cref{sec:synthetic_dataset}), which we will make publicly available to help further research on mobile performance.

\subsection{Kernel Deduction} \label{sec:kernel_deduction}
From the model file, we are able to extract the computational graph of the target neural architecture, which includes information on all operations as well as the data flow between operations. 
As discussed in \cref{sec:multithreading}, these operations are sequentially executed on mobile CPUs and multiple threads can collaborate on the computation of each operation. Hence, for each type of operation and each CPU core combination, we train a machine learning model to predict inference latency.

However, when using mobile GPUs, operations of the computational graph can be fused (\cref{sec:kernel_fusion}) or implemented with optimized algorithms (\cref{sec:kernel_selection}); since our measurements illustrate that kernel fusion and kernel selection have substantial effects on performance, identifying which kernels are actually executed for a given computational graph and target device is critical to obtaining accurate latency predictions.
To avoid the cost of deploying the neural architectures on the physical device (which is impractical for NAS with a huge number of candidate neural architectures), we deduce the kernels executed on a device by simulating the process of kernel fusion and kernel selection, according the principles elicited from the implementation of TFLite.
Specifically, to predict latency on mobile GPUs, we first fuse kernels according to the rules presented in \cref{sec:kernel_fusion}; then, we use the rules presented in \cref{sec:kernel_selection} to select a kernel among $\{\text{Conv2D}, \text{Winograd}, \text{GroupedConv2D}\}$ based on the parameters of each convolution operation (e.g., input size, output size, kernel size) and the target device.

\subsection{Prediction Models} \label{sec:prediction_model}

For each neural architecture, after obtaining the computational graph (and applying kernel fusion/selection estimation for mobile GPUs), we predict the latency of each operation from its configuration parameters.
We use parameters that define the shape of an operation augmented with features associated with both memory access cost (e.g., size of input/output data, parameters) and computational cost (e.g., FLOPs).
Feature space details are given in the Appendix in \cref{table:feature}.
%%%summarizes the features affecting latency of each operation type; we include features associated with both memory access cost (e.g., size of input data, output data and parameters) and computational cost (e.g., FLOPs).

Formally, given feature vectors $\bm{x}_i \in X$ of an operation of the computational graph and corresponding latencies $y_i \in Y$ measured on a specific device, for $i=1\,\dots,N$, where $N$ is the size of the training dataset\footnote{
Technically, each operation has a different training set size, but for clarity of presentation, we use $N$ as a generic size.},
our goal is to train a prediction model $f$ minimizing the mean absolute percentage error (MAPE) $L_{MAPE} = \frac{1}{N} \sum_{i=1}^{N} \left|(f(\bm{x}_i) - y_i)/y_i\right|$.
Since input features can be of different magnitudes, we standardize each feature $j$ based on its mean $\mu_j = (\sum_{i=1}^N x_{i,j})/N$ and standard deviation $\sigma_j = \sqrt{[\sum_{i=1}^N (x_{i,j}-\mu_j)^2]/N}$ in the training set, i.e., ${\hat x}_{i,j} = (x_{i,j} - \mu_j)/\sigma_j$ for all $1 \le i \le N$,
and we minimize the mean square percentage error based on standardized feature vectors $\bm{{\hat x}_i}$ and latencies~$y_i$, $\forall 1 \le i \le N$:
$f^* = \min_{f} \frac{1}{N} \sum_{i=1}^{N} \left|(f(\bm{{\hat x}_i}) - y_i)/y_i \right|^2$. To develop a prediction model, we consider the following representative ML approaches \cite{murphy2012machine} adopted in the literature \cite{bouzidi2021performance,bouhali2021execution,zhang2021nn,geoffrey2021habitat,justus2018predicting}; as will be shown in our evaluations (\cref{sec:result_framework_heterogeneity}), properly accounting for characteristics we identified as significant to end-to-end latency (\cref{sec:background_gpu_performance}), results in similar prediction accuracy across these ML approaches.

\paragraph{Lasso}
We first consider a linear model $f(\bm{x}) = \bm{w}^T \bm{x}$ and estimate the optimal weights $\bm{w}^*$ as
\begin{align}
\bm{w}^* = \min_{\bm{w}} & \quad \frac{1}{N} \sum_{i=1}^{N} \left| \frac{\bm{w}^T \bm{{\hat x}_i} - y_i}{y_i} \right|^2 + \alpha \norm{\bm{w}}_1 \qquad\textrm{subject to} \quad \bm{w} \ge 0\,. \label{eq:constraint}
\end{align}
An L1 regularization term with hyperparameter $\alpha$ is included to control model complexity and to favor a sparse solution.
We use grid search in $[10^{-5},10^2]$ to find the best $\alpha$.
Since each input feature ${\hat x}_{i,j}$ is positively correlated with latency, we constrain weights $w_{i,j}$ to be nonnegative (\cref{eq:constraint}).

\paragraph{Random Forest (RF)} An RF model includes multiple decision trees to reduce overfitting of a single decision tree. We tune hyperparameters including the number of decision trees (1 to 10) and the minimum number of samples to split an internal node (2 to 50) using 5-fold cross-validation. 

\paragraph{Gradient-Boosted Decision Trees (GBDT)} GBDT generates decision trees with gradient boosting on multiple stages. We tune hyperparameters including the number of gradient boosting stages (1~to 200) and the number of examples required to split a node (2 to 7) using 5-fold cross-validation.

\paragraph{Multi-Layer Perceptron (MLP)} An MLP consists of multiple layers of fully-connected layers. We tune the hyperparameters for the number of layers from 1 to 6 and the number of neurons in each layer from $\{64, 128, 256, 512\}$.
Similarly to previous work \cite{geoffrey2021habitat}, we use ReLU activations after each layer and the Adam optimizer with learning rate from $\{5 \times 10^{-3}, 5 \times 10^{-4}, 5 \times 10^{-5}\}$, and weight decay from $\{10^{-3}, 10^{-4}, 10^{-5}\}$.
We use 20\% of training data as the validation set, and stop training after no improvement on the validation error over 50 epochs.

\begin{figure}[t]
    \captionsetup{justification=centering}
	\centering
	\begin{subfigure}[b]{.37\linewidth}
		\centering
		\includegraphics[width=\linewidth]{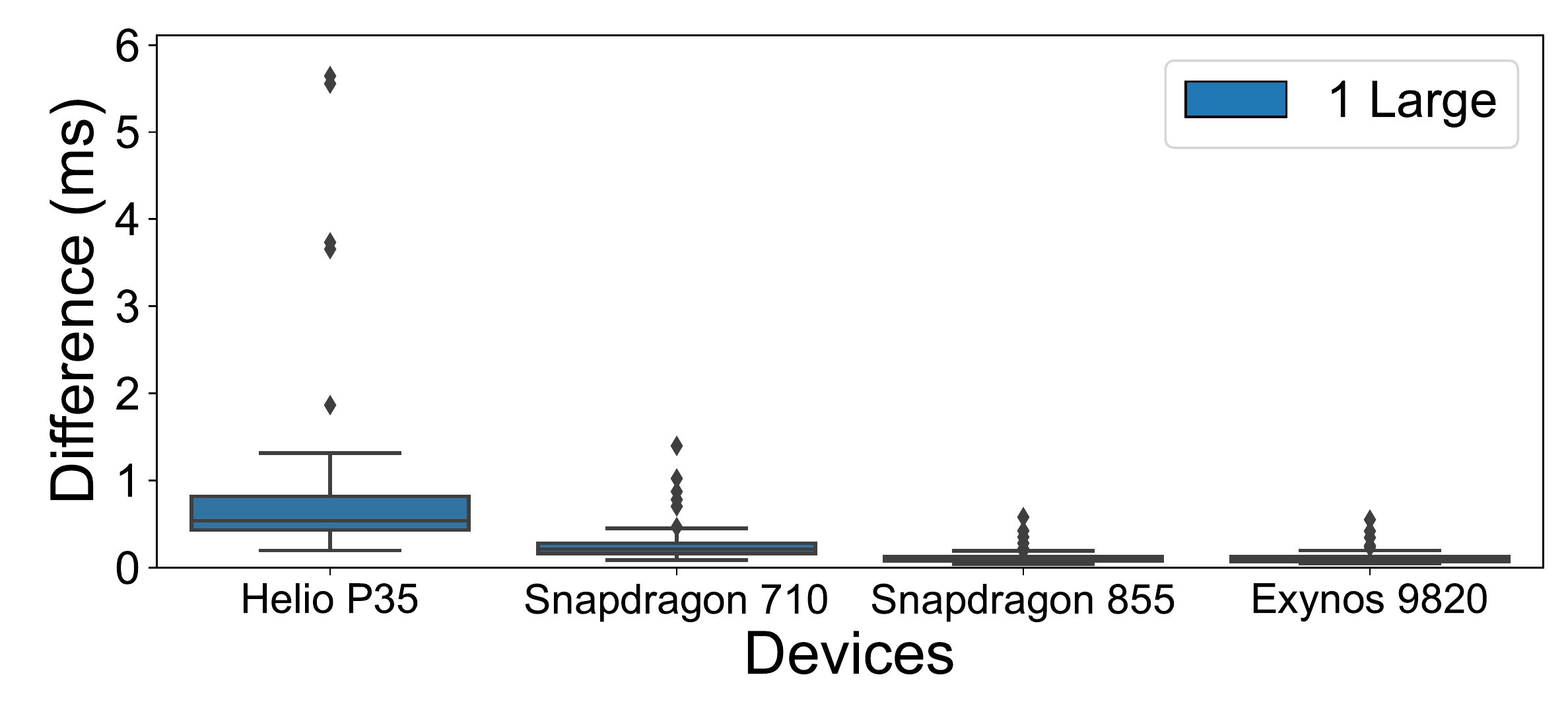}
		\caption{CPUs}\label{fig:cpu_overhead}
	\end{subfigure}
	\begin{subfigure}[b]{.37\linewidth}
		\centering
		\includegraphics[width=\linewidth]{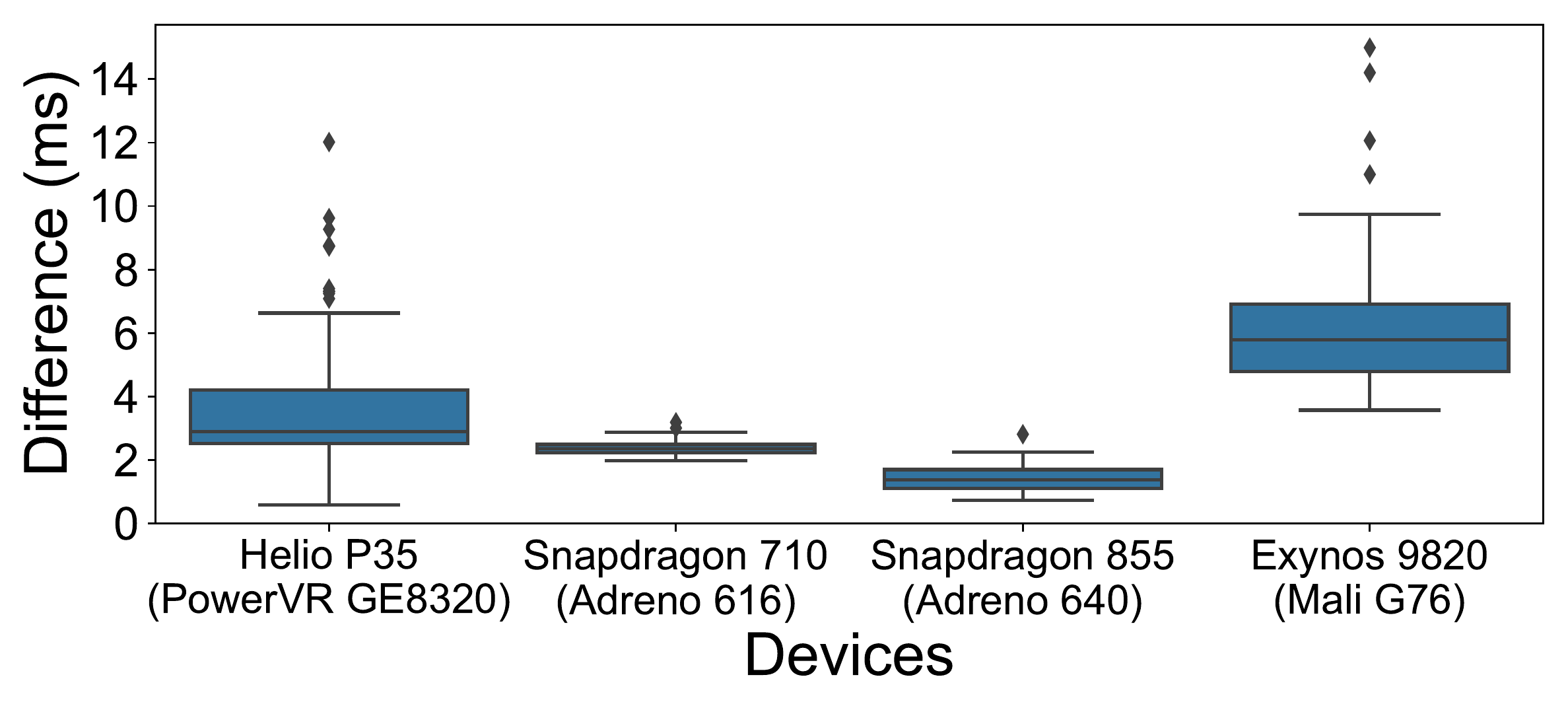}
		\caption{GPUs}\label{fig:gpu_overhead}
	\end{subfigure}
	\caption{Difference between End-to-end Latency and Sum of Operation-wise/Kernel-wise Latency for State-of-the-art Neural Architectures}\label{fig:overhead}\vspace{-1em}
\end{figure}

\smallskip
After predicting execution latency of each operation on CPU cores or on the GPU, we account for additional latency due to overhead and data transfers in TFLite; as shown in \cref{fig:overhead}, the sum of the latencies measured for all operations is consistently lower than the measured end-to-end latency, especially on GPUs (\cref{fig:gpu_overhead}).
Since the difference fluctuates around a constant value for all neural architectures on a specific GPU, we use the average difference between end-to-end latencies and operation-wise latencies in the training dataset to estimate this additional latency $T_{\text{overhead}}$.
Formally, for a neural architecture with the set of operations $C$, we predict end-to-end latency as
$T_{\text{overhead}} + \sum_{c \in C} f^*_c(\bm{{\hat x}_c})$
where $f^*_c$ is the latency predictor trained from measurements of operations with the same type as $c$.

\subsection{Synthetic Dataset} \label{sec:synthetic_dataset}

Next, we present our synthetic dataset consisting of neural architectures from a NAS space, which covers a broad range of operations and building blocks in recent work. We first introduce the technique used to profile each operation, and then we describe the design of the NAS space.

\subsubsection{Profiling Kernel Latency} \label{sec:profiling}

% Profiling: OpenCL command queues enable profiling
% 
% Adreno: Less profiling overhead
% Mali, PowerVR: Substantial profiling overhead
% 
% Solution (inspired by TFLite implementation): Dispatching a kernel K times (K=256), and only profiling the first and last events.

For mobile CPUs, we utilize \textit{TFLite Model Benchmark Tool} \cite{tflitebenchmarktool} to measure the latency of each operation.
However, the tool currently provides no official interface to profile kernels on mobile GPUs.
As a solution, we record the timestamps of each OpenCL kernel by enabling profiling information collection at the OpenCL command queue.
To reduce the overhead of recording timestamps, we dispatch the same kernel multiple times (specifically, 256) and record only the timestamps for the first and last events, thus amortizing profiling overhead.

% \begin{figure}[t]
% \captionsetup{justification=centering}
% \begin{minipage}[t]{0.495\linewidth}
% 	\centering
% 	\begin{subfigure}[b]{.494\linewidth}
% 		\centering
% 		\includegraphics[width=\linewidth]{figures/performance_box_plot/latency_breakdown_common_cpu_pixel4_outliers.pdf}
% 		\caption{Snapdragon 855 \newline (1 Large Core)}\label{fig:latency_breakdown_common_cpu_pixel4}
% 	\end{subfigure}
% 	\begin{subfigure}[b]{.494\linewidth}
% 		\centering
% 		\includegraphics[width=\linewidth]{figures/performance_box_plot/latency_breakdown_common_cpu_s10_outliers.pdf}
% 		\caption{Exynos 9820 \newline (1 Large Core)}\label{fig:latency_breakdown_common_cpu_s10}
% 	\end{subfigure}
% 	\caption{Latency Breakdown of State-of-the-art Neural Architectures on CPUs}
% 	\label{fig:latency_breakdown_common_cpu}
% \end{minipage}
% \hfill
% \begin{minipage}[t]{0.495\linewidth}
% 	\centering
% 	\begin{subfigure}[b]{.494\linewidth}
% 		\centering
% 		\includegraphics[width=\linewidth]{figures/performance_box_plot/latency_breakdown_common_gpu_pixel4_outliers.pdf}
% 		\caption{Snapdragon 855 \newline (Adreno 640)}\label{fig:latency_breakdown_common_gpu_pixel4}
% 	\end{subfigure}
% 	\begin{subfigure}[b]{.494\linewidth}
% 		\centering
% 		\includegraphics[width=\linewidth]{figures/performance_box_plot/latency_breakdown_common_gpu_s10_outliers.pdf}
% 		\caption{Exynos 9820 \newline (Mali G76)}\label{fig:latency_breakdown_common_gpu_s10}
% 	\end{subfigure}
% 	\caption{Latency Breakdown of State-of-the-art Neural Architectures on GPUs}
% 	\label{fig:latency_breakdown_common_gpu}
% \end{minipage}
% \end{figure}

\begin{figure}[t]
	\centering
	\begin{subfigure}[b]{.49\linewidth}
		\centering
		\includegraphics[width=\linewidth]{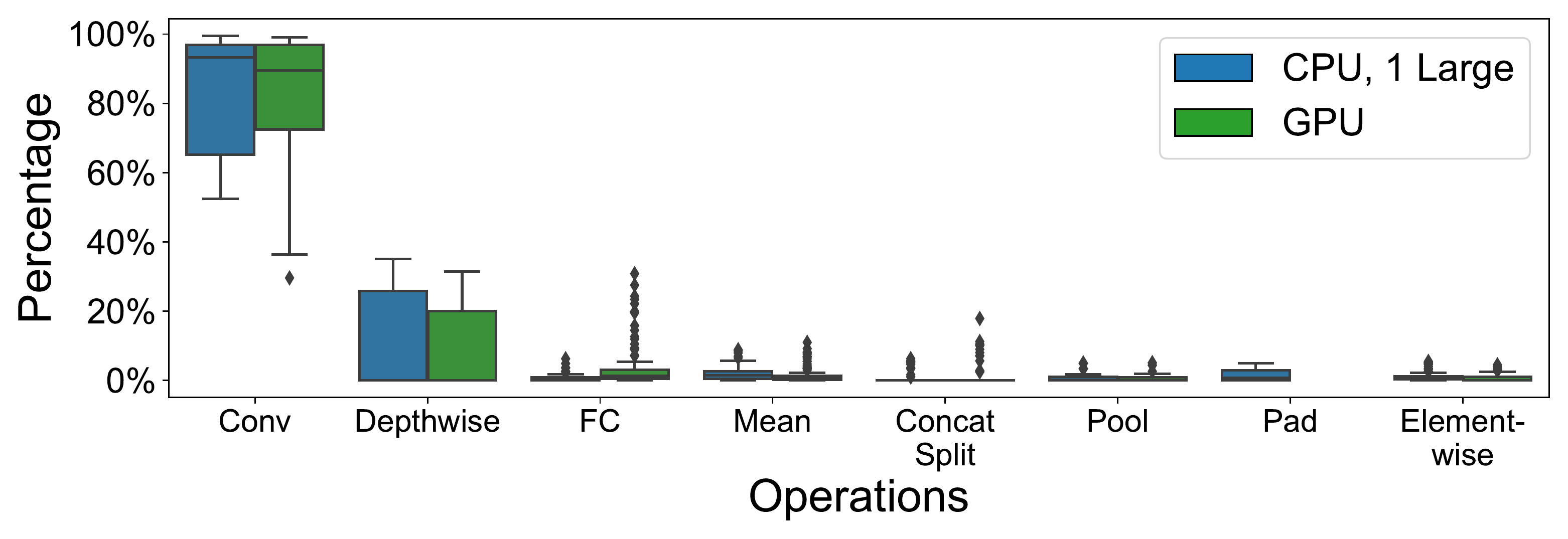}
		\caption{Snapdragon 855 (Adreno 640)}\label{fig:latency_breakdown_common_pixel4}
	\end{subfigure}
	\begin{subfigure}[b]{.49\linewidth}
		\centering
		\includegraphics[width=\linewidth]{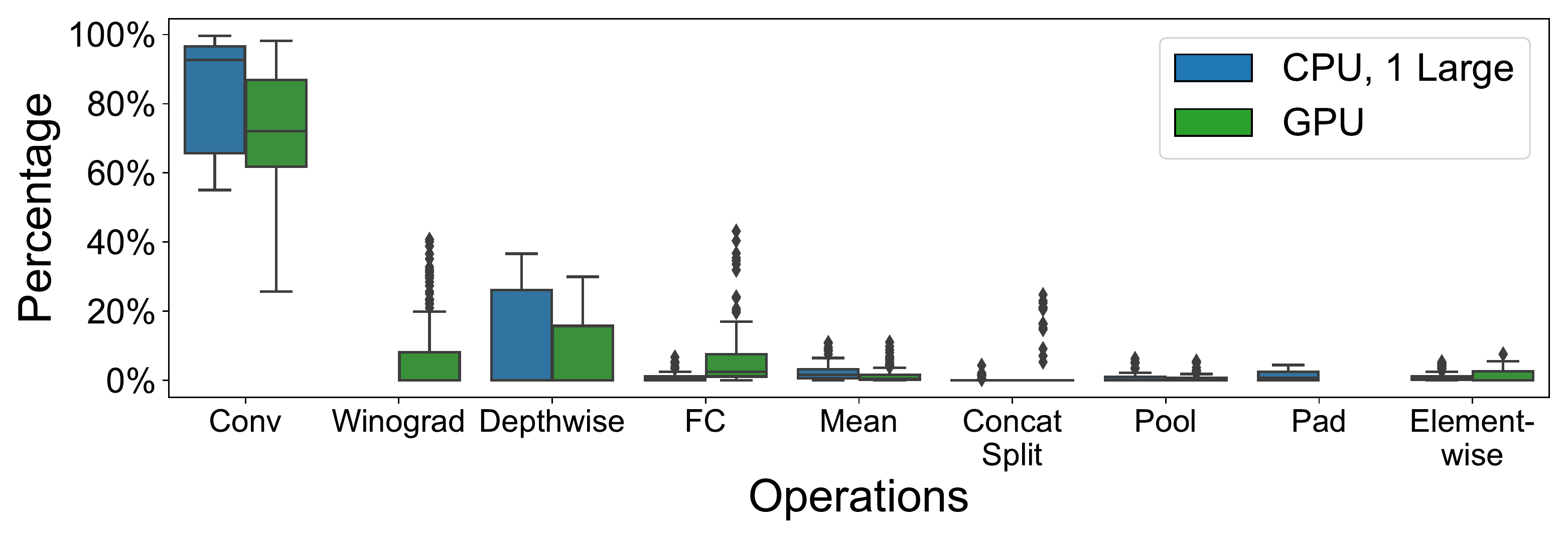}
		\caption{Exynos 9820 (Mali G76)}\label{fig:latency_breakdown_common_s10}
	\end{subfigure}
	\caption{Latency Breakdown of State-of-the-art Neural Architectures}
	\label{fig:latency_breakdown_common}\vspace{-0.5em}
\end{figure}

\cref{fig:latency_breakdown_common} display the average latency breakdown for 102 state-of-the-art neural architectures\footnote{When presenting the percentage of end-to-end latency, we include the results of NAs that may not have all type of operations; e.g., depthwise convolution operations only appear in 58 NAs, so its median across 102 NAs is zero.}.
As can be seen, convolution and depthwise convolution operations account for a significant proportion of the end-to-end latency.
In addition, we observe that, for the same set of neural architectures in our dataset, Winograd kernels are applied on Mali G76 but not on Adreno 640, because the selection of kernels is dependent on the hardware platform, as discussed in \cref{sec:kernel_selection}.

\subsubsection{NAS Space for Sampling Neural Architectures} \label{sec:nas_space}

\begin{figure}[t]
    \centering
	\includegraphics[width=.5\linewidth]{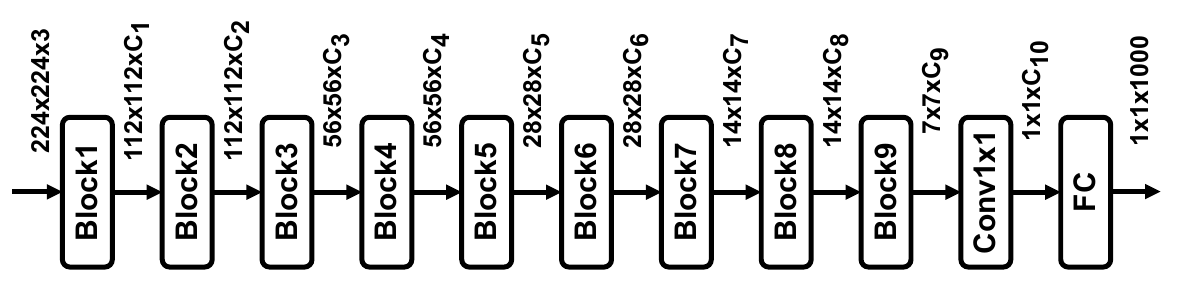}\vspace{-0.8em}
	\caption{Design of the NAS Space for Synthetic Dataset\label{fig:nas_space}\vspace{-1em}}
\end{figure}

\cref{fig:latency_breakdown_common} highlights the importance of studying the performance of convolution and depthwise convolution operations. Consequently, we design a search space to effectively sample various configurations of these operations for the purpose of understanding their performance characteristics.
As illustrated in \cref{fig:nas_space}, synthetic neural architectures of our NAS space use a sequence of 9 blocks halving input width/height after blocks 1, 3, 5, 7, 9; then a convolution with kernel size 1x1 and fully-connected layer produce a output vector of 1000 dimension. The type and parameters of each building block are selected uniformly at random as:
\begin{enumerate}
    \item A convolution layer (with kernel size 3x3, 5x5 or 7x7, optionally group size $4k, \forall 1 \le k \le 16$). 
    \item Depthwise separable convolution \cite{howard2017mobilenets} (with kernel size 3x3, 5x5 or 7x7).
    \item Linear bottleneck \cite{sandler2018mobilenetv2} (with kernel size 3x3, 5x5 or 7x7, expansion rate 1, 3 or 6, optionally including Squeeze-and-Excite as \cite{howard2019searching}).
    \item Average or maximum pooling layer (with pooling size 1x1 or 3x3).
    \item A split layer (with number of splits 2, 3 or 4), followed by element-wise operations performed on each output tensor, and a concatenation layer which merges all output tensors.
\end{enumerate}
Due to the limited memory and computing resources on mobile devices, we sample the output channel sizes of these building blocks (identified as $C_1$ to $C_9$)  with the following constraints:
$\{C_1, ... C_5\}$ are uniformly sampled from $[8, 80]$; $\{C_6, ... C_9\}$ are uniformly sampled from $[80, 400]$; $C_{10}$ is uniformly sampled from $[1200, 1800]$.

% In addition, number of channels $C_i\ (1 \le i \le 9)$ are sampled from \cref{alg:channel_sampling}.
% \begin{algorithm}
% \begin{pseudo}[fullwidth,indent-length=1.1em,font=\small]*
% \toprule
% \hd{SamplingChannels}()\\
% [bol=\midrule]
% $v = []$ \\
% \kw{for} $i=1$ to $5$\\+
%     $v[i]$ = \textsc{Uniform}(8, 80)\quad\ct{Sample from a uniform distribution}\\-
% \kw{for} $i=6$ to $9$\\+
%     $v[i]$ = \textsc{Uniform}(80, 400)\quad\ct{Sample from a uniform distribution}\\-
% $v[10]$ = \textsc{Uniform}(1200, 1800)\quad\ct{Sample from a uniform distribution}\\-
% \kw{return} \textsc{Sort}($v$)\quad\ct{Sort from smallest to largest}\\*
% \bottomrule
% \end{pseudo}
% \caption{Sample Channels} \label{alg:channel_sampling}
% \end{algorithm}

\begin{figure}[t]
	\centering
	\begin{subfigure}[b]{.49\linewidth}
		\centering
		\includegraphics[width=\linewidth]{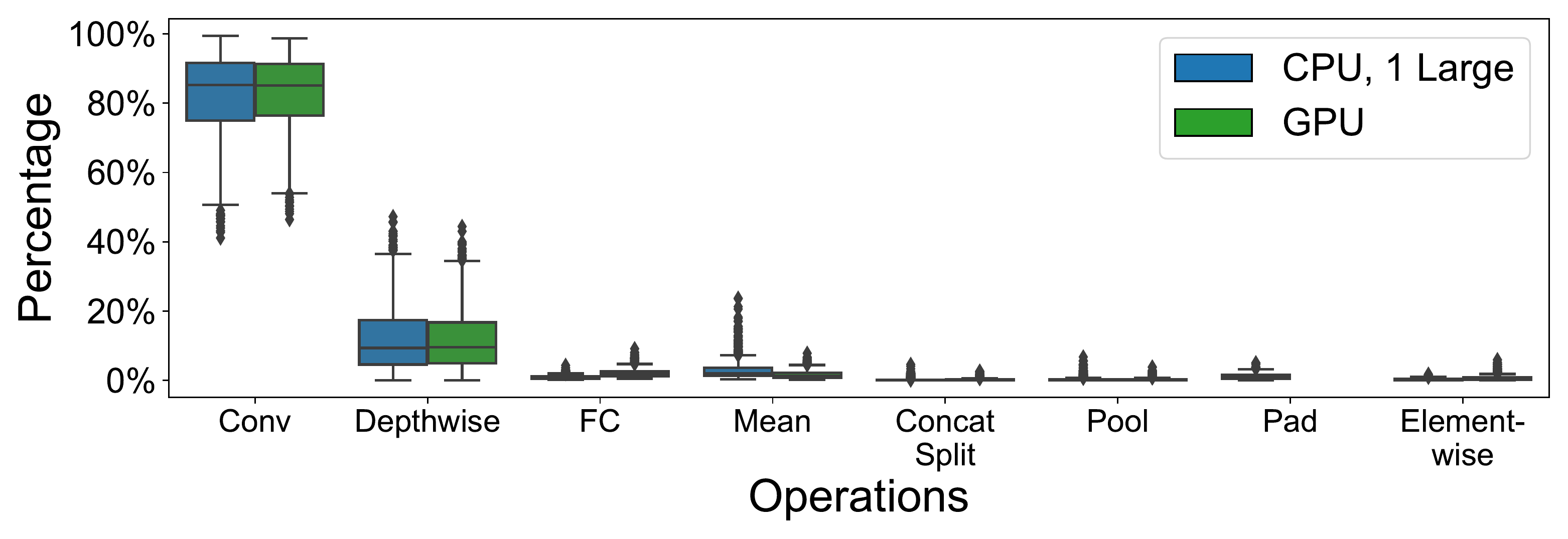}
		\caption{Snapdragon 855 (Adreno 640)}\label{fig:latency_breakdown_nas_pixel4}
	\end{subfigure}
	\begin{subfigure}[b]{.49\linewidth}
		\centering
		\includegraphics[width=\linewidth]{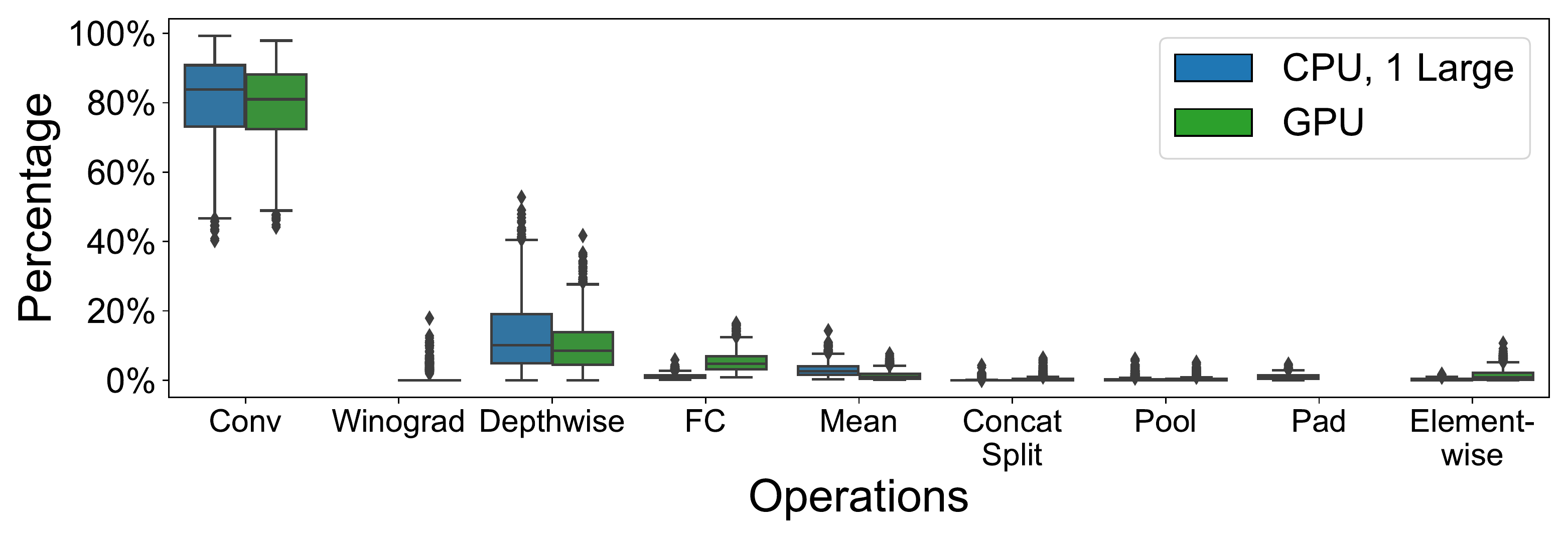}
		\caption{Exynos 9820 (Mali G76)}\label{fig:latency_breakdown_nas_s10}
	\end{subfigure}
	\caption{Latency Breakdown of Synthetic Neural Architectures}\vspace{-1em}
	\label{fig:latency_breakdown_nas}
\end{figure}

We adopt a synthetic dataset including \emph{1000 neural architectures} sampled from this NAS space. For each neural architecture, we collect training measurements on \emph{4 mobile platforms} in \cref{table:mobile_platforms}, for a total of \emph{72 scenarios}, covering (1) combinations of homogeneous or heterogeneous cores, (2) 32-bit floating point and 8-bit integer representations, and (3) mobile GPUs from different manufacturers. \cref{fig:latency_breakdown_nas} illustrate the latency breakdown for neural architectures in our synthetic dataset; the latency distribution over different operations is similar to state-of-the-art neural architectures in \cref{fig:latency_breakdown_common}.

% \begin{figure}[t]
% \captionsetup{justification=centering}
% \begin{minipage}[t]{0.495\linewidth}
% 	\centering
% 	\begin{subfigure}[b]{.494\linewidth}
% 		\centering
% 		\includegraphics[width=\linewidth]{figures/performance_box_plot/latency_breakdown_nas_cpu_pixel4_outliers.pdf}
% 		\caption{Snapdragon 855 \newline (1 Large Core)}\label{fig:latency_breakdown_nas_cpu_pixel4}
% 	\end{subfigure}
% 	\begin{subfigure}[b]{.494\linewidth}
% 		\centering
% 		\includegraphics[width=\linewidth]{figures/performance_box_plot/latency_breakdown_nas_cpu_s10_outliers.pdf}
% 		\caption{Exynos 9820 \newline (1 Large Core)}\label{fig:latency_breakdown_nas_cpu_s10}
% 	\end{subfigure}
% 	\caption{Latency Breakdown of Synthetic Neural Architectures on CPUs}
% 	\label{fig:latency_breakdown_nas_cpu}
% \end{minipage}
% \hfill
% \begin{minipage}[t]{0.495\linewidth}
% 	\centering
% 	\begin{subfigure}[b]{.494\linewidth}
% 		\centering
% 		\includegraphics[width=\linewidth]{figures/performance_box_plot/latency_breakdown_nas_gpu_pixel4_outliers.pdf}
% 		\caption{Snapdragon 855 \newline (Adreno 640)}\label{fig:latency_breakdown_nas_gpu_pixel4}
% 	\end{subfigure}
% 	\begin{subfigure}[b]{.494\linewidth}
% 		\centering
% 		\includegraphics[width=\linewidth]{figures/performance_box_plot/latency_breakdown_nas_gpu_s10_outliers.pdf}
% 		\caption{Exynos 9820 \newline (Mali G76)}\label{fig:latency_breakdown_nas_gpu_s10}
% 	\end{subfigure}
% 	\caption{Latency Breakdown of Synthetic Neural Architectures on GPUs}
% 	\label{fig:latency_breakdown_nas_gpu}
% \end{minipage}\vspace{-1em}
% \end{figure}

\section{Results} \label{sec:result}

This section presents a comprehensive evaluation of our latency prediction framework across a broad range of scenarios: first, we present results on the default setting of NAS (\cref{sec:result_default_setting}), and then we evaluate the impact of hardware heterogeneity (\cref{sec:result_hardware_heterogeneity}), neural architecture diversity (\cref{sec:result_neural_architecture_heterogeneity}), and ML framework optimizations (\cref{sec:result_framework_heterogeneity}). In addition, to address a common criticism of cost of training data collection, we present results using a small number of training examples and a simple linear model (\cref{sec:result_limited_training_data}).

\subsection{Default Setting: Evaluation on Neural Architectures from NAS Space}\label{sec:result_default_setting}

We first test our framework in a common scenario of applying our latency prediction model during NAS: we sample test neural architectures (the candidate architectures during NAS) and the training neural architectures (the profiling architectures to train our latency prediction model) uniformly at random from the same search space (\cref{sec:nas_space}).
These sampled neural architectures constitute our synthetic dataset of 1000 samples. Here, 900 of these are used for training and 100 for testing.

\begin{figure}[t]
	\centering
	\begin{subfigure}[b]{.49\linewidth}
		\centering
		\includegraphics[width=\linewidth]{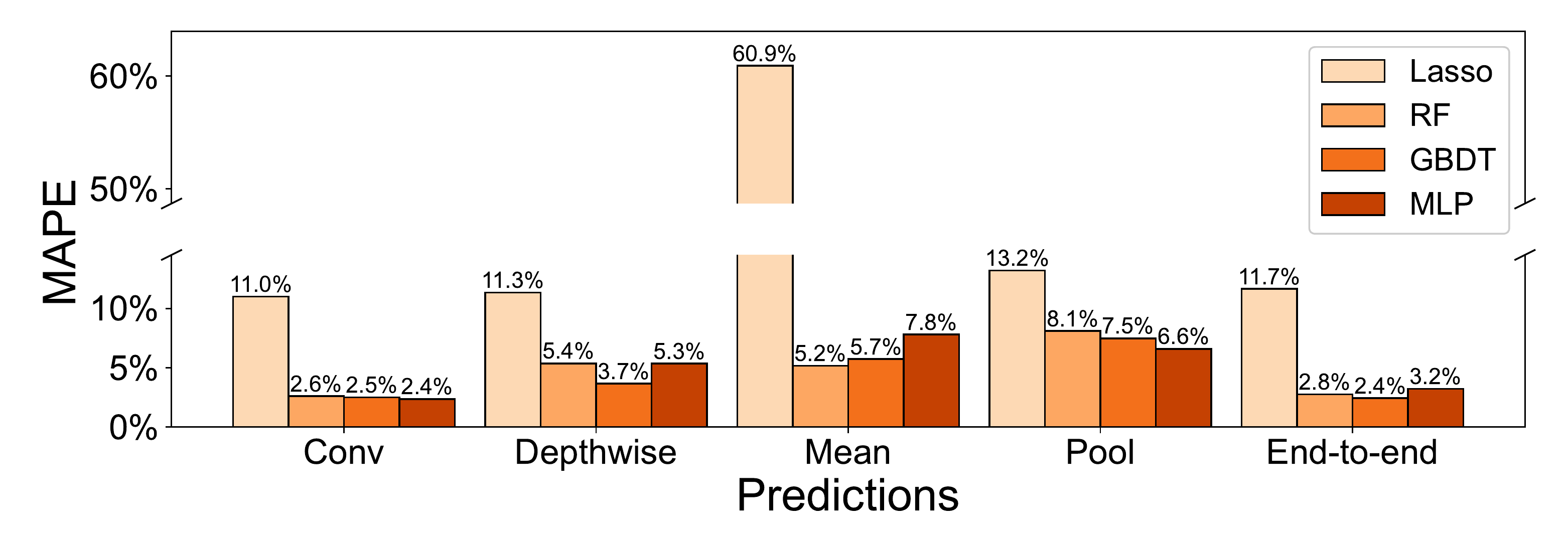}
		\caption{CPU (One Large Core)}\label{fig:comparison_nas_900_cpu_all}
	\end{subfigure}
	\begin{subfigure}[b]{.49\linewidth}
		\centering
		\includegraphics[width=\linewidth]{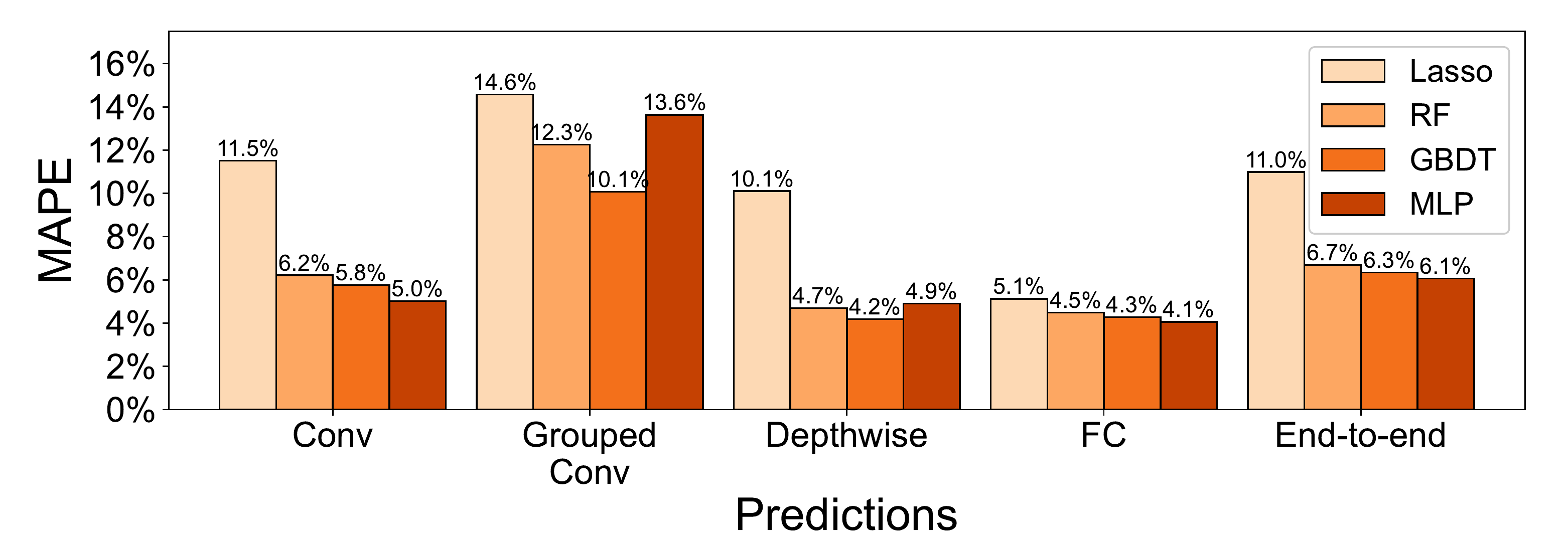}
		\caption{GPU}\label{fig:comparison_nas_900_gpu_all}
	\end{subfigure}
	\caption{Latency Predictions of Different ML Models (Synthetic Neural Architectures)}
	\label{fig:comparison_nas_900}\vspace{-1em}
\end{figure}

\cref{fig:comparison_nas_900} presents average (across 4 platforms) MAPE under different ML approaches when predicting end-to-end latency, as well as latency of the 4 operation types accounting for most of end-to-end latency (convolution, depthwise convolution, mean, pooling); due to lack of space, MAPE of each platform is reported in \cref{table:end_to_end_predictions_real_world} of the Appendix. 
Based on the latency breakdown of synthetic neural architectures on CPUs and GPUs (\cref{fig:latency_breakdown_nas}), convolution operations typically account for the largest proportion of end-to-end latency; consequently, prediction error of convolution dominates the error of end-to-end latency prediction for all four ML approaches on both CPUs and GPUs. For example, Lasso has a large MAPE (60.9\%) on mean operations on CPU, while its MAPE for end-to-end latency is only 11.7\%, because, as shown in \cref{fig:comparison_nas_900_cpu_all}, 75\% of mean operations contribute to less than 4.9\% of the end-to-end latency (on the platforms in \cref{fig:latency_breakdown_nas}).

As can be seen, in our default setting, all nonlinear ML approaches (RF, GDBT, MLP) achieve comparable accuracy on end-to-end latency predictions, with average MAPE across four platforms below 3.2\% for CPU predictions and below 6.7\% for GPU predictions;
Lasso achieves less accurate predictions (11.7\% on CPUs and 11.0\% on GPUs), because its linear model cannot represent non-linear relationships between latency and operation features, as identified by previous work \cite{tang2021bridge,zhang2021nn}.

\subsection{Case Study: Hardware Heterogeneity}\label{sec:result_hardware_heterogeneity}

% \begin{figure}[t]
% 	\centering
% 	\begin{subfigure}[b]{.49\linewidth}
% 		\centering
% 		\includegraphics[width=\linewidth]{figures/results/nas_cpu_s10_conv_mape.pdf}
% 		\caption{Exynos 9820}\label{fig:nas_cpu_s10_conv_mape}
% 	\end{subfigure}
% 	\begin{subfigure}[b]{.49\linewidth}
% 		\centering
% 		\includegraphics[width=\linewidth]{figures/results/nas_cpu_pixel4_conv_mape.pdf}
% 		\caption{Snapdragon 855}\label{fig:nas_cpu_pixel4_conv_mape}
% 	\end{subfigure}
% 	\begin{subfigure}[b]{.49\linewidth}
% 		\centering
% 		\includegraphics[width=\linewidth]{figures/results/nas_cpu_a03s_conv_mape.pdf}
% 		\caption{Helio P35}\label{fig:nas_cpu_a03s_conv_mape}
% 	\end{subfigure}
% 	\begin{subfigure}[b]{.49\linewidth}
% 		\centering
% 		\includegraphics[width=\linewidth]{figures/results/nas_cpu_mi8se_conv_mape.pdf}
% 		\caption{Snapdragon 710}\label{fig:nas_cpu_mi8se_conv_mape}
% 	\end{subfigure}
% 	\caption{Predictions of GBDT on Convolution with Multiple CPU Cores (Synthetic Neural Architectures)}
% 	\label{fig:nas_cpu_conv_mape}
% \end{figure}

\begin{figure}[t]
	\centering
	
    \begin{subfigure}[b]{.49\linewidth}
		\centering
		\includegraphics[width=\linewidth]{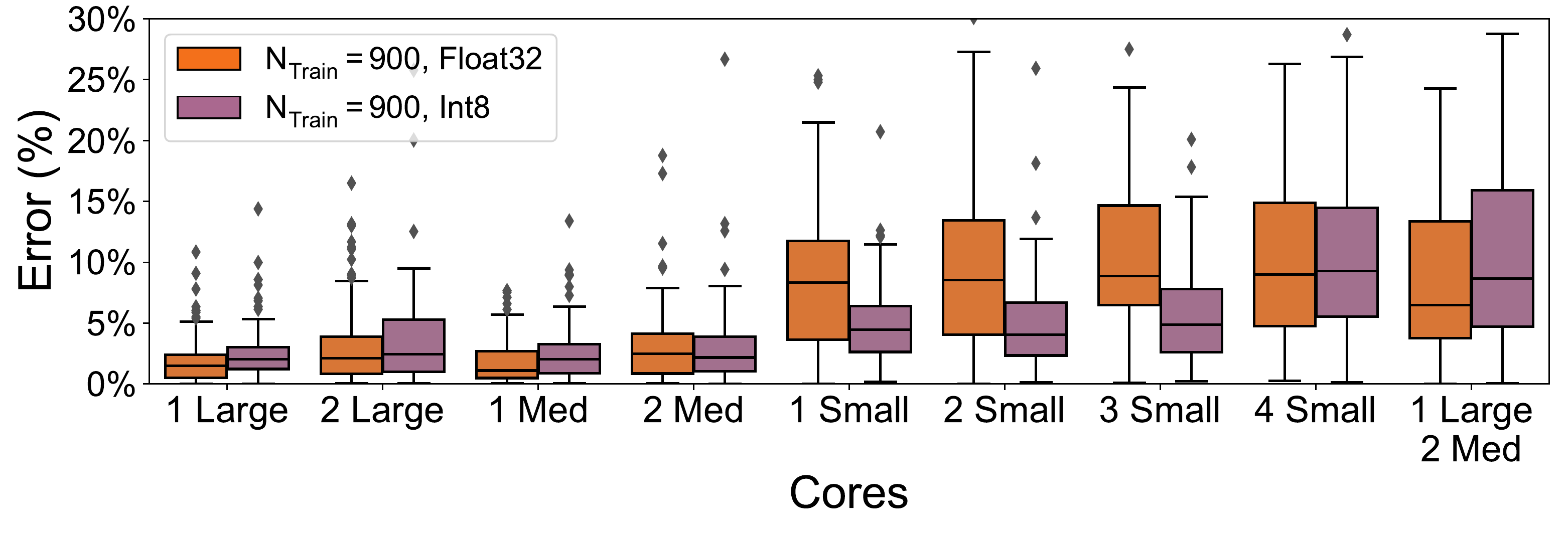}\vspace{-0.5em}
		\caption{Exynos 9820}\label{fig:result_nas_900_GBDT_cpu_s10}
	\end{subfigure}
	\begin{subfigure}[b]{.49\linewidth}
		\centering
		\includegraphics[width=\linewidth]{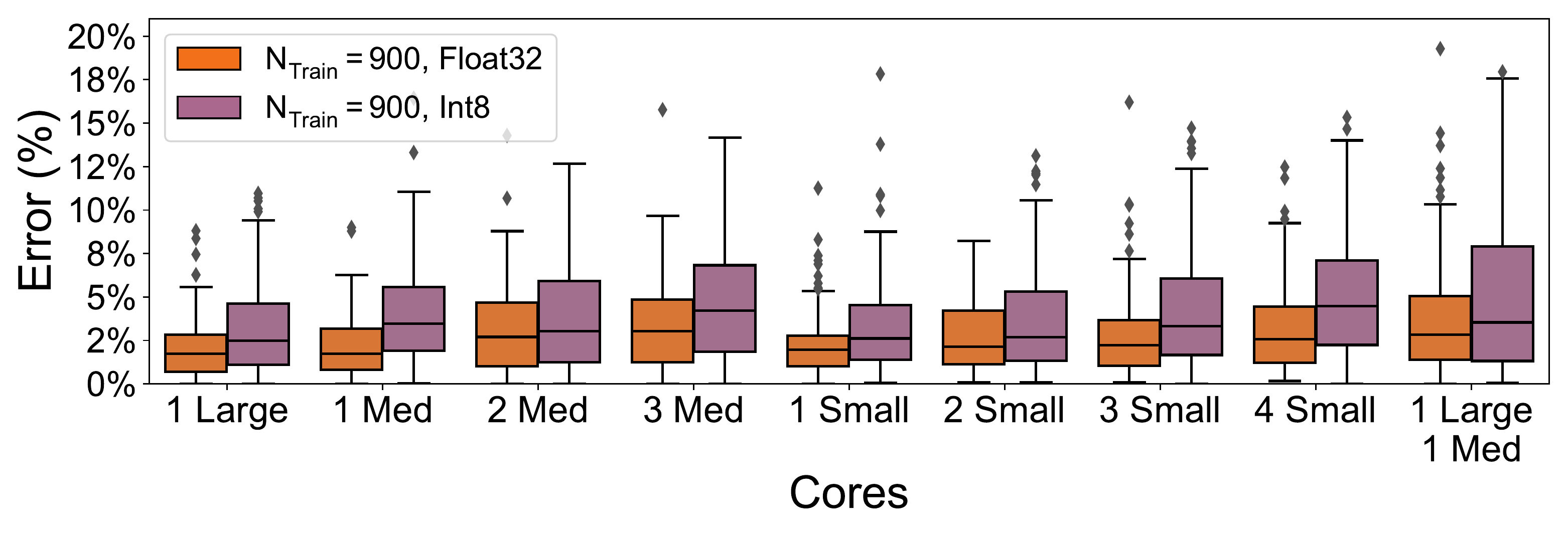}\vspace{-0.5em}
		\caption{Snapdragon 855}\label{fig:result_nas_900_GBDT_cpu_pixel4}
	\end{subfigure}
	\begin{subfigure}[b]{.49\linewidth}
		\centering
		\includegraphics[width=\linewidth]{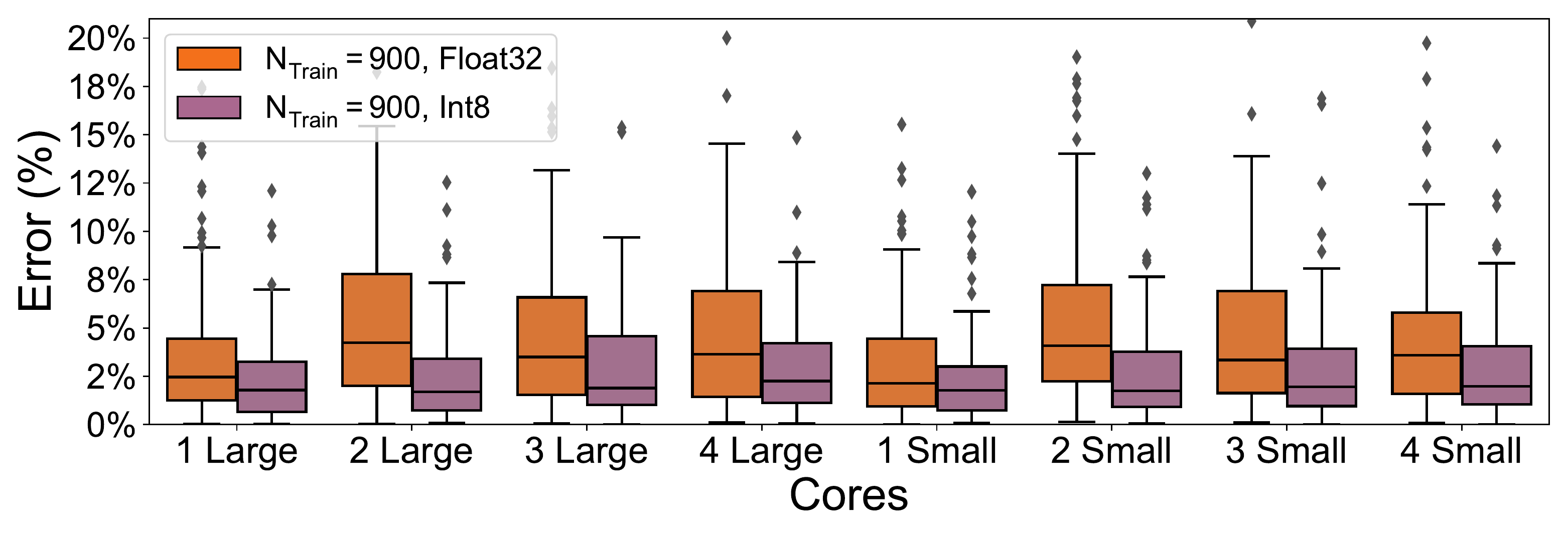}\vspace{-0.5em}
		\caption{Helio P35}\label{fig:result_nas_900_GBDT_cpu_a03s}
	\end{subfigure}
	\begin{subfigure}[b]{.49\linewidth}
		\centering
		\includegraphics[width=\linewidth]{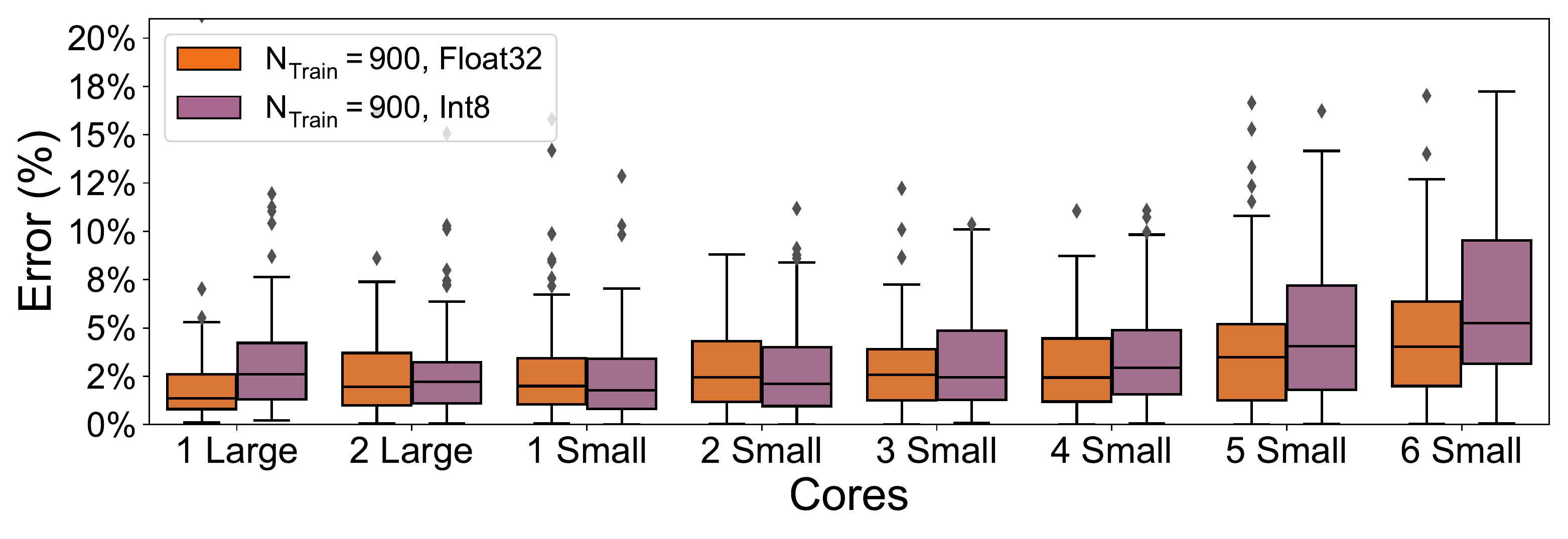}\vspace{-0.5em}
		\caption{Snapdragon 710}\label{fig:result_nas_900_GBDT_cpu_mi8se}
	\end{subfigure}
	\caption{GBDT Predictions of End-to-end Latency using Multiple CPU Cores (Synthetic Neural Architectures)}
	\label{fig:nas_cpu_end_to_end_mape}\vspace{-.5em}
\end{figure}

Next, we evaluate our prediction framework under hardware heterogeneity, including scenarios with different CPU core combinations and with both floating-point and integer representations. We select GBDT as a representative ML approach in this section, since it shows comparable or slightly better predictions than RF and MLP in the case of a large CPU core (\cref{fig:comparison_nas_900_cpu_all}).

\cref{fig:nas_cpu_end_to_end_mape} illustrates GBDT predictions of end-to-end latency for various core combinations; for clarity of presentation, we omit some outliers (<9\% data points for 1 large and 2 medium cores of Exynos 9820, and <4\% data points for all other configurations), and report plots with all data points in the Appendix (\cref{fig:appendix_nas_cpu_end_to_end_mape}).
We observe that an increasing number of homogeneous cores typically leads to higher prediction errors. 
%%%As discussed in the Appendix (\cref{fig:appendix_nas_cv_cpu}), 
Using more cores can result in larger measurement variance, due to background jobs running on mobile devices (e.g., camera, sensors, and networking services); measurement variance can impair the quality of profiling results and thus affect prediction accuracy.
For example, from the results on Snapdragon 710 shown in \cref{fig:result_nas_900_GBDT_cpu_mi8se}, the MAPE on 6 small cores (5.2\% for floating-point and 6.4\% for integer quantization) is significantly higher than on 1 small core (2.0\% and 3.2\%, respectively), due to the substantial interference of background jobs when an inference task attempts to make use of all the efficient cores on the device.
(Additional supporting data is included in \cref{fig:appendix_nas_cv_cpu} in the Appendix.)
Overall, GBDT achieves accurate predictions across all platforms: the worst MAPE for homogeneous cores is 10.5\% on Exynos 9820, 5.8\% on Snapdragon 855, 6.0\% on Helio P35, and 6.4\% on Snapdragon 710.

Note that using heterogeneous cores results in even higher variability of latency measurements due to inter-cluster communication \cite{wang2019high}. In addition, as explained in \cref{sec:multithreading}, operations without multithreading implementations can be scheduled on arbitrary cores, complicating prediction accuracy; for example, when using 1 large and 1 medium core on Snapdragon 855, prediction errors (MAPEs of 3.9\% for floating-point and 5.5\% for integer quantization) are higher with respect to using 2 medium cores (3.2\% and 3.9\%, respectively).

\begin{figure}[t]
	\centering
	\begin{subfigure}[b]{.4\linewidth}
		\centering
		\includegraphics[width=\linewidth]{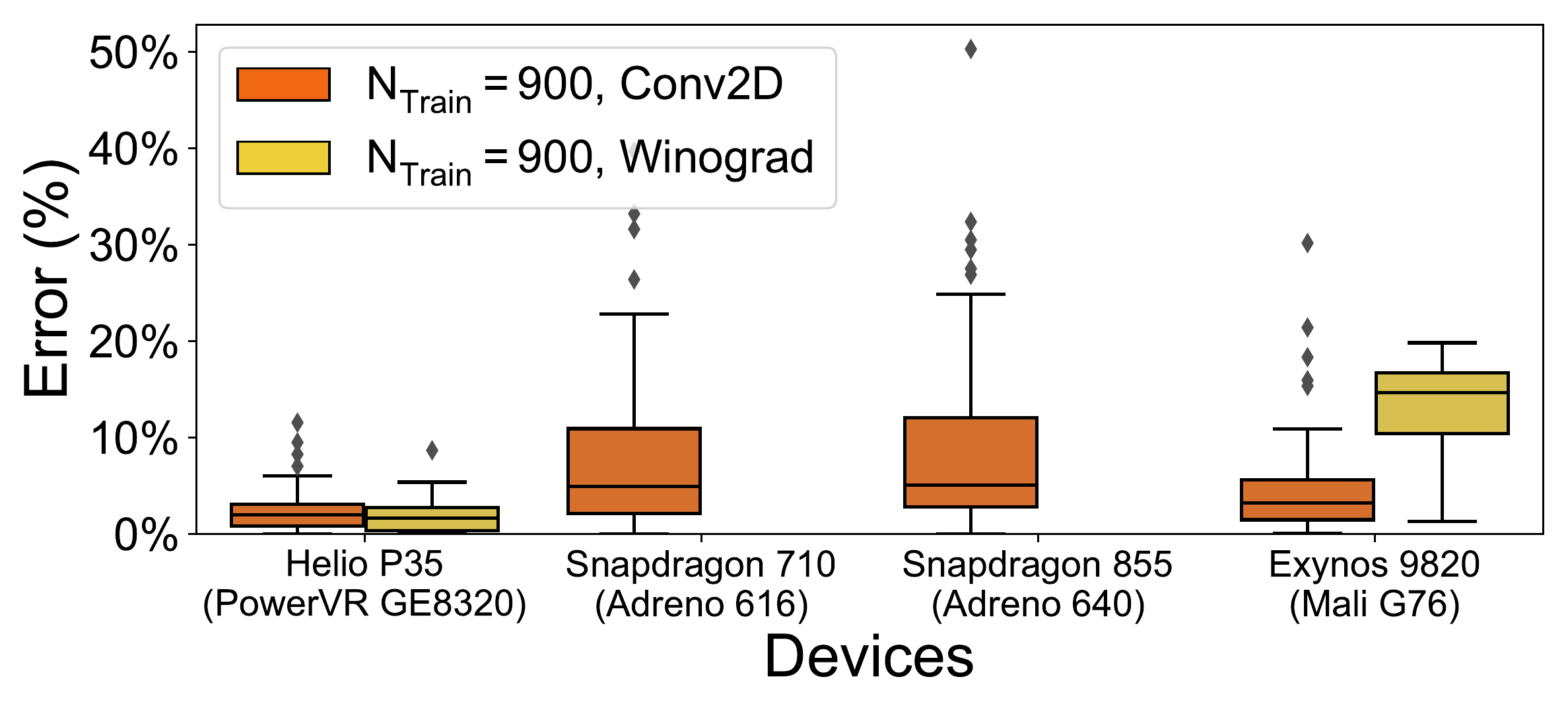}
		\caption{Convolution}\label{fig:result_nas_900_GBDT_gpu_ops}
	\end{subfigure}
	\begin{subfigure}[b]{.4\linewidth}
		\centering
		\includegraphics[width=\linewidth]{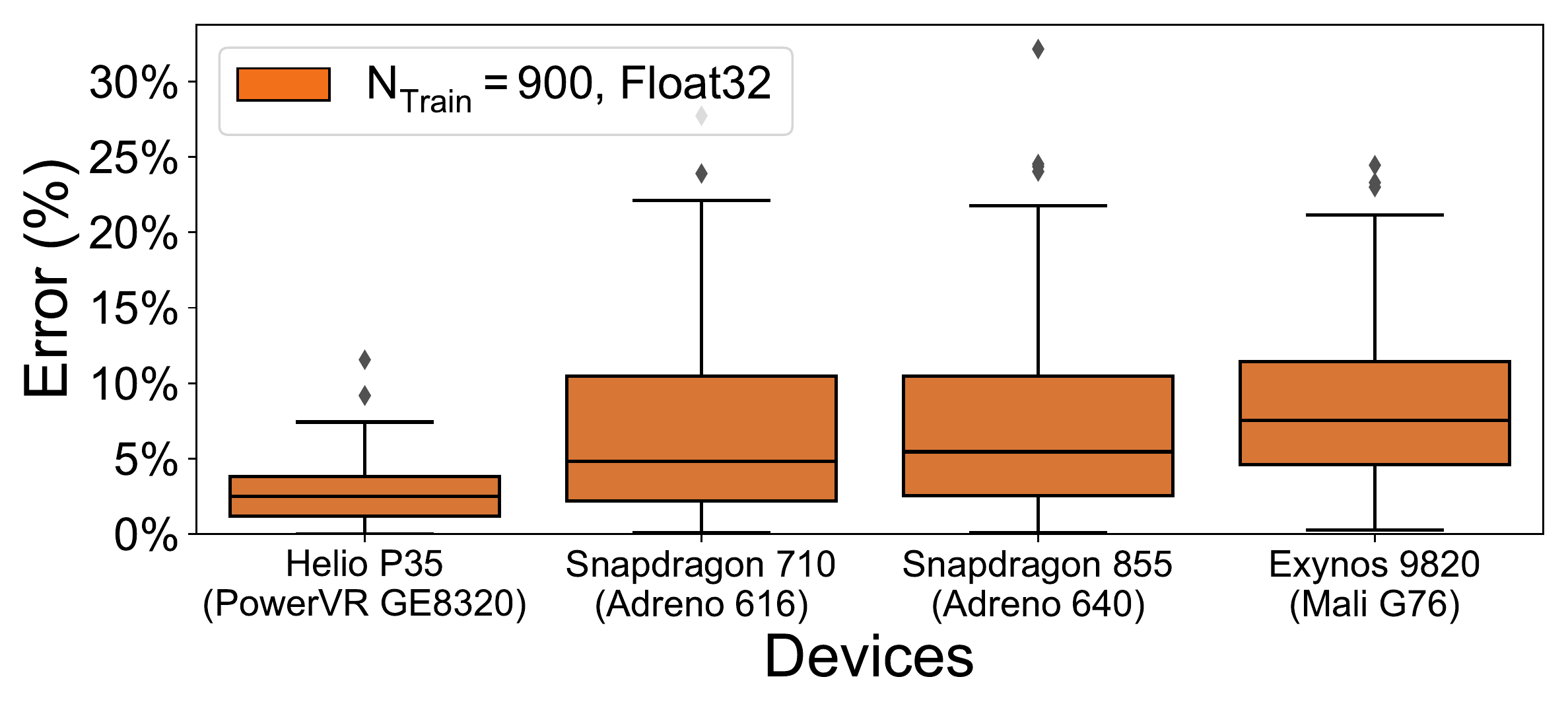}
		\caption{End-to-end}\label{fig:result_nas_900_GBDT_gpu}
	\end{subfigure}

	\caption{Predictions of GBDT on GPUs (Synthetic Neural Architectures)}
	\label{fig:nas_gpu_mape}\vspace{-1.2em}
\end{figure}

\cref{fig:nas_gpu_mape} presents predictions of GBDT on multiple GPUs.
For convolution operations, we split the results of Conv2D and Winograd kernels in \cref{fig:result_nas_900_GBDT_gpu_ops} because separate latency predictors are trained for each kernel; no Winograd kernel is used on Adreno 640 and 616 due to the rules of kernel selection presented in \cref{sec:kernel_selection}.
Overall, GBDT achieves good end-to-end predictions across all four GPUs, with worst MAPE of 8.2\% corresponding to Exynos 9820.

\subsection{Case Study: Neural Architecture Diversity} \label{sec:result_neural_architecture_heterogeneity}

\begin{figure}[t]
	\centering
	\begin{subfigure}[b]{.49\linewidth}
		\centering
		\includegraphics[width=\linewidth]{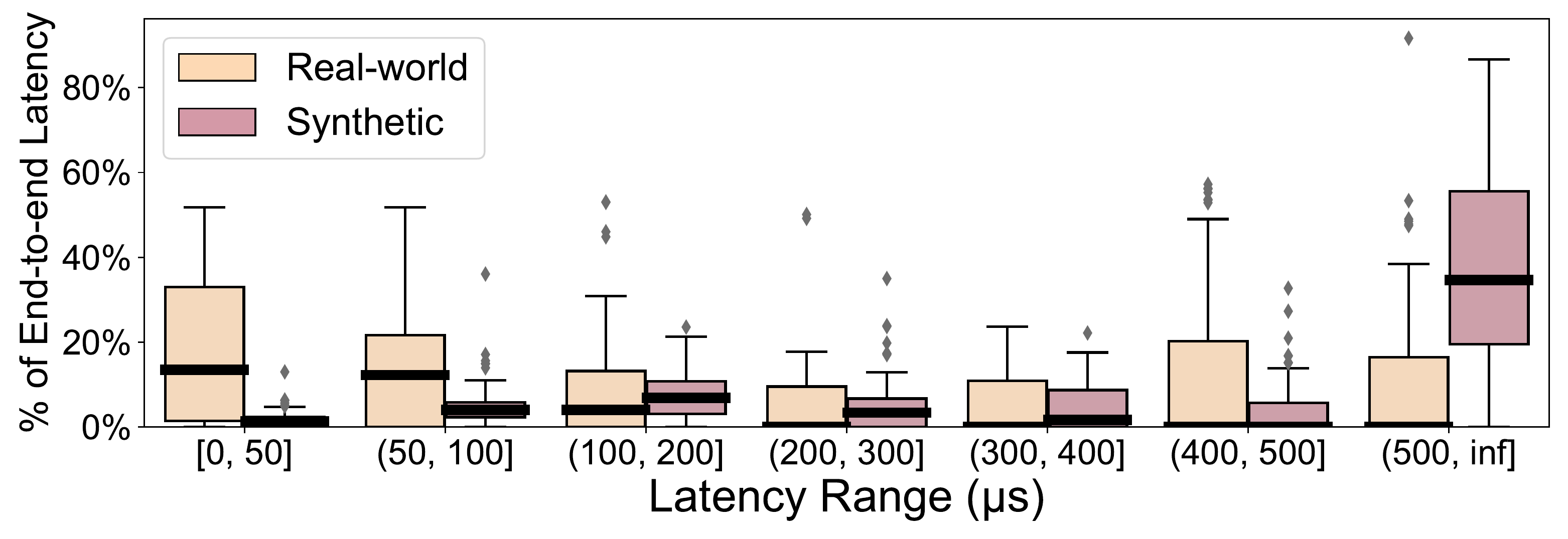}
		\caption{Percentage of End-to-end Latency}\label{fig:result_analysis_lasso_latency_cpu_a03s}
	\end{subfigure}
	\begin{subfigure}[b]{.49\linewidth}
		\centering
		\includegraphics[width=\linewidth]{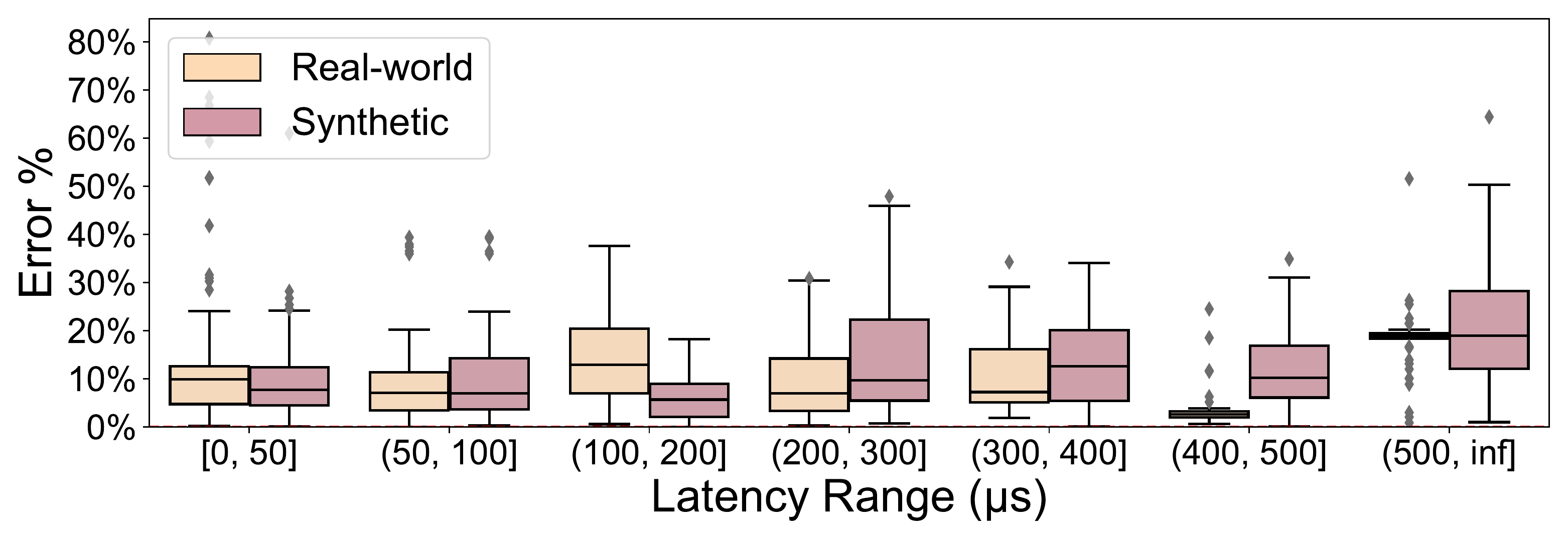}
		\caption{Predictions from Lasso}\label{fig:result_analysis_lasso_error_cpu_a03s}
	\end{subfigure}
	\vspace{-0.5em}
	\caption{Convolution Operations with Latencies from Different Ranges (on 1 Large CPU Core of Helio P35)}
	\label{fig:result_analysis_lasso_cpu_a03s}\vspace{-0.5em}
\end{figure}

Next, we evaluate our framework under diverse neural architectures: we consider a scenario where training data include only a small number of neural architectures sampled \emph{at the early stages of NAS}, while test data are high-accuracy neural architectures generated \emph{at the end of NAS.}
In our evaluation, we use 1000 synthetic neural architectures as training data and 102 real-world neural architectures (from existing literature) as test data. 
The two sets of neural architectures have \emph{different distributions} (i.e., we introduce a dataset shift): we observe that the latency of convolution operations in real-world neural architectures is generally lower than in synthetic neural architectures. \cref{fig:result_analysis_lasso_latency_cpu_a03s} shows percentage of end-to-end latency attributed to convolution operations (split by range) on Helio P35 (with a single large core): convolutions greater than 500 ms dominate end-to-end latency in our synthetic neural architectures, while faster convolutions are more important in real-world neural architectures.

\begin{figure}[t]
	\centering
	\begin{subfigure}[b]{.49\linewidth}
		\centering
		\includegraphics[width=\linewidth]{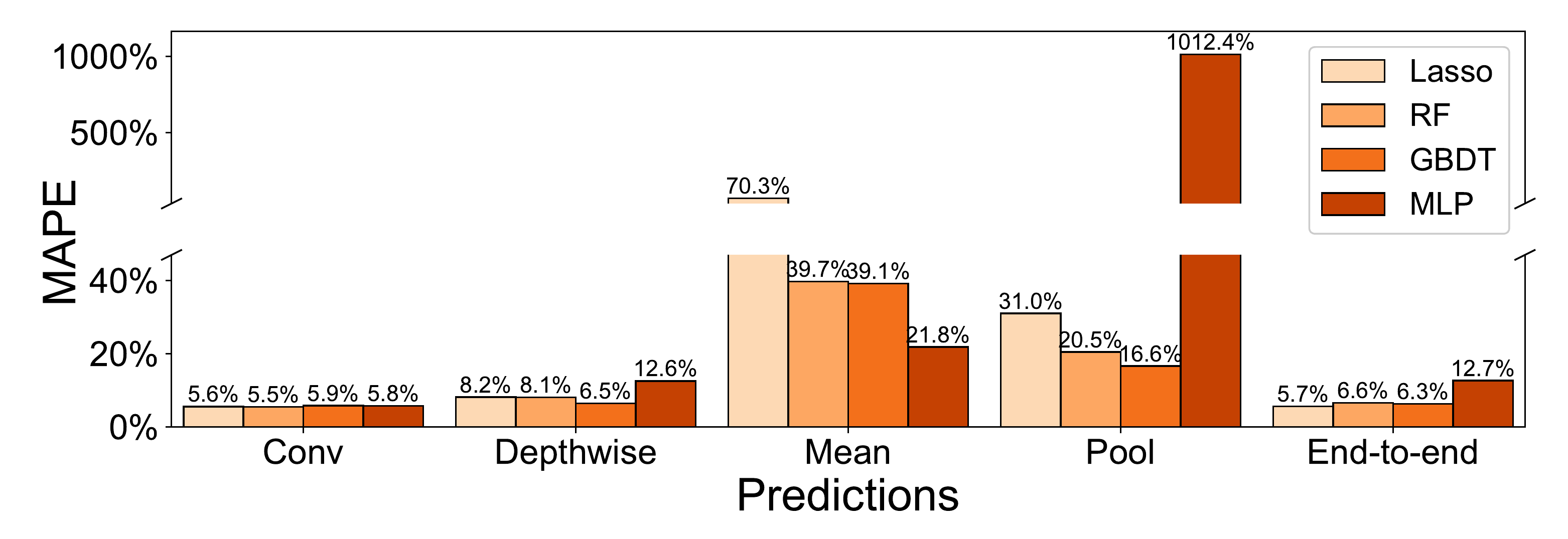}
		\caption{CPU (One Large Core)}\label{fig:comparison_common_1000_cpu_all}
	\end{subfigure}
	\begin{subfigure}[b]{.49\linewidth}
		\centering
		\includegraphics[width=\linewidth]{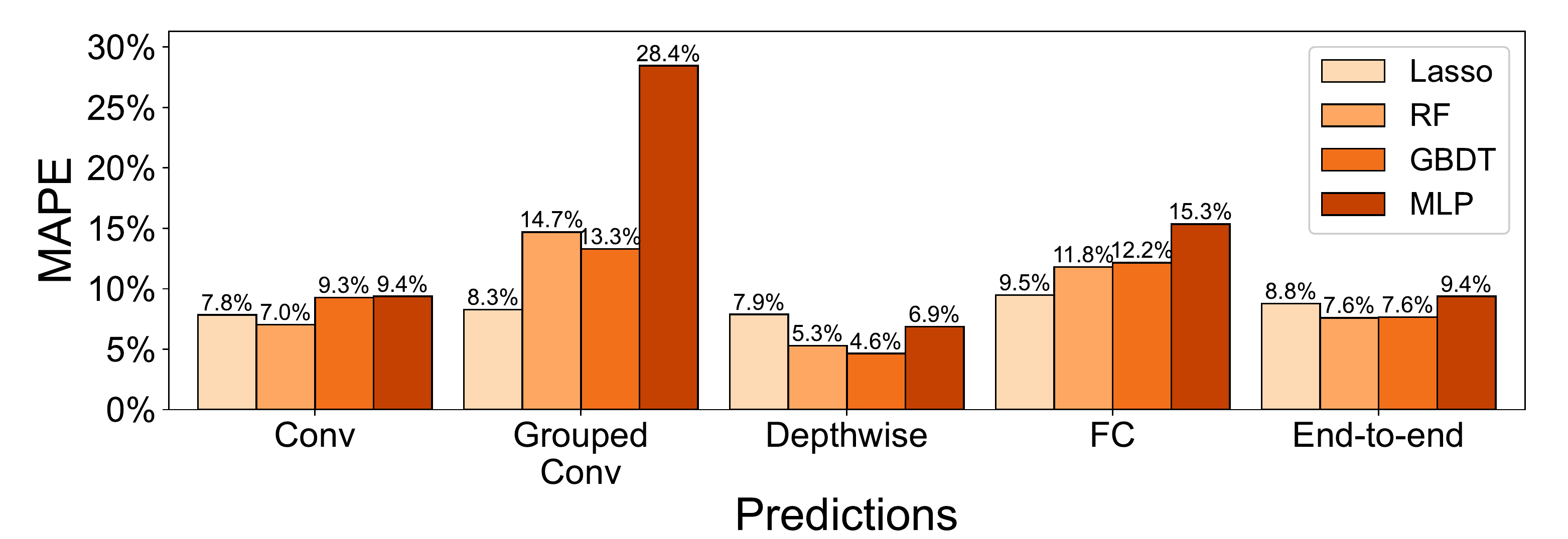}
		\caption{GPU}\label{fig:comparison_common_1000_gpu_all}
	\end{subfigure}
	\caption{Predictions of ML Models (Real-world Neural Architectures)}
	\label{fig:comparison_common_1000}\vspace{-1em}
\end{figure}

\cref{fig:comparison_common_1000_cpu_all} shows the average MAPE across four devices for the real-world neural architectures on CPUs.
For most ML approaches trained on synthetic neural architectures, prediction errors are higher for real-world neural architectures than synthetic neural architectures (\cref{fig:comparison_nas_900}), which are generated from the same distribution as the training data.
The only exception is Lasso, which has better performance on real-world neural architectures, achieving the lowest end-to-end MAPE on CPUs (5.7\%).
We attribute this anomaly to the better accuracy of Lasso predictions on fast operations (< 500 ms) due to higher weights assigned to faster operations (in \cref{eq:constraint}), which we observe in both synthetic and real-world architectures (\cref{fig:result_analysis_lasso_error_cpu_a03s}); since real-world architectures include a larger proportion of fast operations, average accuracy is better on this test set.

\cref{fig:comparison_common_1000_gpu_all} presents predictions on mobile GPUs.
We observe that, for some small real-world neural architectures, the overhead of TFLite is significant. Since the overhead has high runtime variability (in particular, on PowerVR GE8320 and Mali G76), it can affect the accuracy of end-to-end latency predictions, especially for neural architectures with low latency, such as MobileNets.

\subsection{Case Study: ML Framework Optimizations}\label{sec:result_framework_heterogeneity}

Next, we illustrate the improvements of GPU predictions resulting from accounting for ML framework optimizations such as kernel fusion and kernel selection.

\paragraph{Kernel Fusion}

\begin{figure}[t]
	\centering
	\begin{subfigure}[b]{.23\linewidth}
		\centering
		\includegraphics[width=\linewidth]{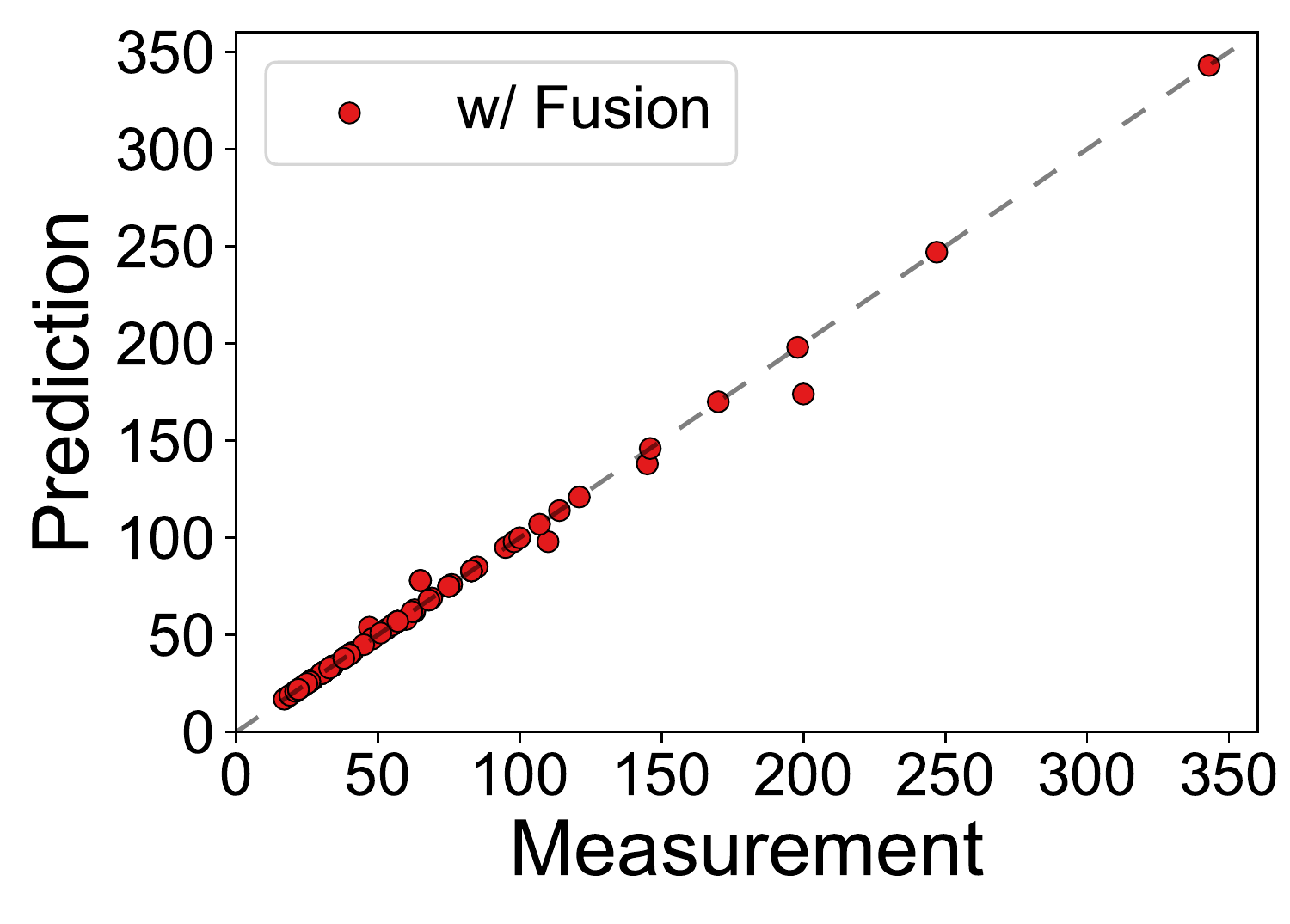}
		\caption{Number of Kernels}\label{fig:kernel_fusion_number_diagonal}
	\end{subfigure}
	\begin{subfigure}[b]{.37\linewidth}
		\centering
		\includegraphics[width=\linewidth]{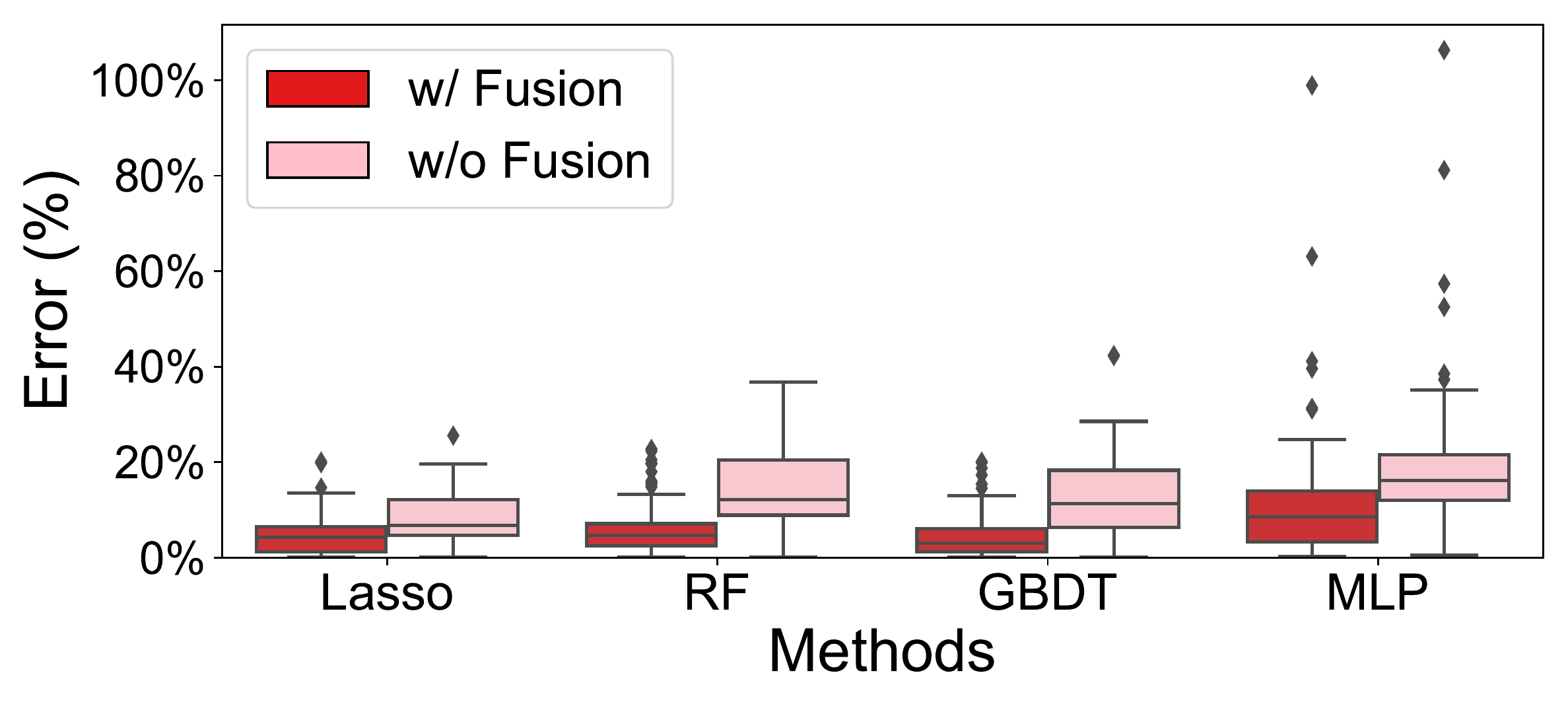}
		\caption{Snapdragon 710 (Adreno 616)}\label{fig:effect_kernel_fusion_end-to-end_mi8se}
	\end{subfigure}
	\begin{subfigure}[b]{.37\linewidth}
		\centering
		\includegraphics[width=\linewidth]{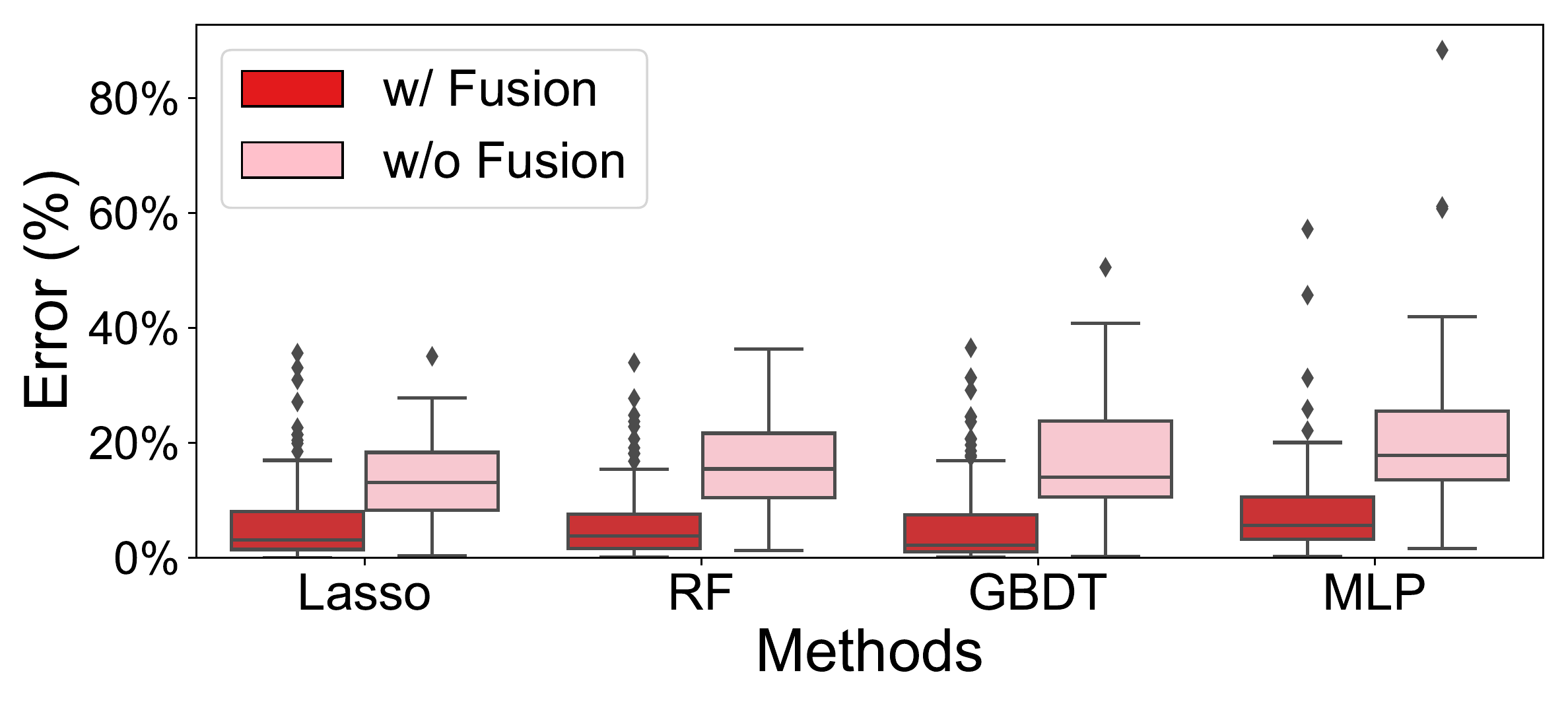}
		\caption{Helio P35 (PowerVR GE8320)}\label{fig:effect_kernel_fusion_end-to-end_a03s}
	\end{subfigure}
	\caption{End-to-end Latency Predictions for Whether Considering Kernel Fusion}
	\label{fig:effect_kernel_fusion}\vspace{-1em}
\end{figure}

In \cref{sec:kernel_fusion}, we show that kernel fusion considerably reduces the number of kernels and leads to improvements in end-to-end latency. 
\cref{fig:kernel_fusion_number_diagonal} shows that, after applying our algorithm (\cref{alg:kernel_fusion} detailed in the Appendix) for estimating which kernels will be fused by TFLite (\cref{sec:kernel_fusion}), we obtain a number of kernels close to actual measurements collected on 102 real-world neural architectures.
\cref{fig:effect_kernel_fusion_end-to-end_mi8se,fig:effect_kernel_fusion_end-to-end_a03s} illustrate that we obtain substantial error reduction in end-to-end latency prediction with respect to ML models which do not consider kernel fusion (labeled as ``w/o Fusion'').

\paragraph{Kernel Selection}

\begin{figure}[t]
	\centering
	\begin{subfigure}[b]{.4\linewidth}
		\centering
		\includegraphics[width=\linewidth]{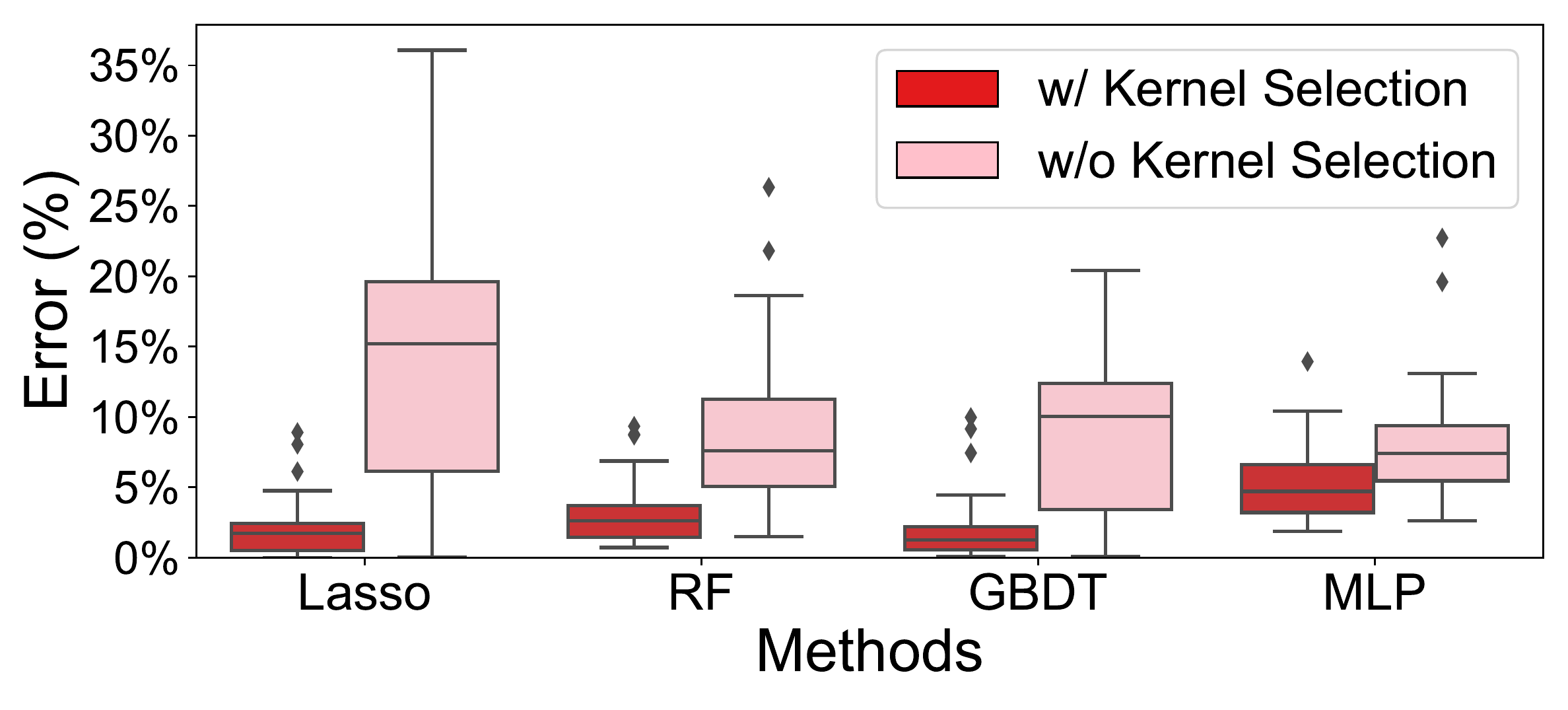}
		\caption{End-to-end Latency}\label{fig:effect_kernel_selection_end-to-end_a03s}
	\end{subfigure}
	\begin{subfigure}[b]{.4\linewidth}
		\centering
		\includegraphics[width=\linewidth]{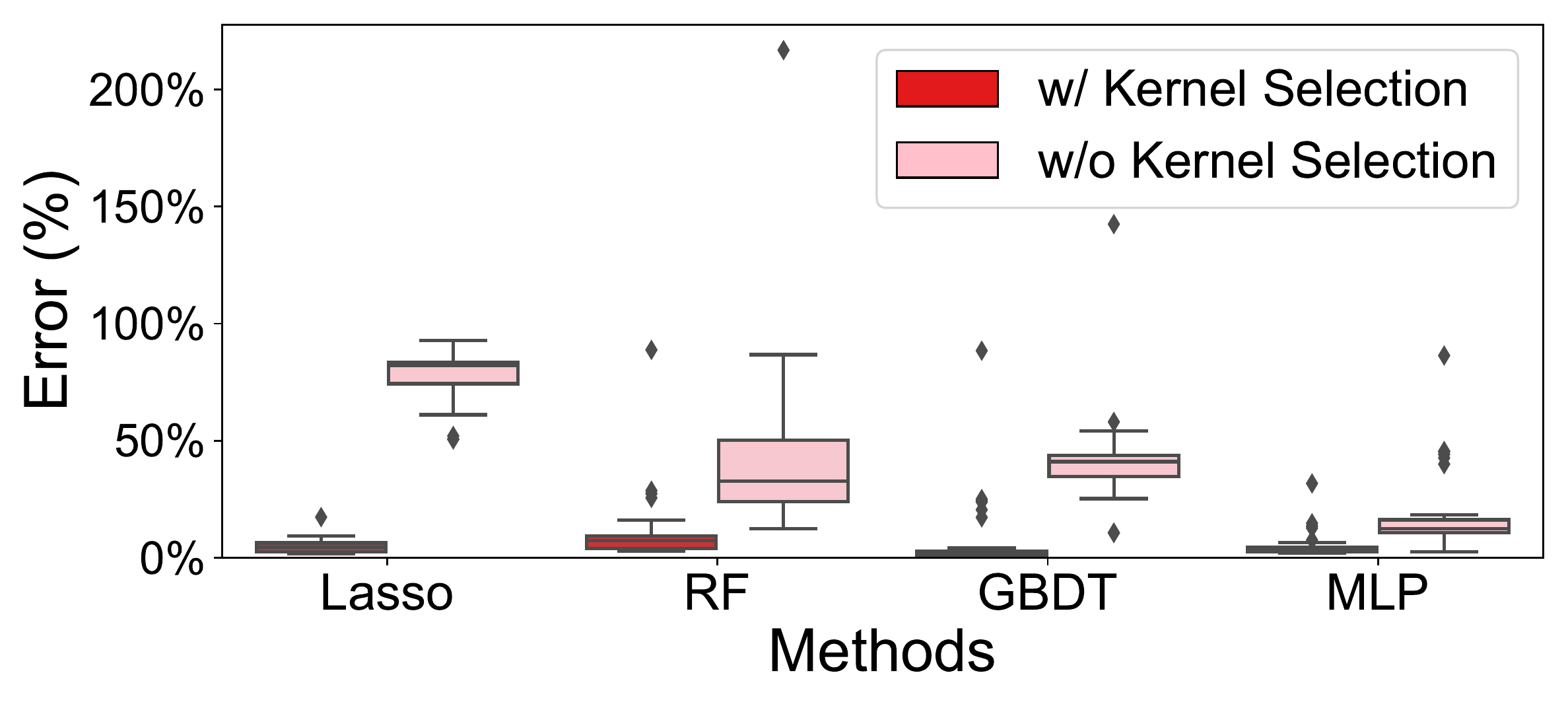}
		\caption{Winograd Kernel Latency}\label{fig:effect_kernel_selection_winograd_a03s}
	\end{subfigure}
	\caption{Prediction Error Reduction on Helio P35 (PowerVR GE8320) by Accounting for Kernel Selection}
	\label{fig:effect_kernel_selection_a03s}
\end{figure}

As introduced in \cref{sec:kernel_selection}, a convolution operation in the computational graph can be evaluated by TFLite using different kernel implementations compatible with the target device and convolution parameters.
We deduce the actual kernels selected by TFLite for convolution operations (specifically, Conv2D and Winograd) and train separate predictors for each (since they have different performance characteristics).
\cref{fig:effect_kernel_selection_end-to-end_a03s} shows the considerable error reduction achieved by accounting for kernel selection on PowerVR GE8320, for real-world neural architectures that support Winograd kernels; \cref{fig:effect_kernel_selection_winograd_a03s} confirms that this reduction is due to more accurate predictions of the latency of Winograd kernels.

\subsection{Case Study: Limited Training Data} \label{sec:result_limited_training_data}

The high cost of collecting sufficient training data is a common criticism of ML approaches to predict latency of neural architectures during NAS \cite{lu2021one}. In this section, we study the effects of training set size on different ML approaches, illustrating the benefits of a simple model when training data is limited.

\subsubsection{Comparison of ML Approaches} \label{sec:result_limited_training_data_comparison_ML_approaches}

\begin{figure}[t]
	\centering
	\begin{subfigure}[b]{.4\linewidth}
		\centering
		\includegraphics[width=\linewidth]{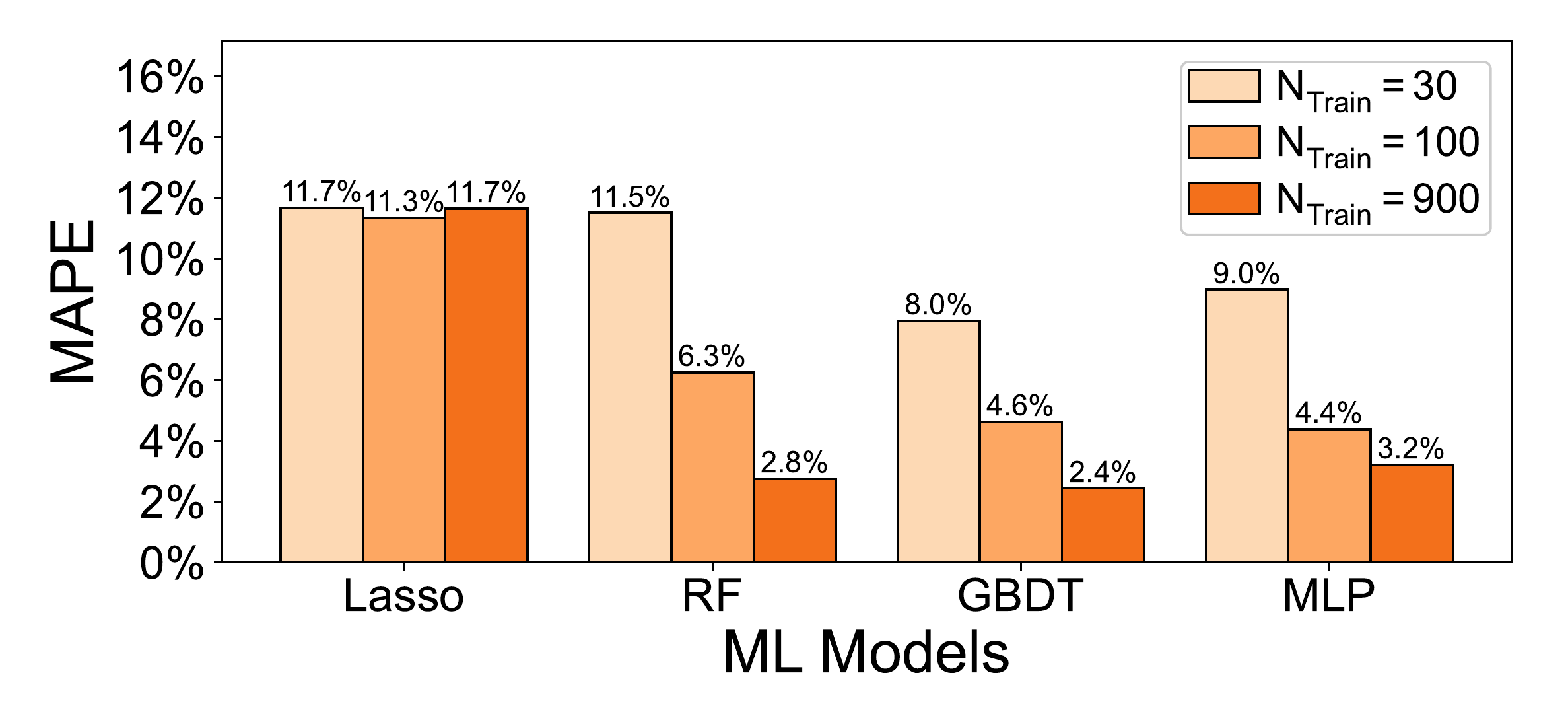}
		\caption{CPU (a Large Core)}\label{fig:comparison_nas_cpu_end_to_end}
	\end{subfigure}
	\begin{subfigure}[b]{.4\linewidth}
		\centering
		\includegraphics[width=\linewidth]{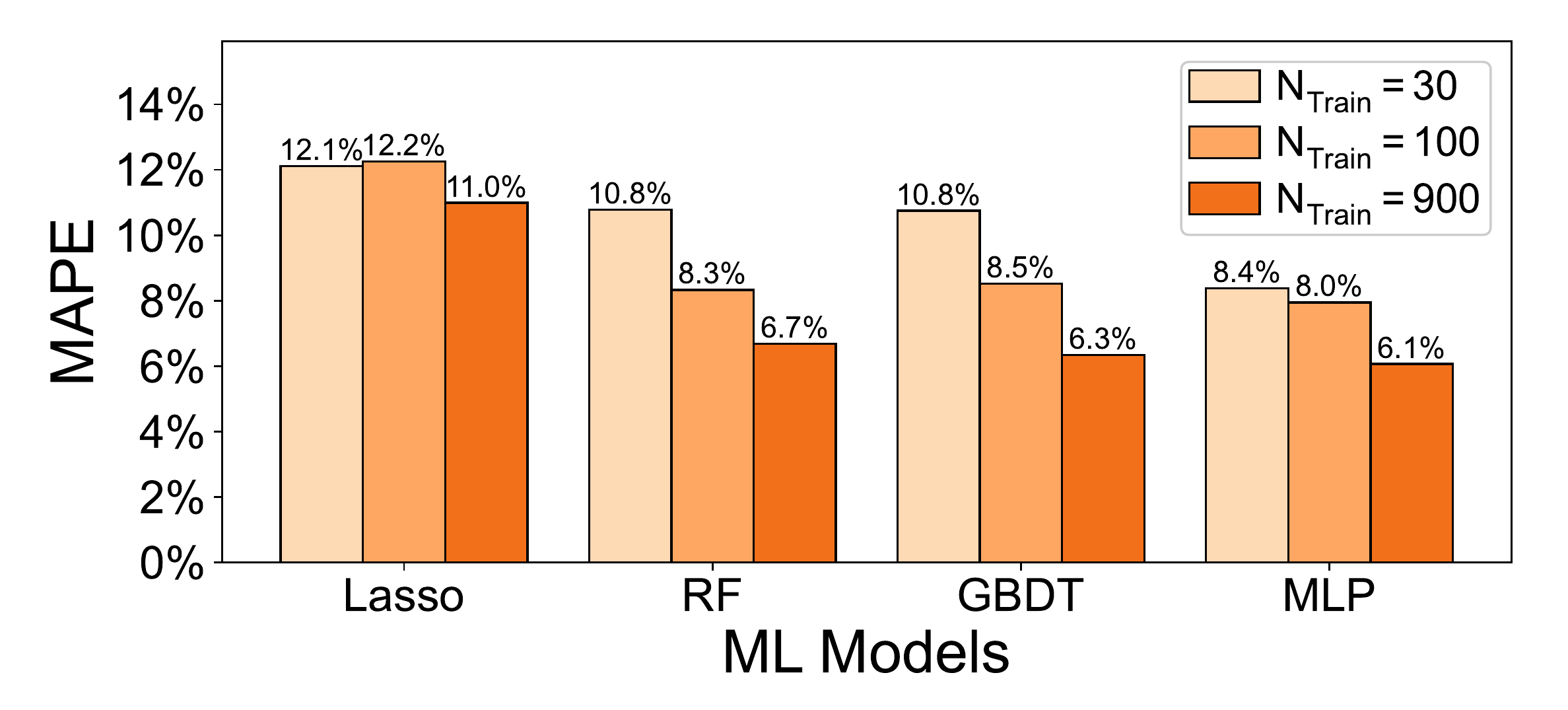}
		\caption{GPU}\label{fig:comparison_nas_gpu_end_to_end}
	\end{subfigure}
	\vspace{-0.5em}
	\caption{Predictions of End-to-end Latency with Different Training Set Sizes (Synthetic Neural Architectures)}
	\label{fig:comparison_nas}\vspace{-0.5em}
\end{figure}

\begin{figure}[t]
	\centering
	\begin{subfigure}[b]{.4\linewidth}
		\centering
		\includegraphics[width=\linewidth]{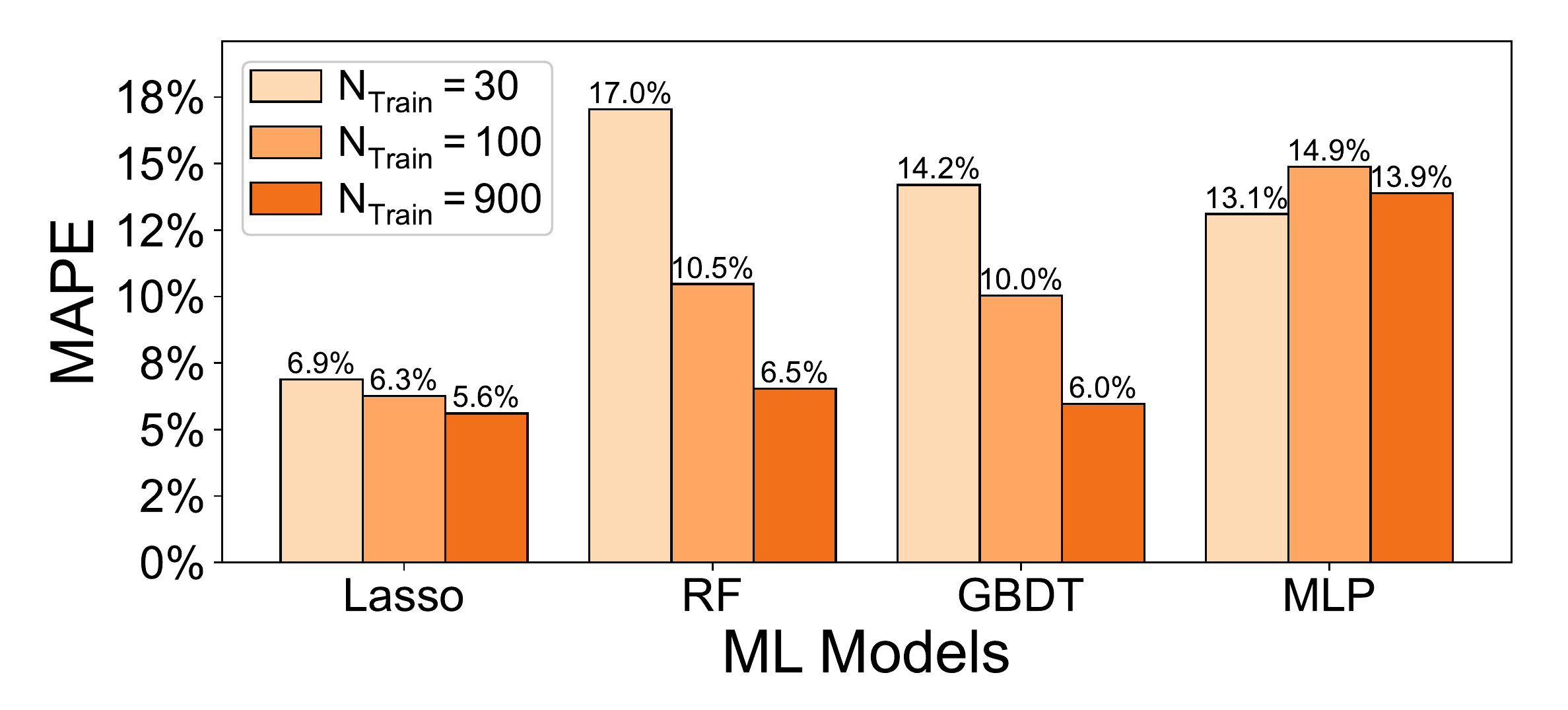}
		\caption{CPU (a Large Core)}\label{fig:comparison_common_cpu_end_to_end}
	\end{subfigure}
	\begin{subfigure}[b]{.4\linewidth}
		\centering
		\includegraphics[width=\linewidth]{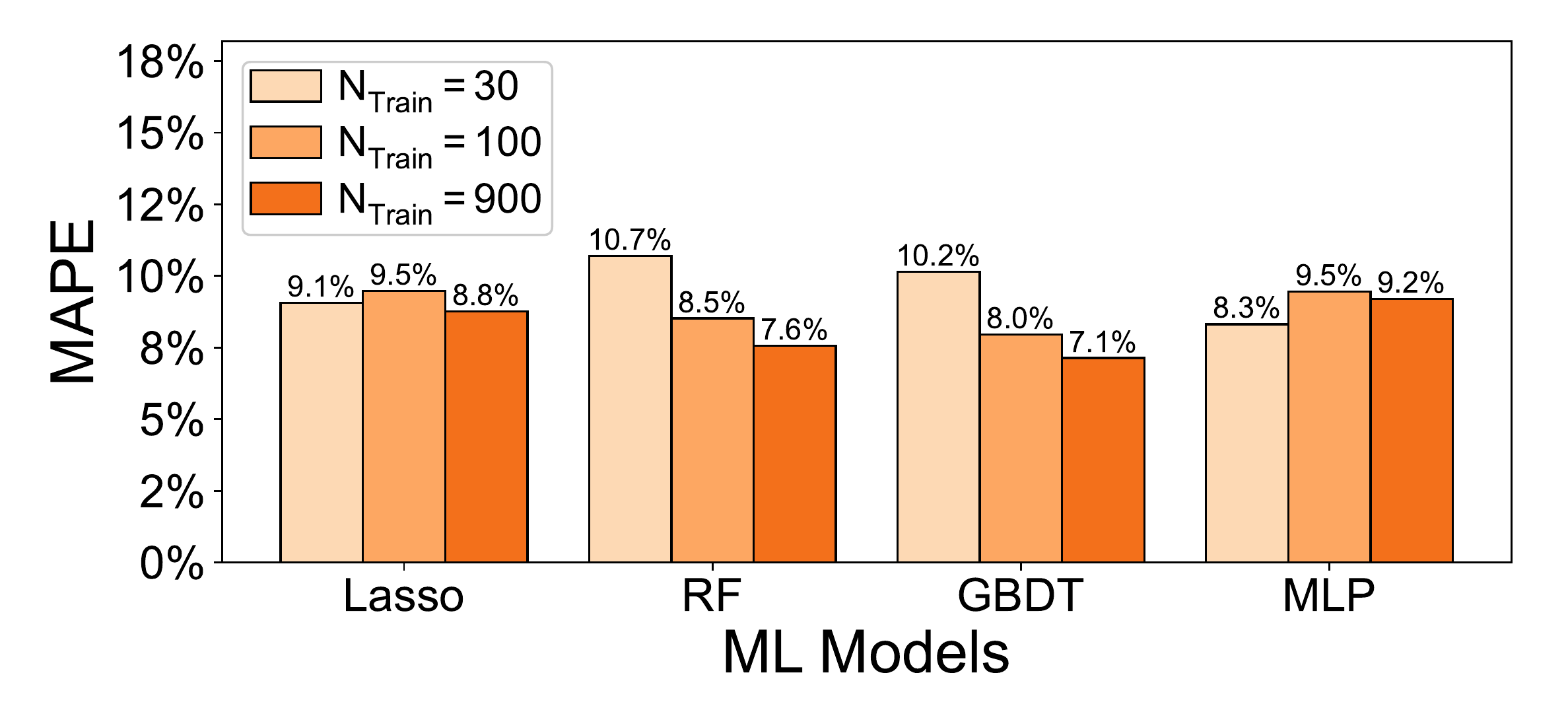}
		\caption{GPU}\label{fig:comparison_common_gpu_end_to_end}
	\end{subfigure}
	\vspace{-0.5em}
	\caption{Predictions of End-to-end Latency with Different Training Set Sizes (Real-world Neural Architectures)}
	\label{fig:comparison_common}\vspace{-0.5em}
\end{figure}

\cref{fig:comparison_nas,fig:comparison_common} show prediction errors of different ML approaches for varying training set sizes $N_{\text{Train}}$, on synthetic neural architectures (presented in \cref{sec:result_default_setting}) and real-world neural architectures (presented in \cref{sec:result_neural_architecture_heterogeneity}), respectively (errors are average MAPE across 4 platforms; MAPEs for each platform are reported in \cref{table:end_to_end_predictions_synthetic,table:end_to_end_predictions_real_world} in the Appendix).
Predictions of Lasso are less sensitive to the size of training data, while other more complex approaches achieve lower error when the training set size is increased from 30 to 900.
Notably, MLP achieves lower prediction errors with a smaller training set of size 30. This is due to severe prediction errors on concatenation/split operations: on Pixel 4 (one large CPU core), MAPEs on concatenation/split operations are 56.7\%, 1400.4\% and 1068.7\%, after training on 30, 100 and 900 neural architectures, respectively; due to lack of space, more detailed data is in the Appendix (\cref{fig:appendix_comparison_common_cpu_mlp_pixel4}). This anomaly is due to the very small amount of training data (only 5, 25 and 312 concatenation/split operations from training data of 30, 100 and 900 neural architectures, respectively). Instead, for convolution operations we have sufficient data and the prediction errors are 7.8\%, 5.1\% and 4.6\% with training set of size 30, 100 and 900, respectively, on the same platform.

Notably, for real-world neural architectures, using only 30 training examples, Lasso considerably outperforms other ML approaches on CPUs with a large core (\cref{fig:comparison_common_cpu_end_to_end}), with the average MAPE of 6.9\% across four platforms.
As pointed out by prior work \cite{lu2021one}, the cost of profiling only 30 neural architectures on each target device is negligible compared to the time-consuming process of NAS.

\subsubsection{Predictions of Lasso on Limited Training Data} \label{sec:result_limited_training_data_lasso_predictions}

Next, we thoroughly evaluate the predictions of Lasso with limited training set size (i.e., 30 neural architectures) on real-world neural architectures, across a broad range of scenarios with hardware heterogeneity.

% \begin{figure}[t]
% 	\centering
% 	\begin{subfigure}[b]{.49\linewidth}
% 		\centering
% 		\includegraphics[width=\linewidth]{figures/results/common_30_cpu_s10_conv_mape.pdf}
% 		\caption{Exynos 9820}\label{fig:common_cpu_s10_conv_mape}
% 	\end{subfigure}
% 	\begin{subfigure}[b]{.49\linewidth}
% 		\centering
% 		\includegraphics[width=\linewidth]{figures/results/common_30_cpu_pixel4_conv_mape.pdf}
% 		\caption{Snapdragon 855}\label{fig:common_cpu_pixel4_conv_mape}
% 	\end{subfigure}
% 	\begin{subfigure}[b]{.49\linewidth}
% 		\centering
% 		\includegraphics[width=\linewidth]{figures/results/common_30_cpu_a03s_conv_mape.pdf}
% 		\caption{Helio P35}\label{fig:common_cpu_a03s_conv_mape}
% 	\end{subfigure}
% 	\begin{subfigure}[b]{.49\linewidth}
% 		\centering
% 		\includegraphics[width=\linewidth]{figures/results/common_30_cpu_mi8se_conv_mape.pdf}
% 		\caption{Snapdragon 710}\label{fig:common_cpu_mi8se_conv_mape}
% 	\end{subfigure}
% 	\caption{Predictions of Lasso on Convolution with Multiple CPU Cores (Real-world Neural Architectures)}
% 	\label{fig:common_cpu_conv_mape}
% \end{figure}

\begin{figure}[t]
	\centering
	\begin{subfigure}[b]{.48\linewidth}
		\centering
		\includegraphics[width=\linewidth]{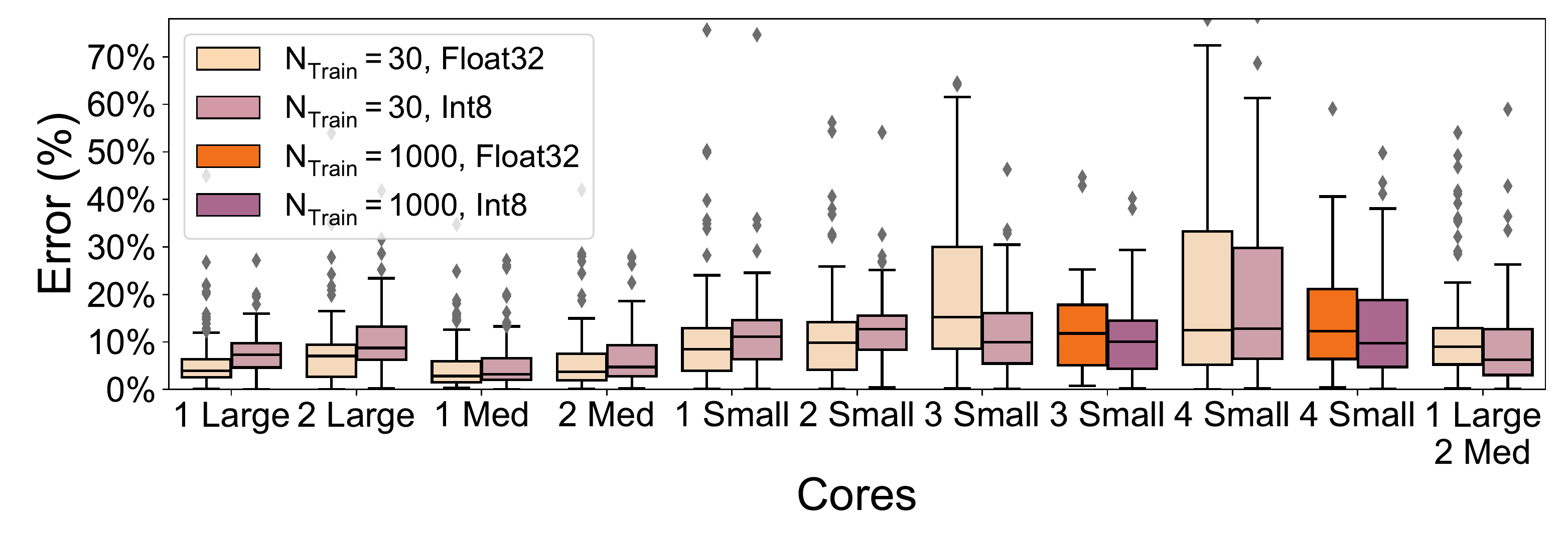}\vspace{-0.5em}
		\caption{Exynos 9820}\label{fig:result_common_30_lasso_cpu_s10}
	\end{subfigure}
	\begin{subfigure}[b]{.49\linewidth}
		\centering
		\includegraphics[width=\linewidth]{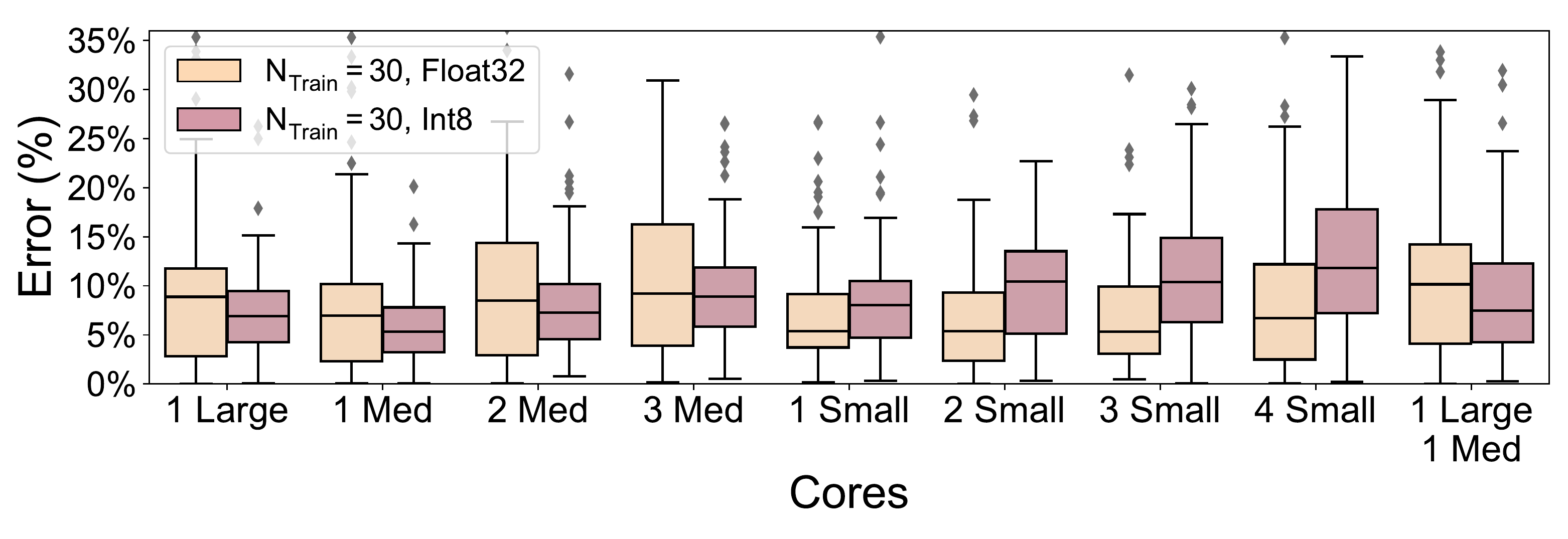}\vspace{-0.5em}
		\caption{Snapdragon 855}\label{fig:result_common_30_lasso_cpu_pixel4}
	\end{subfigure}
	\begin{subfigure}[b]{.49\linewidth}
		\centering
		\includegraphics[width=\linewidth]{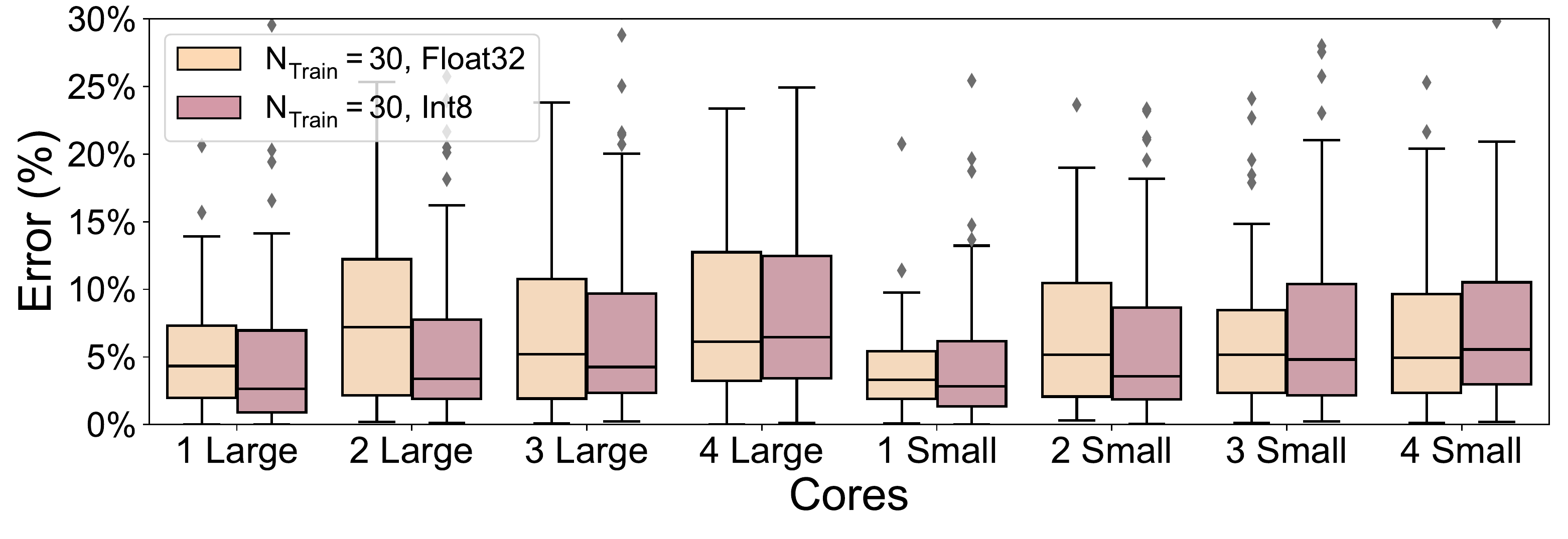}\vspace{-0.5em}
		\caption{Helio P35}\label{fig:result_common_30_lasso_cpu_a03s}\vspace{-.5em}
	\end{subfigure}
	\begin{subfigure}[b]{.49\linewidth}
		\centering
		\includegraphics[width=\linewidth]{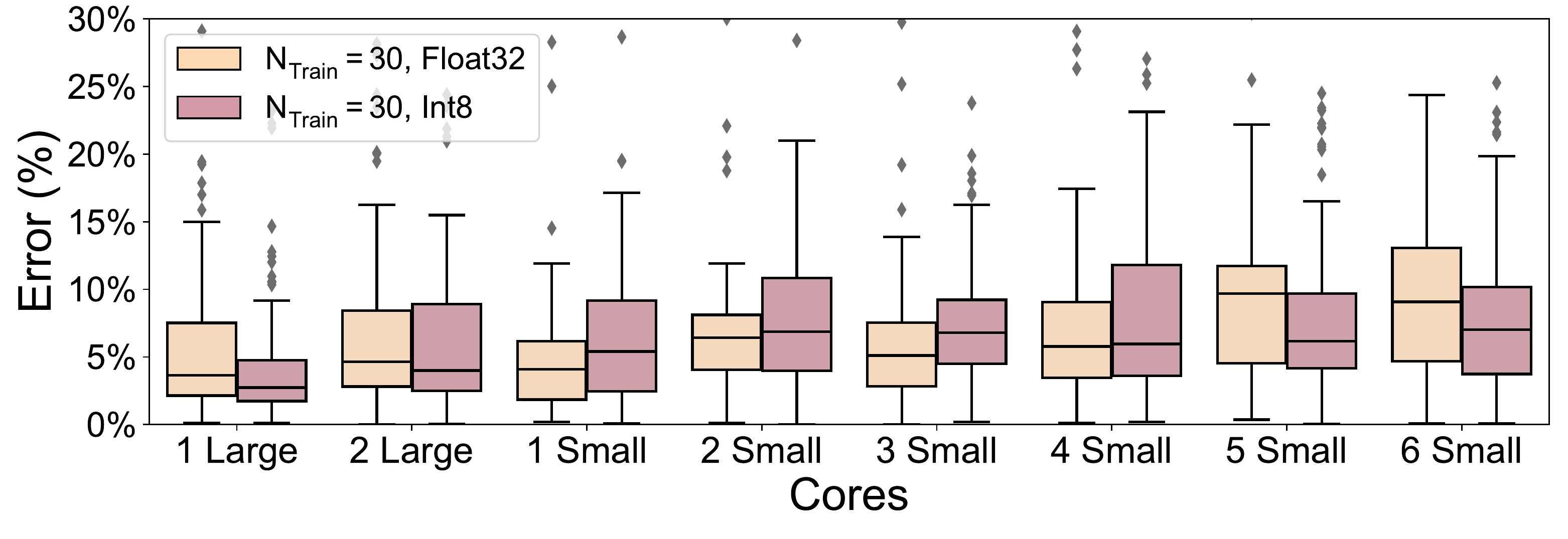}\vspace{-0.5em}
		\caption{Snapdragon 710}\label{fig:result_common_30_lasso_cpu_mi8se}\vspace{-.5em}
	\end{subfigure}
	\caption{End-to-end Latency Predictions of Lasso with Multiple CPU Cores (Real-world Neural Architectures)\label{fig:common_cpu_end_to_end_mape}\vspace{-1em}}
\end{figure}

\cref{fig:common_cpu_end_to_end_mape} displays the predictions of Lasso on real-world neural architectures, across various combinations of cores and data representations; for clarity of presentation, we omit some outliers (<4\% data points per configuration), and report plots with all data points in the Appendix (\cref{fig:appendix_common_cpu_end_to_end_mape}).
Generally, the trend of prediction errors for homogeneous and heterogeneous clusters are similar to the results in \cref{fig:nas_cpu_end_to_end_mape}.
The maximum MAPE for combinations of homogeneous cores is 22.9\% on Exynos 9820, 13.5\% on Snapdragon 855, 9.6\% on Helio P35, and 10.9\% on Snapdragon 710. We believe the large prediction errors on Exynos 9820 are due to the variance of measurements collected with many small efficient cores, which can affect the quality of training data for this limited dataset. By adding more training data,  MAPEs can be reduced to less than 14.8\%.

Similarly, for all devices, the worst case typically appears in the case of utilizing all small cores, due to the interference from the background jobs scheduled on the cluster of efficient cores. One exception is Helio P35, where the predictions on large and small cores show similar errors; we believe that the reason is related to the similarity between the two core clusters on Helio P35, since both core clusters are Cortex-A53 but running at different clock speeds (as shown in \cref{table:feature}). 

\begin{figure}[t]
	\centering
	\begin{subfigure}[b]{.4\linewidth}
		\centering
		\includegraphics[width=\linewidth]{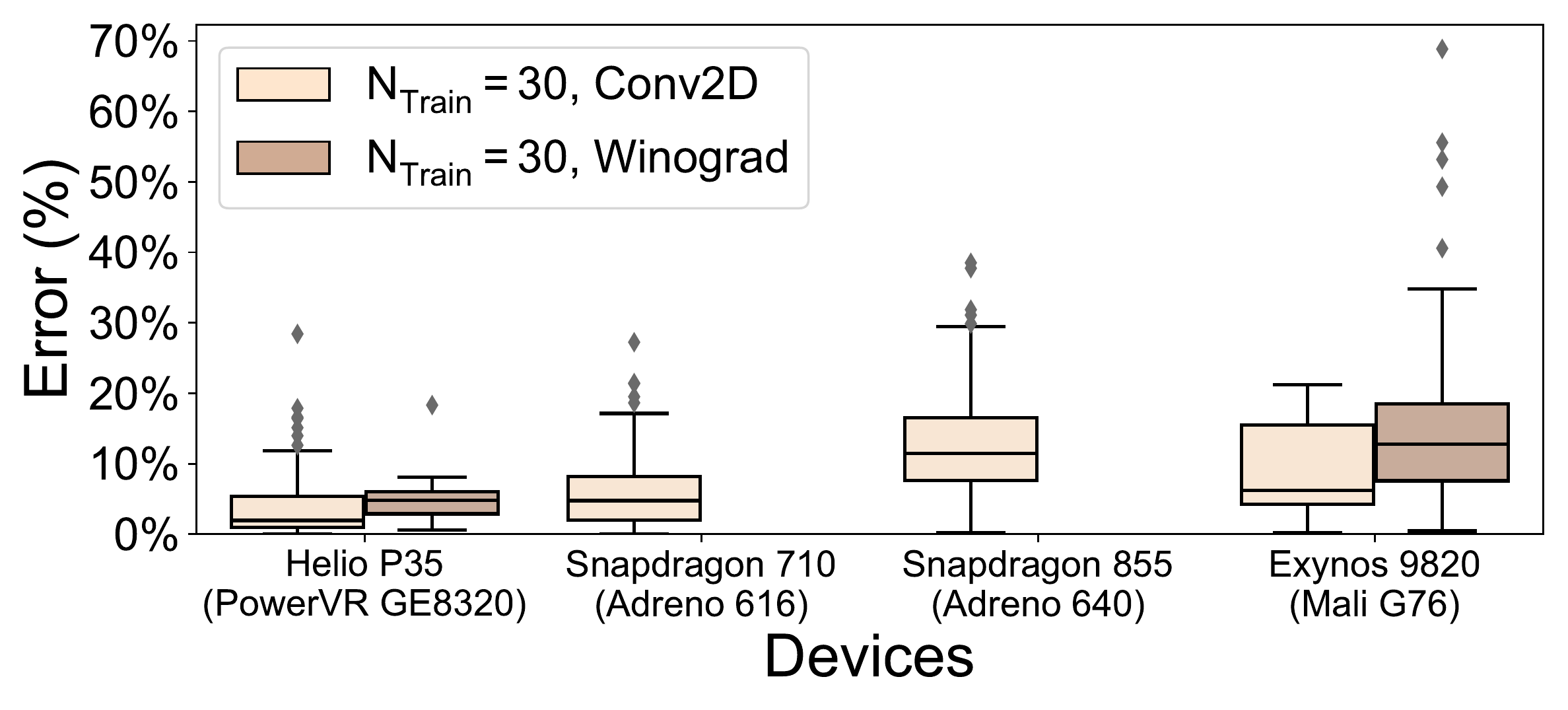}\vspace{-.3em}
		\caption{Convolution}\label{fig:result_common_30_lasso_gpu_ops}
	\end{subfigure}
	\begin{subfigure}[b]{.4\linewidth}
		\centering
		\includegraphics[width=\linewidth]{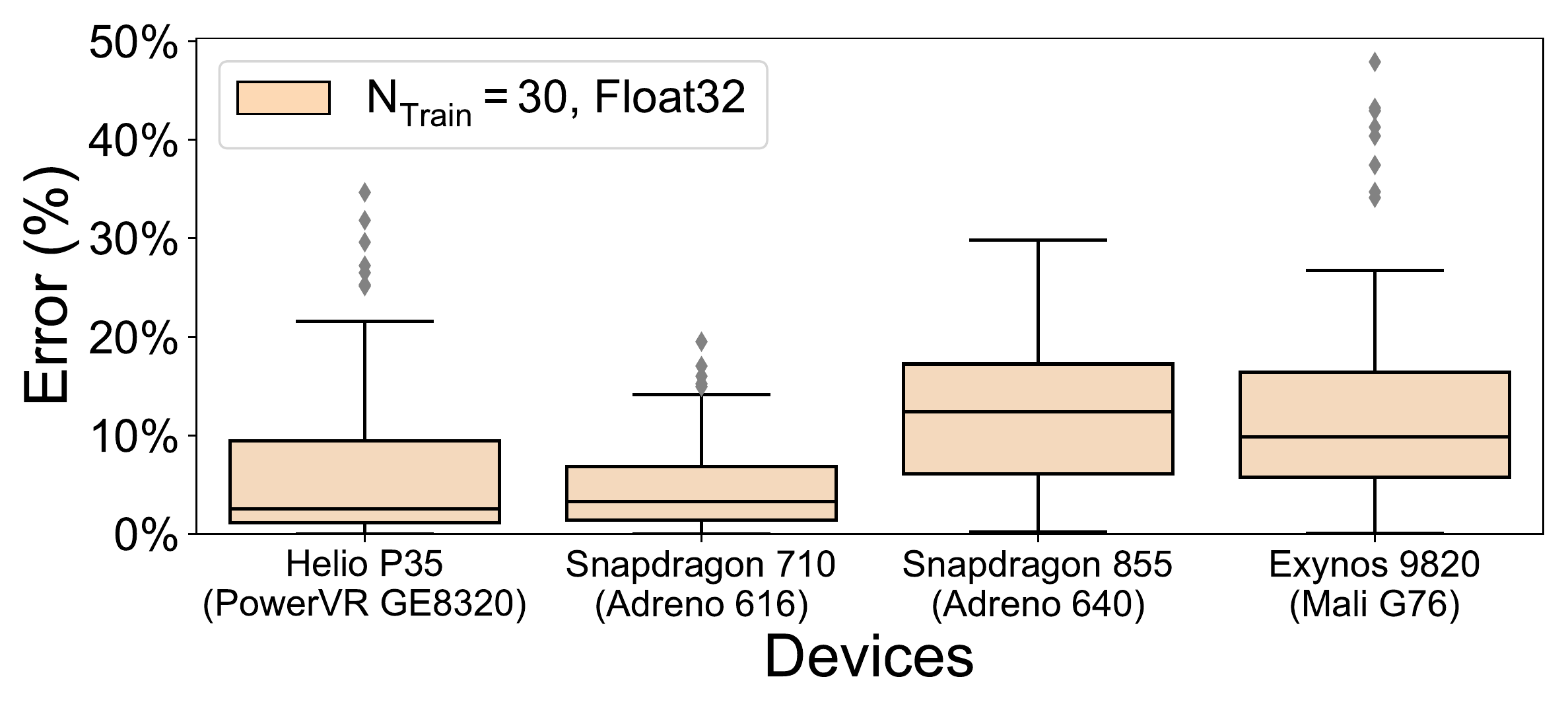}\vspace{-.3em}
		\caption{End-to-end}\label{fig:result_common_30_lasso_gpu}
	\end{subfigure}
	\vspace{-.7em}
	\caption{Predictions of Lasso on GPUs (Real-world Neural Architectures)\label{fig:common_gpu_mape}\vspace{-1.5em}}
\end{figure}

\cref{fig:common_gpu_mape} shows the predictions of Lasso across multiple mobile GPUs.
In general, the end-to-end predictions on the slower GPUs (MAPEs of 5.0\% on PowerVR GE830 and 5.4\% on Adreno 616) are better than on faster GPUs (MAPEs of 11.0\% on Mali G76 and 10.7\% on Adreno 640), since we observe smaller variance on slower GPUs over the longer execution time.

Since all the features are already standardized, we use the magnitude of weights in the Lasso model to analyze the importance of different features. On all devices, we find the most critical features (those with largest weights) of convolution operations on both CPUs and GPUs to be FLOPs and kernel size, which are strongly correlated with the costs of computation and memory access, respectively. However, as noted earlier, FLOPs alone are not an accurate proxy metric for the actual latency.
Different from standard convolution operations, the top two critical features of depthwise convolution operations are FLOPs and input size.
Input size can dominate the cost of memory access for depthwise convolutions since their kernel sizes are substantially smaller than those of standard convolutions.

\section{Related Work}\label{section:related_work}

\paragraph{Hardware heterogeneity}
Most existing work aims at latency predictions of training or inference tasks on cloud GPUs \cite{abbasi2021maple,geoffrey2021habitat,justus2018predicting,hafeez2020empirical,gao2021runtime} or edge GPUs \cite{bouzidi2021performance,bouhali2021execution}, where Nvidia GPUs dominate the market for ML workloads. However, the heterogeneity of mobile platforms makes performance prediction more difficult, particularly when using heterogeneous cores.
Our paper studies multiple mainstream mobile devices from different manufacturers, and tackles hardware heterogeneity across these platforms.
Recent works \cite{lu2021one,zhang2021nn,cai2019proxylessnas} focus on performance predictions of mobile CPUs, but only limited to a single core with floating-point computations. Instead, our work evaluates inference latency of mobile CPUs across a broad range of realistic scenarios, including the utilization of \emph{multiple heterogeneous CPU} cores, and both \emph{floating point and integer data representations}.

\vspace{-1mm}\paragraph{ML Framework Optimizations}
The majority of existing work \cite{geoffrey2021habitat,justus2018predicting,hafeez2020empirical,cai2019proxylessnas} proposes to predict latency based on the features extracted from neural architectures and hardware, but neglects the effects of ML framework optimizations.
As identified by our results, accounting for these optimizations results in significant improvements of the predictions for real-world neural architectures across multiple ML approaches.
Since ML framework optimizations cannot be analyzed on Nvidia cloud and edge GPUs (cuDNN is not open-source \cite{chetlur2014cudnn}), recent work \cite{zhang2021nn} proposes a black-box approach to learn their policies (i.e., the algorithms for kernel fusion).
In contrast, on mobile platforms, ML frameworks (e.g., TFLite) use open-source algorithms and OpenCL kernels to support a broad range of heterogeneous GPUs; we highlight their optimizations, accurately inferring the actual kernels used after compilation without deploying and compiling NN models on actual devices.

\vspace{-1mm}\paragraph{Prediction Approaches}
Existing works \cite{abbasi2021maple,gao2021runtime,dudziak2020brp} adopt ML approaches to predict end-to-end latency of neural architectures by encoding \emph{the entire neural architecture} as a single vector of input features; this approach, however, requires complicated ML techniques as well as large amounts of training data. 
In contrast, we make latency predictions for each component of the neural architecture, allowing simple ML approaches that require less training data  and are easier to interpret (e.g., in the case of Lasso) for understanding and development. 
Similarly to our work, component-wise approaches are used by \cite{zhang2021nn, cai2019proxylessnas} for latency prediction on mobile devices, but only limited to a single core for CPUs; ML framework optimizations are considered only by \cite{zhang2021nn} but with a black-box approach.
Analytical performance models also exist in the literature, accounting for the computational cost of operations \cite{qi17paleo} and memory access traffic of GEMM-based convolution \cite{li2021towards,lym2019delta}, but these works only target Nvidia cloud GPUs and their models do not account for the diverse hardware accelerators of mobile platforms, nor for optimizations applied by ML frameworks. 

\section{Conclusions}\label{section:conclusions}

Using measurements collected on 4 mobile devices for a number of neural architectures (1000 synthetic NAS architectures and 102 real-world architectures), we showed the impact of different factors on inference latency, including optimizations applied by ML frameworks for mobile GPUs (kernel fusion and kernel selection), scheduling over heterogeneous subsets of CPU cores and integer representations after quantization, often neglected by related work.
Based on this experimental evaluation, we proposed an approach to estimate end-to-end inference latency by training ML models to predict latency of each component type of neural architectures. 
Our approach can accurately predict latency of novel neural architectures on a given device using limited profiling data (e.g., from 30 architectures); notably, we achieve good accuracy also when the test dataset has different characteristics from training data, a common scenario in NAS.
In future work, we plan to extend our evaluation and prediction approach to other efficiency metrics (e.g., power consumption) and to different classes of specialized hardware accelerators for inference tasks (e.g., Apple Neural Engine).

% \input{content/acknowledgment}

% \begin{acks}
% \end{acks}

\bibliographystyle{ACM-Reference-Format}
\bibliography{main}

\clearpage
\appendix

\section{Details of the Real-world Neural Architectures}\label{appendix:common_models}

In this appendix, we include details of the 102 state-of-the-art neural architectures used in our evaluation.
Due to the limited resources available on mobile devices, we restricted our selection to neural architectures with up to 18 million parameters, and in particular to the architectures proposed (by manual design or NAS) in the following 25 articles:
BagNet~\cite{brendel2019approximating},
BN-Inception~\cite{ioffe2015batch},
DenseNet~\cite{huang2017densely},
DiracNetV2~\cite{zagoruyko2017diracnets},
DLA~\cite{yu2018deep},
EfficientNet~\cite{tan2019efficientnet},
FBNet~\cite{wu2019fbnet},
FD-MobileNet~\cite{qin2018fd},
GhostNet~\cite{han2020ghostnet},
HarDNet~\cite{chao2019hardnet},
HRNet~\cite{wang2020deep},
MnasNet~\cite{tan2019mnasnet},
MobileNet~\cite{howard2017mobilenets},
MobileNetV2~\cite{sandler2018mobilenetv2},
MobileNetV3~\cite{howard2019searching},
PeleeNet~\cite{wang2018pelee},
PreResNet~\cite{he2016identity},
ProxylessNAS~\cite{cai2019proxylessnas},
RegNet~\cite{radosavovic2020designing},
ResNet~\cite{he2016deep},
ResNeXt~\cite{xie2017aggregated},
SE-ResNet/SE-PreResNet~\cite{hu2018squeeze},
SPNASNet~\cite{stamoulis2019single},
SqueezeNet/SqueezeResNet~\cite{iandola2016squeezenet},
VoVNet~\cite{lee2019energy}.

The TensorFlow implementations of these neural architectures are in \cite{imgclsmob}, which also provides pre-trained parameters as well as the Top-1 and Top-5 test errors on the ImageNet-1K dataset. For each architecture, we first generated a TensorFlow model and then converted it to a \texttt{.tflite} model file (using either floating-point or 8-bit integers) that can be compiled for each mobile platform.

\begin{figure}[t]
    \centering
	\includegraphics[width=.3\linewidth]{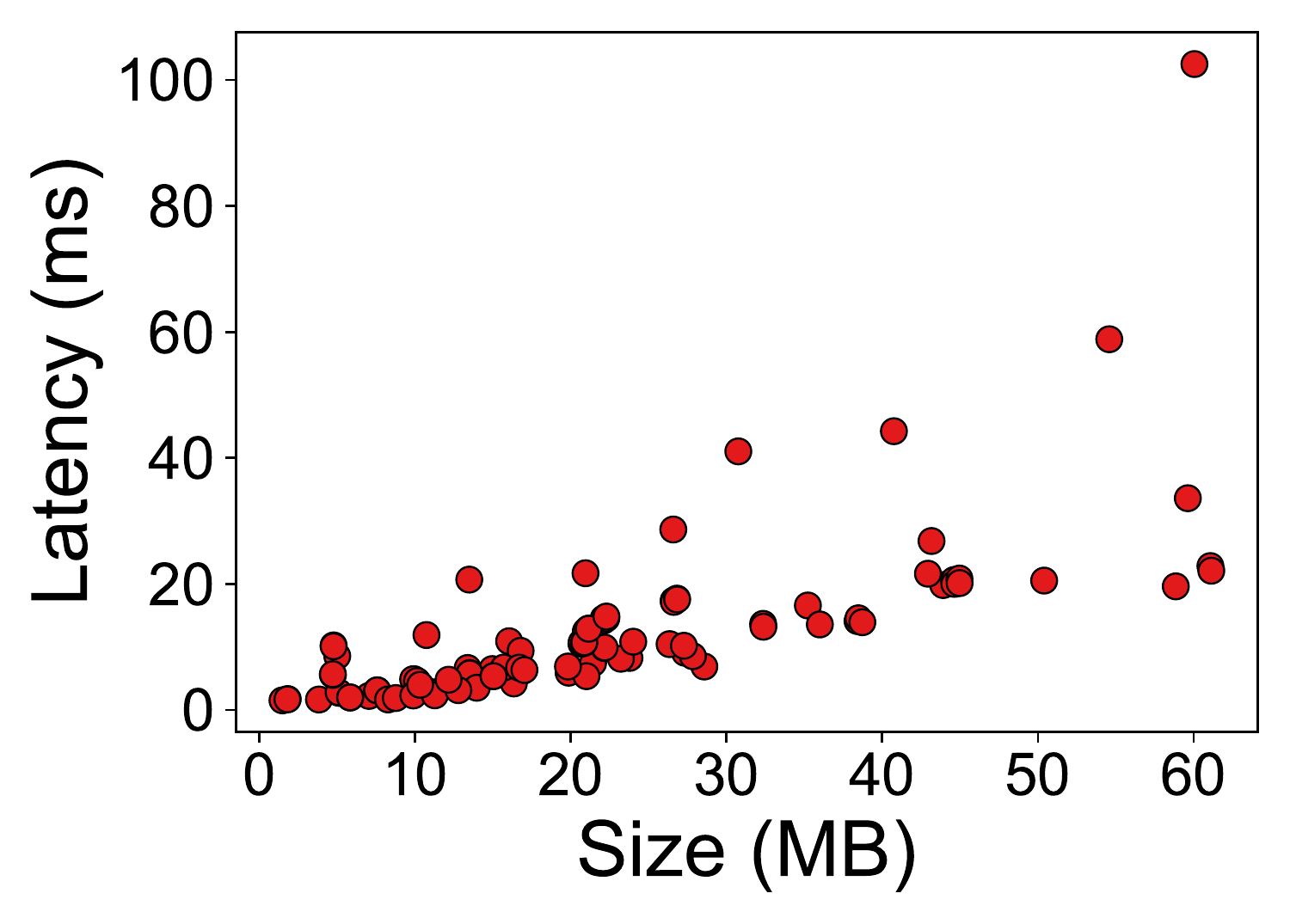}\vspace{-1em}
	\caption{State-of-the-art Neural Architectures}
	\label{fig:model_size_latency}
\end{figure}

\cref{fig:model_size_latency} illustrates the range of model sizes and end-to-end latencies of these real-world neural architectures (on Adreno~640).
\cref{table:feature} presents a summary of the operations that can be found in these architectures; for each operation, we provide the list of parameters used as input features to train our latency predictors.

%In order to construct ML models predicting the latency of each operation, we select features including (1) configuration parameters which describing the shape of the operations, and (2) augmented features related to both memory access (e.g., input size, output size, kernel size) and computation cost (e.g., FLOPs).

\begin{table}[h!]
\renewcommand{\arraystretch}{1.32}
\footnotesize
\begin{tabular}{p{.18\linewidth} p{.72\linewidth}}
\toprule
Operation / Kernel & Features \\
\midrule
Conv2D, Winograd, DepthwiseConv2D & Input height (width), input channel, output height (width), stride, kernel height (width), filters, input size, output size, kernel size, FLOPs \\
GroupedConv2D & Input height (width), input channel, output height (width), stride, kernel height (width), filters, input size, output size, kernel size, group number, FLOPs \\
% DepthwiseConv2D & Input height (width), input channel, output height (width), stride, kernel height (width), filters, FLOPS, input size, output size, kernel size \\
FullyConnected & Input channel, filters, parameter size, FLOPs \\
Mean & Input height (width), input channel, kernel height (width), input size, FLOPs \\
Concat, Split & Input height (width), input channel, kernel height (width), output channel, input size, output size \\
Pooling & Input height (width), input channel, output height (width), stride, kernel height (width), input size, output size, FLOPs \\
Padding & Input height (width), input channel, output height (width), padding size, output size \\
Element-wise & Input height (width), input channel, input size \\
\bottomrule
\end{tabular}
\caption{Feature Space for Each Category of Operations}
\label{table:feature}\vspace{-2em}
\end{table}

\section{Supplementary Data}\label{appendix:supplementary_results}

\begin{figure}[t]
	\centering
	\begin{subfigure}[b]{.49\linewidth}
		\centering
		\includegraphics[width=\linewidth]{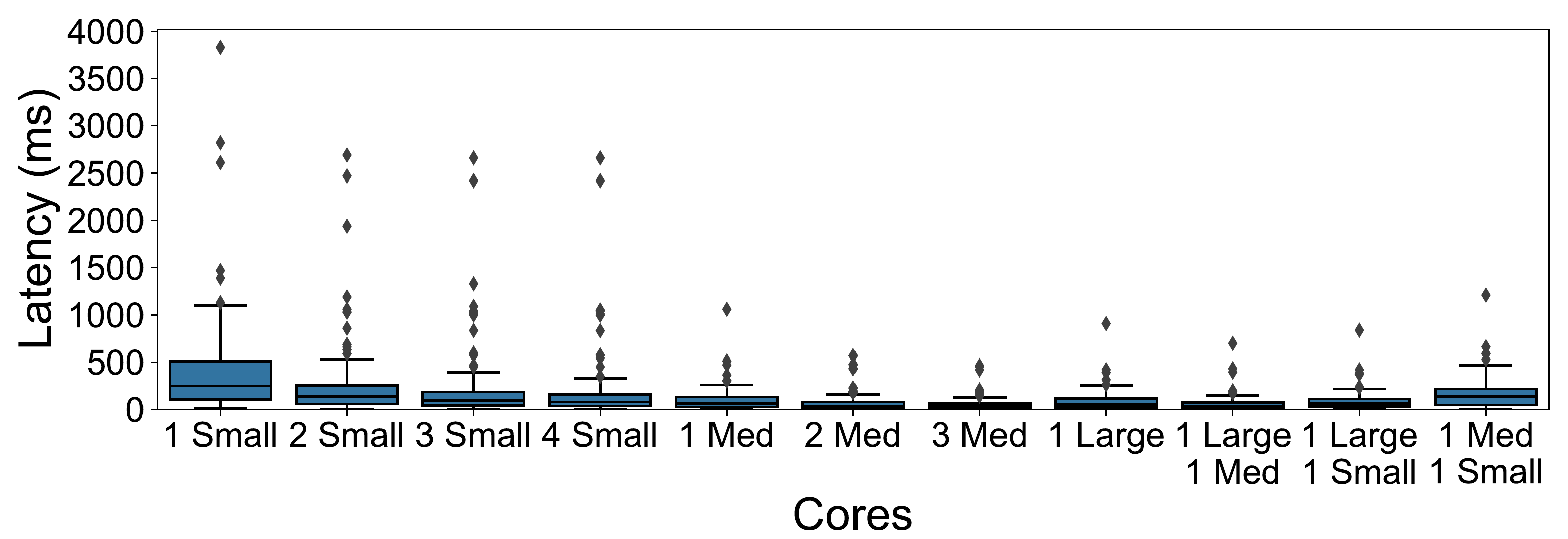}
		\caption{Snapdragon 855}\label{fig:appendix_multithread_pixel4}
	\end{subfigure}
	\begin{subfigure}[b]{.49\linewidth}
		\centering
		\includegraphics[width=\linewidth]{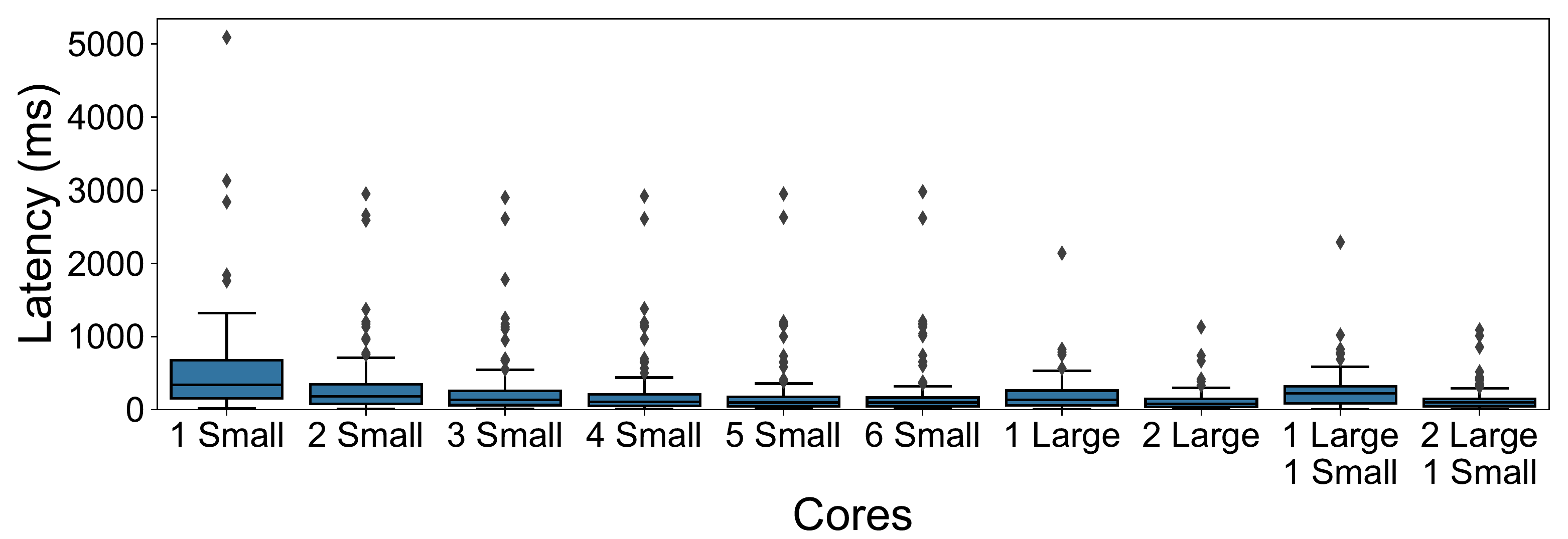}
		\caption{Snapdragon 710}\label{fig:appendix_multithread_mi8se}
	\end{subfigure}
	\begin{subfigure}[b]{.49\linewidth}
		\centering
		\includegraphics[width=\linewidth]{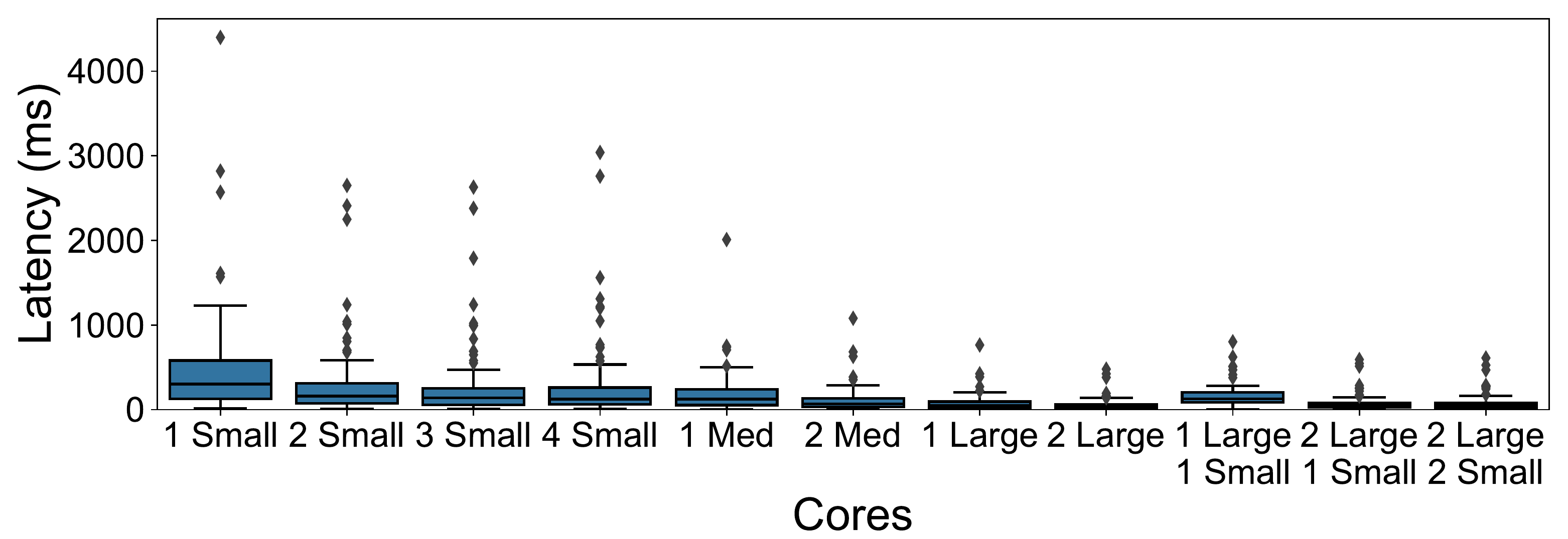}
		\caption{Exynos 9820}\label{fig:appendix_multithread_s10}
	\end{subfigure}
	\begin{subfigure}[b]{.49\linewidth}
		\centering
		\includegraphics[width=\linewidth]{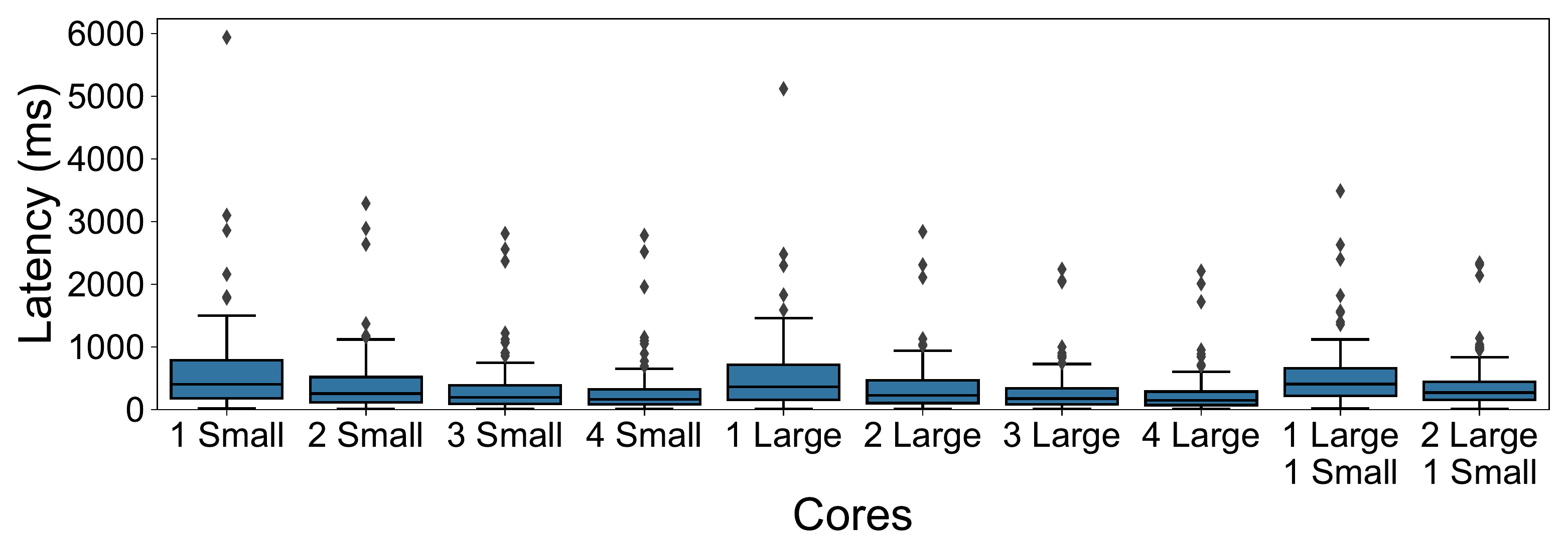}
		\caption{Helio P35}\label{fig:appendix_multithread_a03s}
	\end{subfigure}
	\caption{Effects of Multithreading on End-to-end Latency}
	\label{fig:appendix_multithread}
\end{figure}

\begin{figure}[t]
	\centering

	\begin{subfigure}[b]{.49\linewidth}
		\centering
		\includegraphics[width=\linewidth]{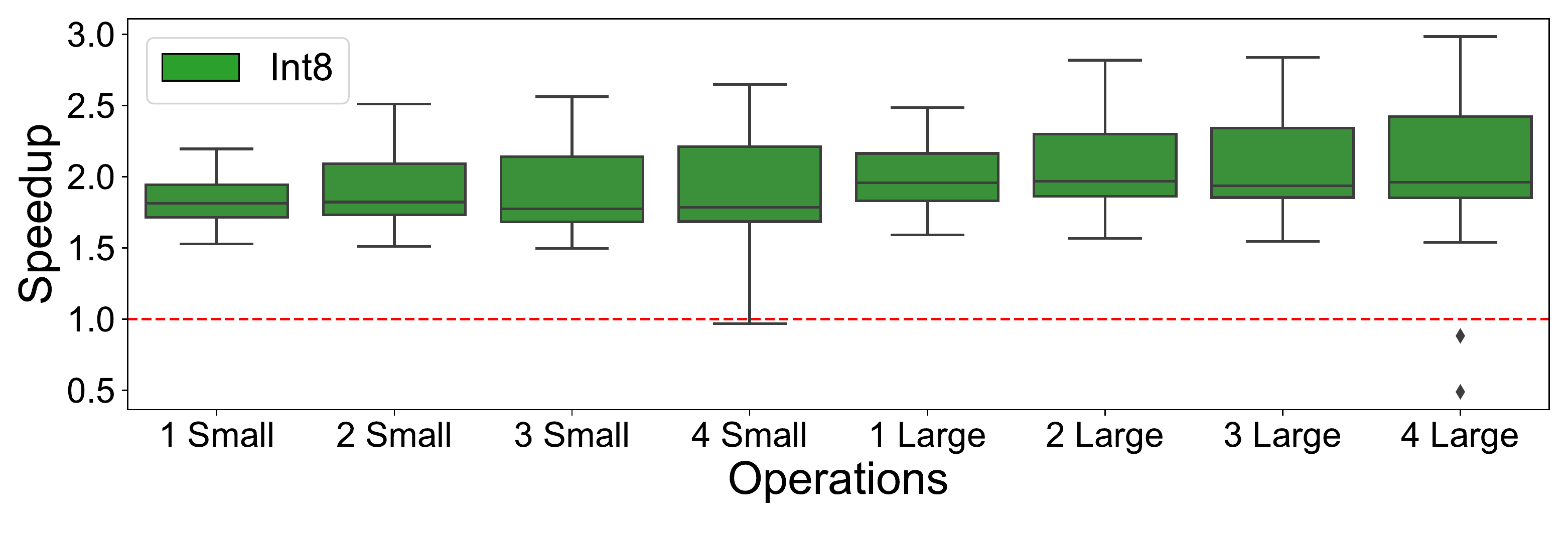}
		\caption{Helio P35}\label{fig:quantization_a03s_1}
	\end{subfigure}
	\begin{subfigure}[b]{.49\linewidth}
		\centering
		\includegraphics[width=\linewidth]{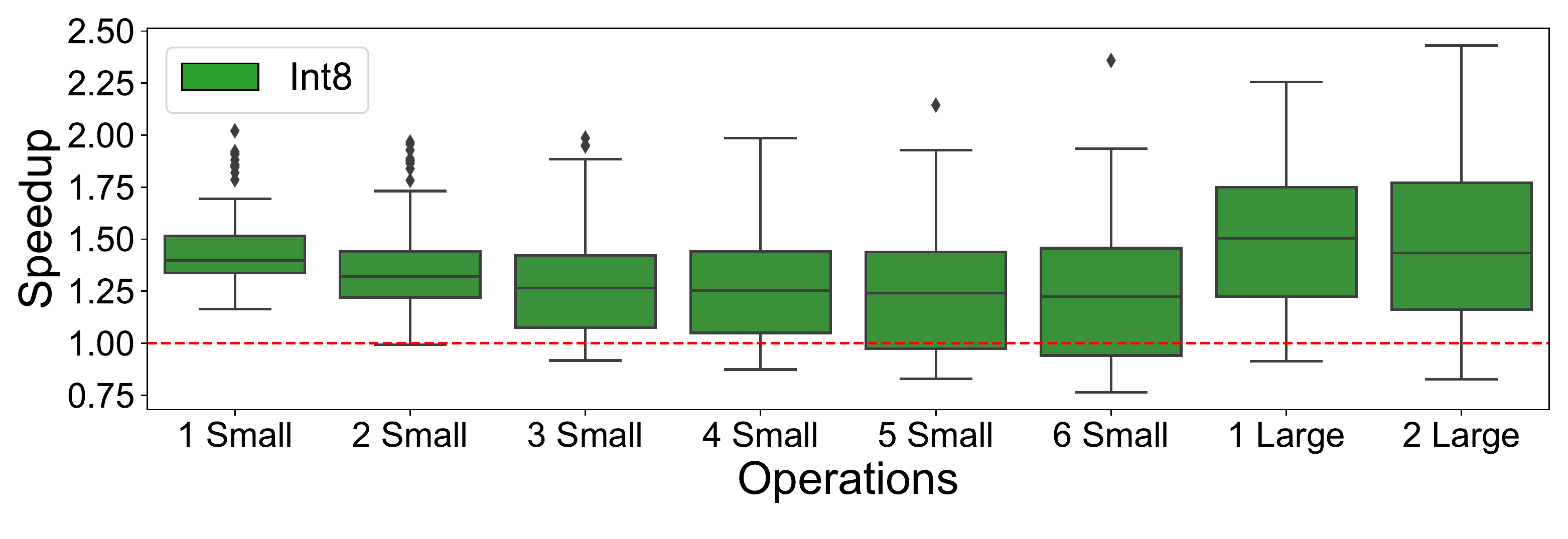}
		\caption{Snapdragon 710}\label{fig:quantization_mi8se_1}
	\end{subfigure}

	\begin{subfigure}[b]{.49\linewidth}
		\centering
		\includegraphics[width=\linewidth]{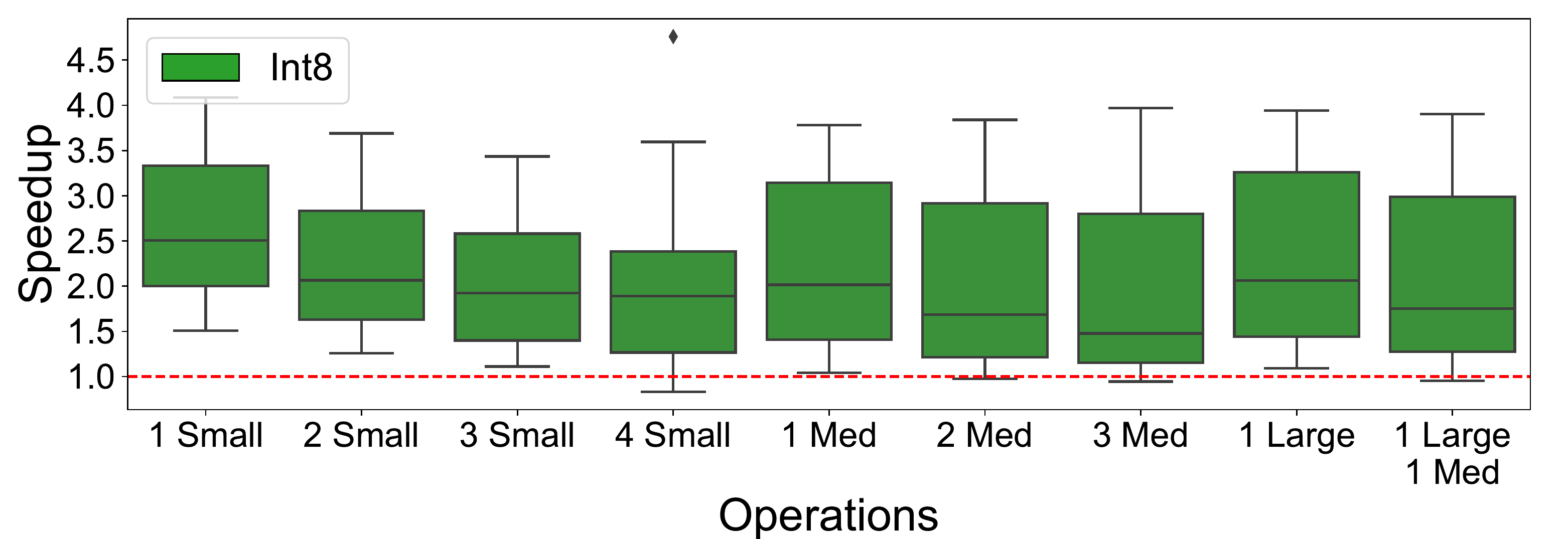}
		\caption{Snapdragon 855}\label{fig:quantization_pixel4_1}
	\end{subfigure}
	\begin{subfigure}[b]{.49\linewidth}
		\centering
		\includegraphics[width=\linewidth]{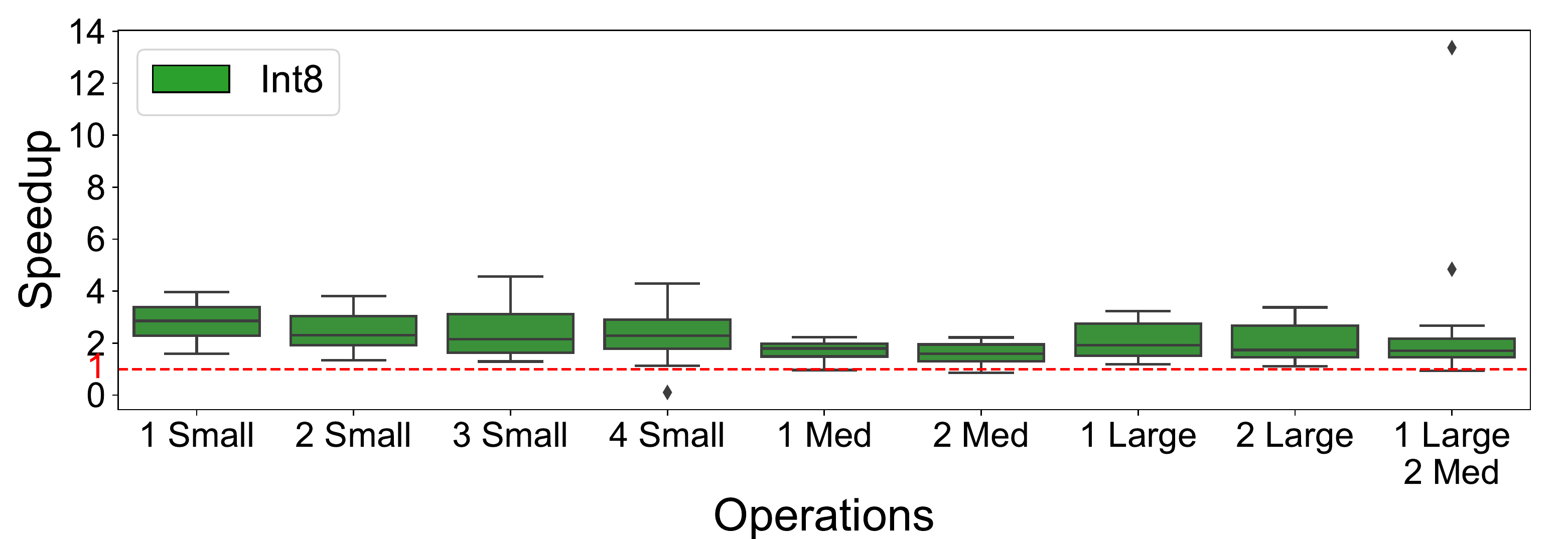}
		\caption{Exynos 9820}\label{fig:quantization_s10_1}
	\end{subfigure}

	\caption{Effects of Quantization on End-to-end Latency}
	\label{fig:appendix_quantization}
\end{figure}

\begin{figure}[t]
	\centering
	\includegraphics[width=.4\linewidth]{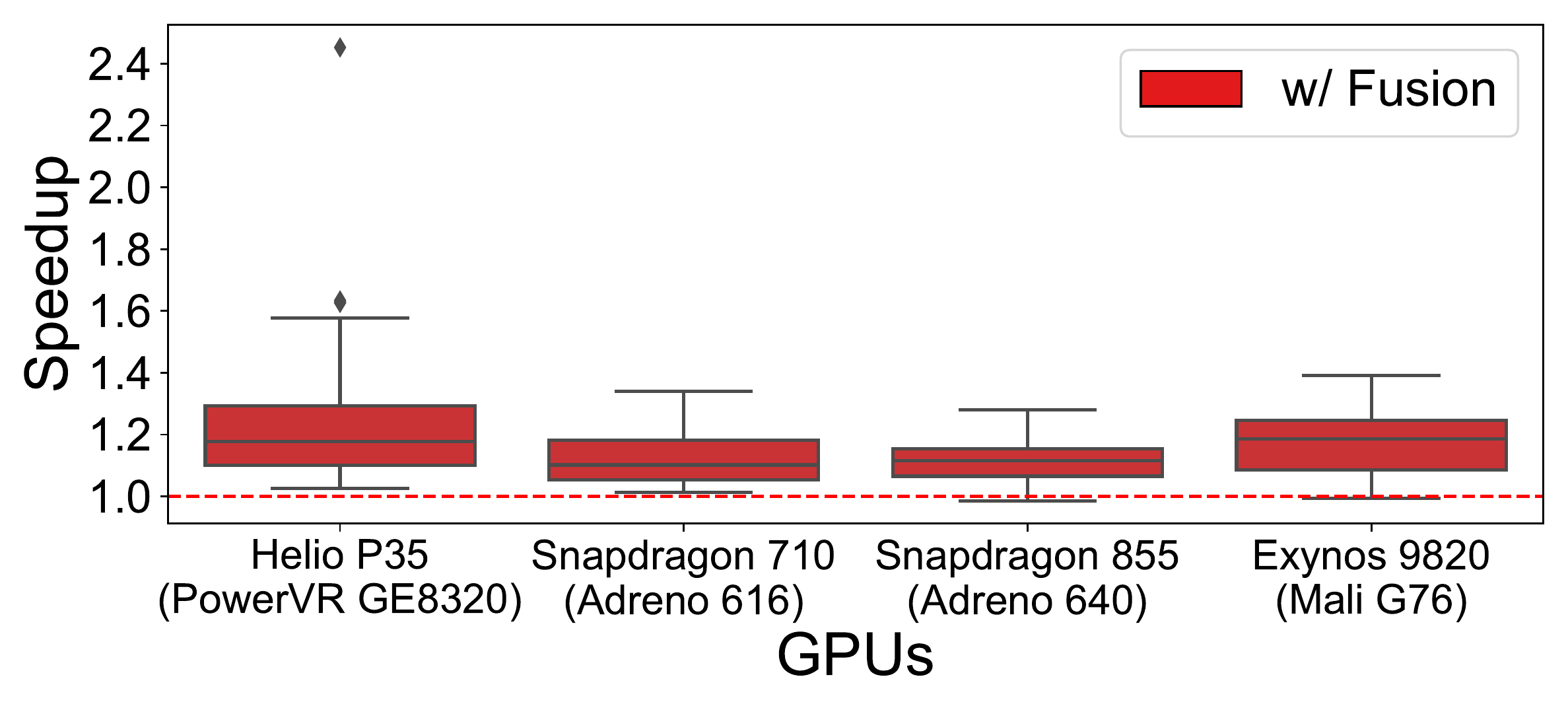}
	\caption{Effects of Kernel Fusion on End-to-end Latency}
	\label{fig:appendix_fusion}
\end{figure}

\begin{figure}[t]
	\centering

	\begin{subfigure}[b]{.49\linewidth}
		\centering
		\includegraphics[width=\linewidth]{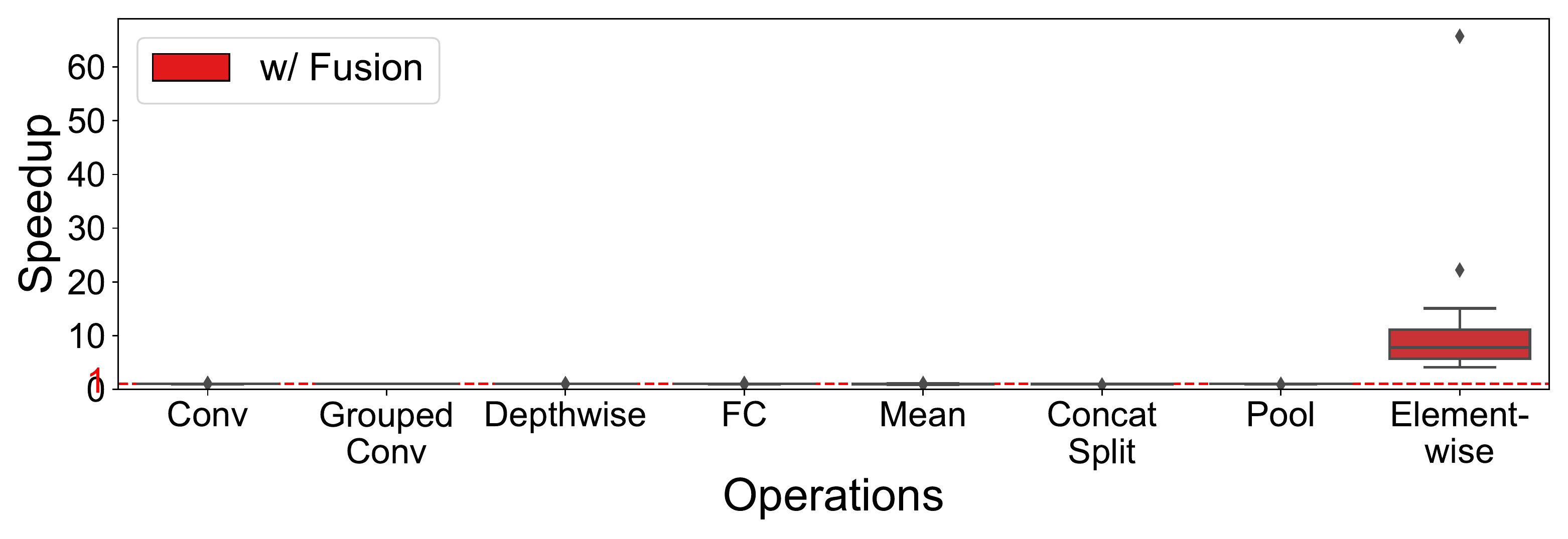}
		\caption{Exynos 9820 (Mali G76)}\label{fig:kernel_fusion_ops_pixel4_outliers}
	\end{subfigure}
	\begin{subfigure}[b]{.49\linewidth}
		\centering
		\includegraphics[width=\linewidth]{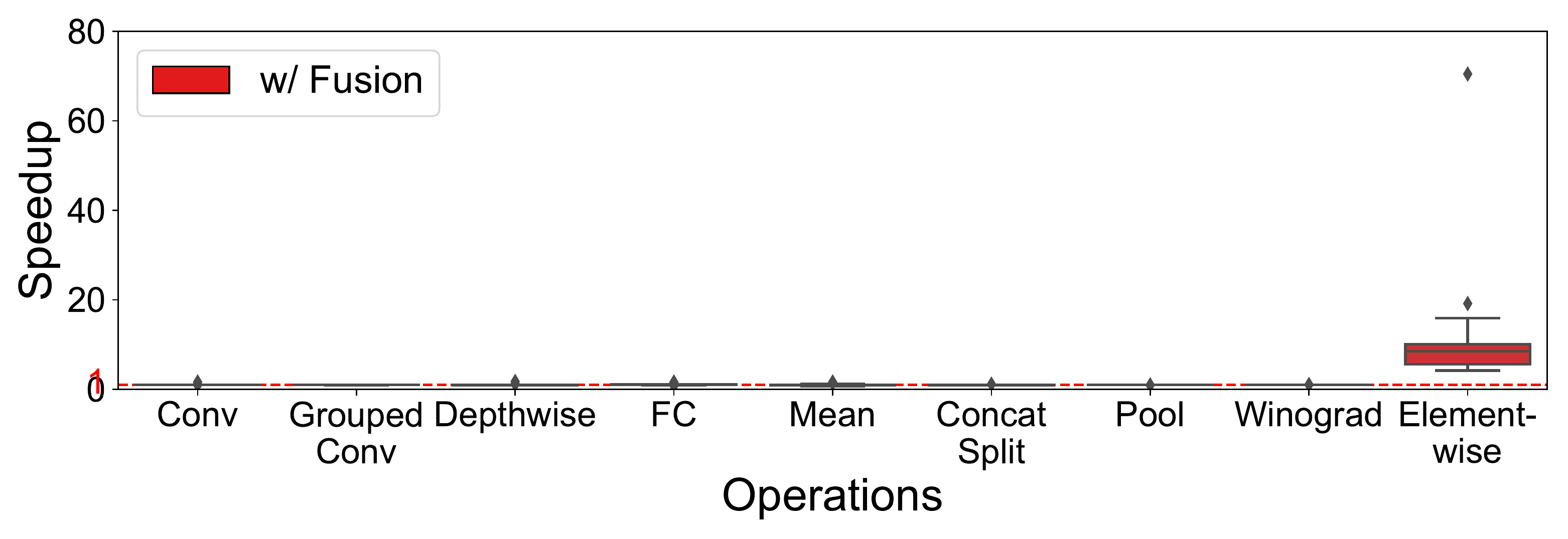}
		\caption{Snapdragon 710 (Adreno 616)}\label{fig:kernel_fusion_ops_a03s_outliers}
	\end{subfigure}

% 	\begin{subfigure}[b]{.49\linewidth}
% 		\centering
% 		\includegraphics[width=\linewidth]{figures/performance_box_plot/kernel_fusion_ops_s10_outliers.pdf}
% 		\caption{Exynos 9820 (Mali G76)}\label{fig:kernel_fusion_ops_s10_outliers}
% 	\end{subfigure}
% 	\begin{subfigure}[b]{.49\linewidth}
% 		\centering
% 		\includegraphics[width=\linewidth]{figures/performance_box_plot/kernel_fusion_ops_mi8se_outliers.pdf}
% 		\caption{Snapdragon 710 (Adreno 616)}\label{fig:kernel_fusion_ops_mi8se_outliers}
% 	\end{subfigure}
	
% 	\begin{subfigure}[b]{.49\linewidth}
% 		\centering
% 		\includegraphics[width=\linewidth]{figures/performance_box_plot/kernel_fusion_ops_s10.pdf}
% 		\caption{Exynos 9820 (Mali G76)}\label{fig:kernel_fusion_ops_s10_2}
% 	\end{subfigure}
% 	\begin{subfigure}[b]{.49\linewidth}
% 		\centering
% 		\includegraphics[width=\linewidth]{figures/performance_box_plot/kernel_fusion_ops_mi8se.pdf}
% 		\caption{Snapdragon 710 (Adreno 616)}\label{fig:kernel_fusion_ops_mi8se_2}
% 	\end{subfigure}

	\caption{Effects of Kernel Fusion on Operation-wise Latency}
	\label{fig:appendix_fusion_ops}
\end{figure}

% \begin{figure}[t]
% 	\centering
% 	\begin{subfigure}[b]{.49\linewidth}
% 		\centering
% 		\includegraphics[width=\linewidth]{figures/performance_box_plot/kernel_optimized_winograd_pixel4.pdf}
% 		\caption{Snapdragon 855 (Adreno 640)}\label{fig:optimized_kernel_winograd_pixel4_1}
% 	\end{subfigure}
% 	\begin{subfigure}[b]{.49\linewidth}
% 		\centering
% 		\includegraphics[width=\linewidth]{figures/performance_box_plot/kernel_optimized_winograd_mi8se.pdf}
% 		\caption{Snapdragon 710 (Adreno 616)}\label{fig:optimized_kernel_winograd_mi8se_1}
% 	\end{subfigure}

% 	\caption{Effects of Using Winograd Kernels on End-to-end Latency}
% 	\label{fig:appendix_optimized_kernel_winograd}
% \end{figure}

In this appendix, we include supplementary data from our measurements and prediction results. This data is provided here for completeness, e.g., to include the full set of outliers that were omitted in some of the figures of the main text due to lack of space and for clarity of presentation.
\cref{fig:appendix_multithread} depicts end-to-end latency of state-of-the-art neural architectures on 4 platforms for different multi-core configurations, including the outliers omitted in \cref{fig:multithread} for clarity of presentation (\cref{sec:multithreading}).
\cref{fig:appendix_quantization} depicts the speedup from quantization, including the small set of outliers omitted in \cref{fig:quantization} (\cref{sec:quantization}).
\cref{fig:appendix_fusion,fig:appendix_fusion_ops} present the speedup of kernel fusion on end-to-end latency and on each type of operations, respectively, including the small set of outliers omitted in \cref{fig:kernel_fusion_speedup,fig:kernel_fusion_ops} (\cref{sec:kernel_fusion}).

\begin{figure}[t]
	\centering
	\begin{subfigure}[b]{.49\linewidth}
		\centering
		\includegraphics[width=\linewidth]{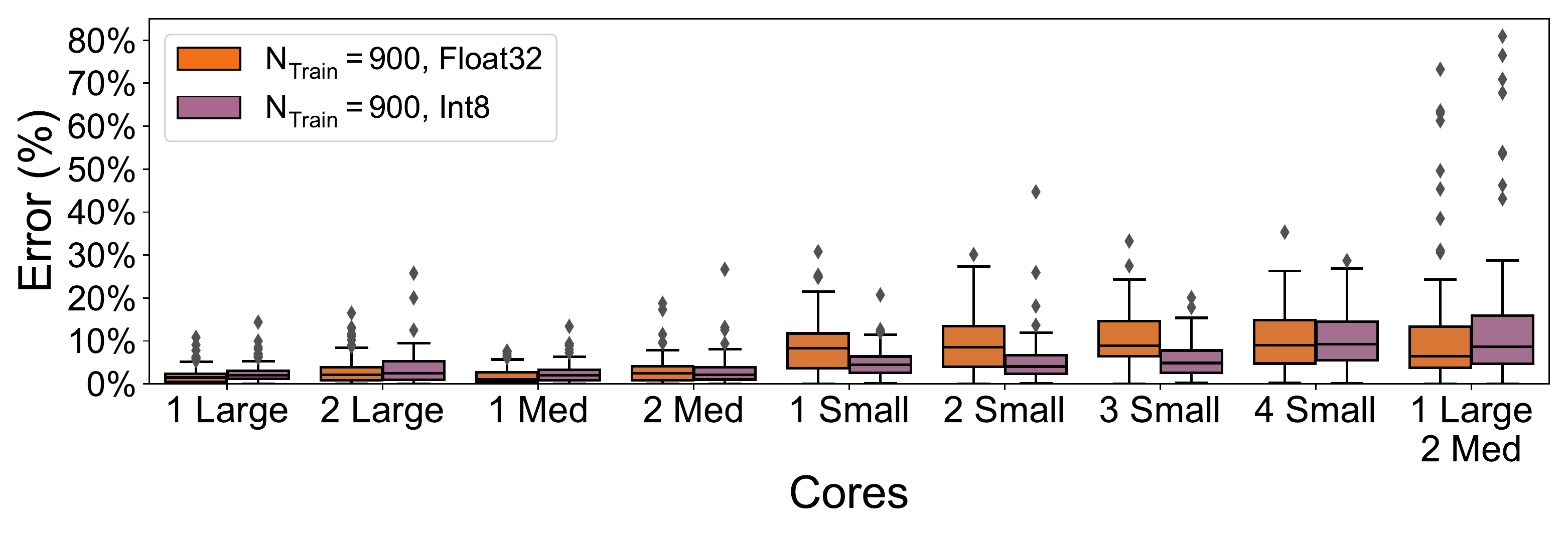}
		\caption{Exynos 9820}\label{fig:result_nas_900_GBDT_cpu_s10_outliers}
	\end{subfigure}
	\begin{subfigure}[b]{.49\linewidth}
		\centering
		\includegraphics[width=\linewidth]{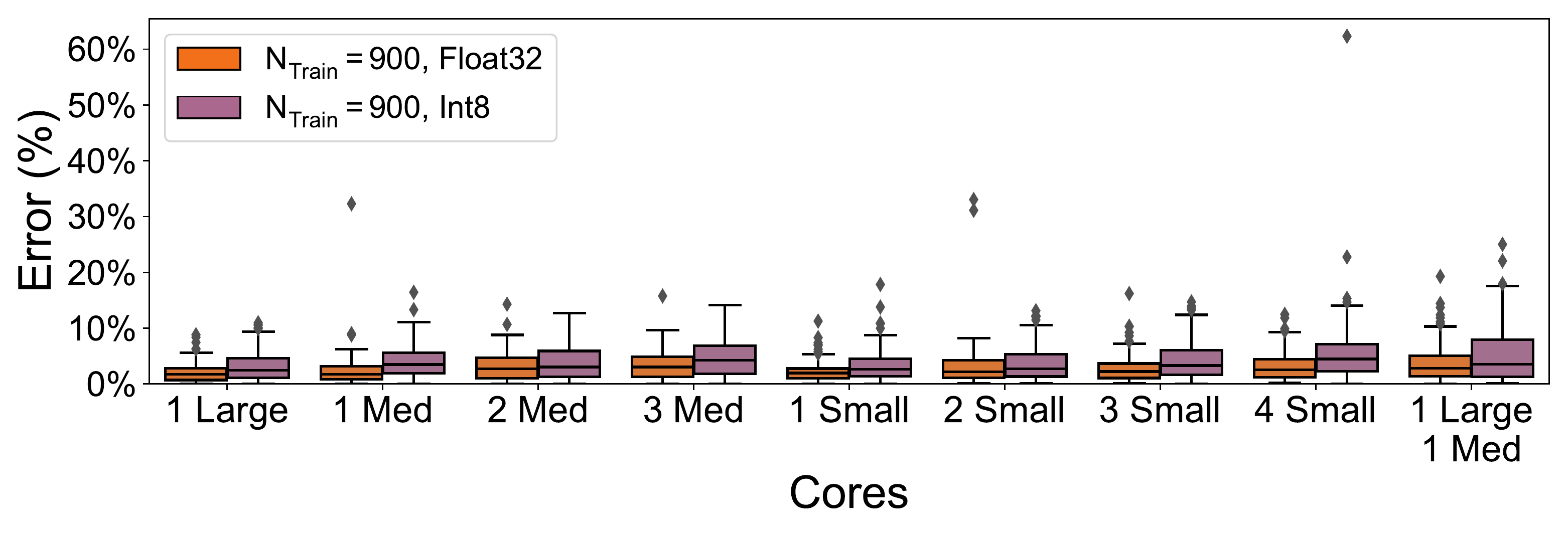}
		\caption{Snapdragon 855}\label{fig:result_nas_900_GBDT_cpu_pixel4_outliers}
	\end{subfigure}
	\begin{subfigure}[b]{.49\linewidth}
		\centering
		\includegraphics[width=\linewidth]{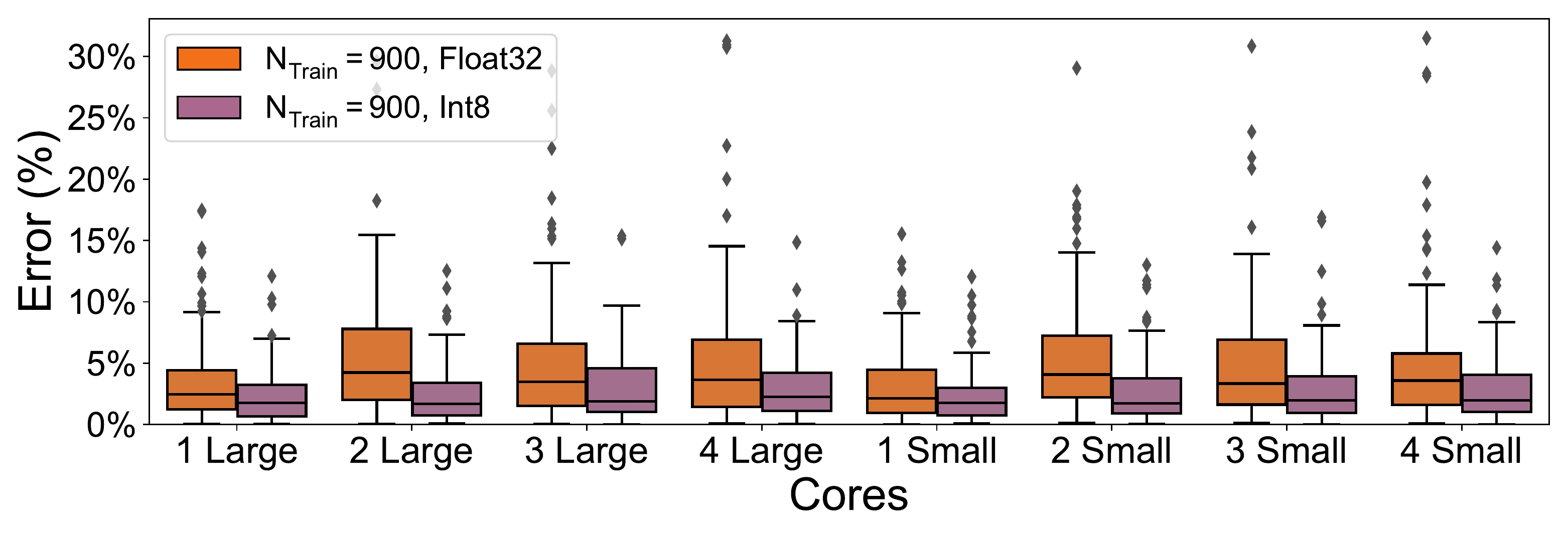}
		\caption{Helio P35}\label{fig:result_nas_900_GBDT_cpu_a03s_outliers}
	\end{subfigure}
	\begin{subfigure}[b]{.49\linewidth}
		\centering
		\includegraphics[width=\linewidth]{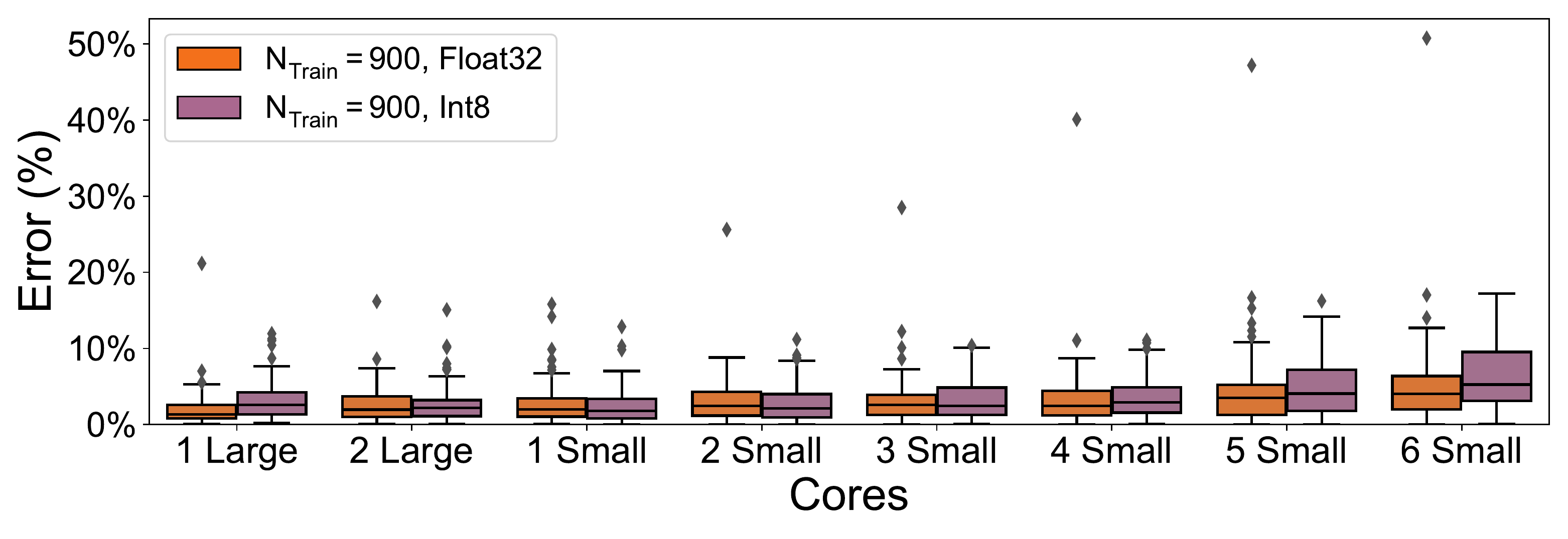}
		\caption{Snapdragon 710}\label{fig:result_nas_900_GBDT_cpu_mi8se_outliers}
	\end{subfigure}

	\caption{Predictions of GBDT on End-to-end Latency with Multiple CPU Cores (Synthetic Neural Architectures)}
	\label{fig:appendix_nas_cpu_end_to_end_mape}
\end{figure}

\begin{figure}[t]
	\centering
	\begin{subfigure}[b]{.49\linewidth}
		\centering
		\includegraphics[width=\linewidth]{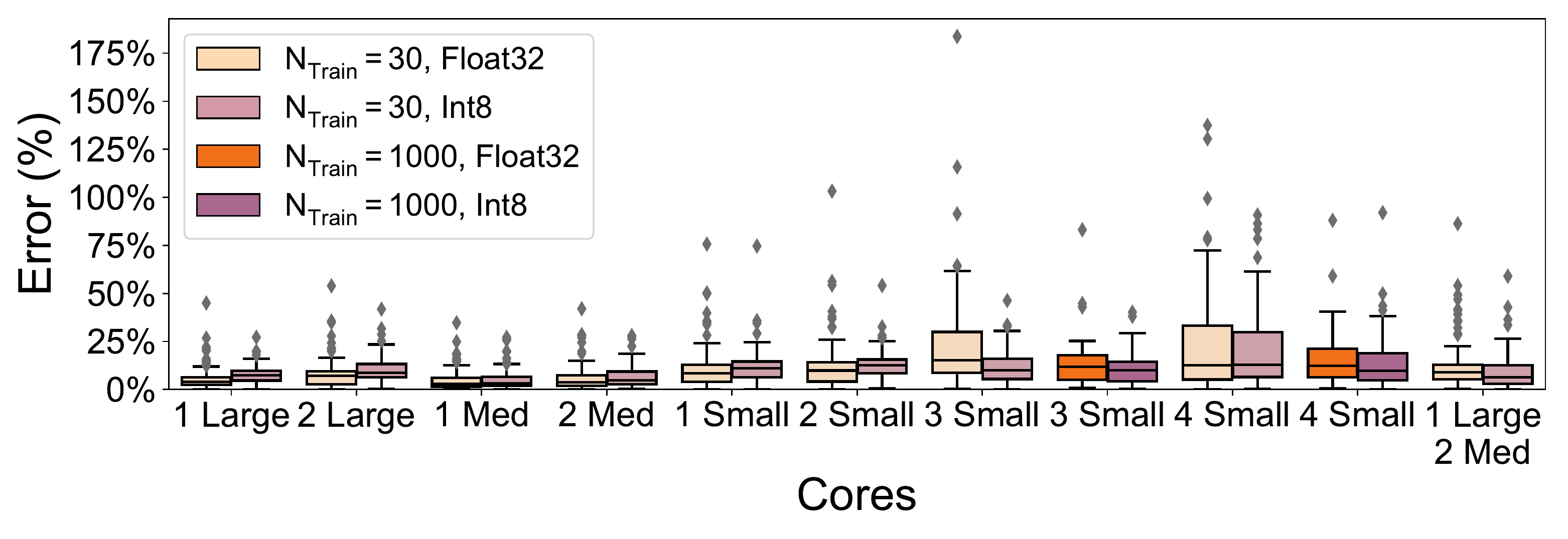}
		\caption{Exynos 9820}\label{fig:result_common_30_lasso_cpu_s10_outliers}
	\end{subfigure}
	\begin{subfigure}[b]{.49\linewidth}
		\centering
		\includegraphics[width=\linewidth]{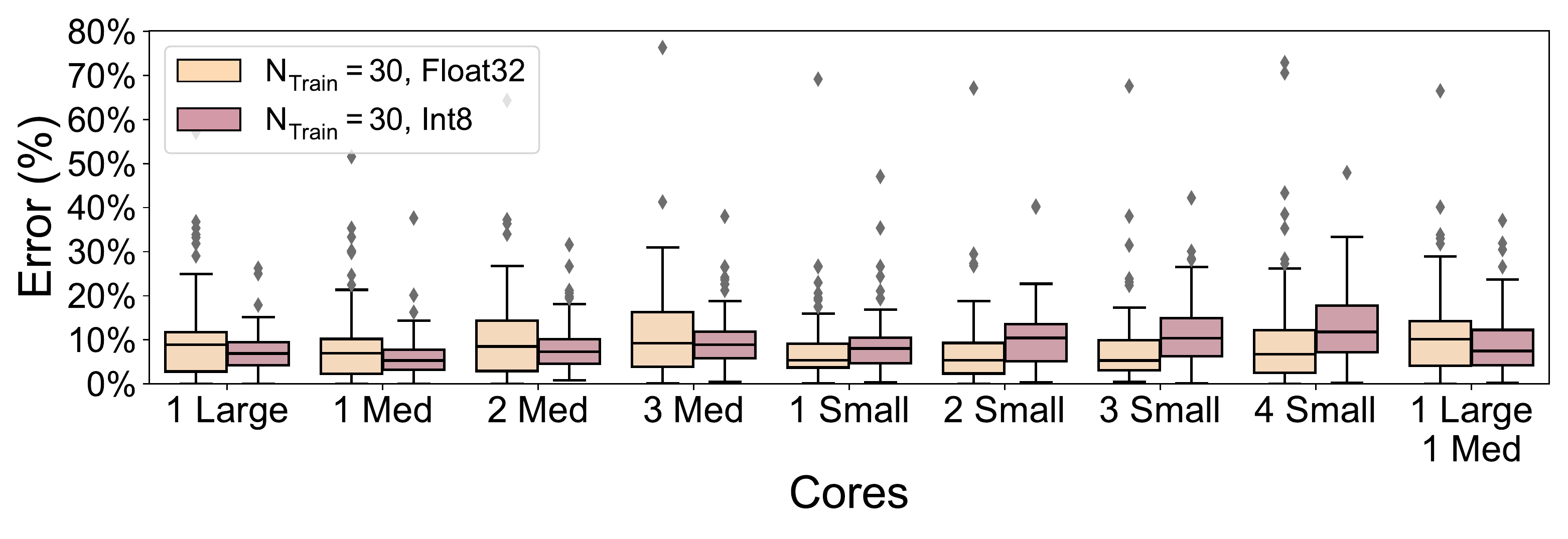}
		\caption{Snapdragon 855}\label{fig:result_common_30_lasso_cpu_pixel4_outliers}
	\end{subfigure}
	\begin{subfigure}[b]{.49\linewidth}
		\centering
		\includegraphics[width=\linewidth]{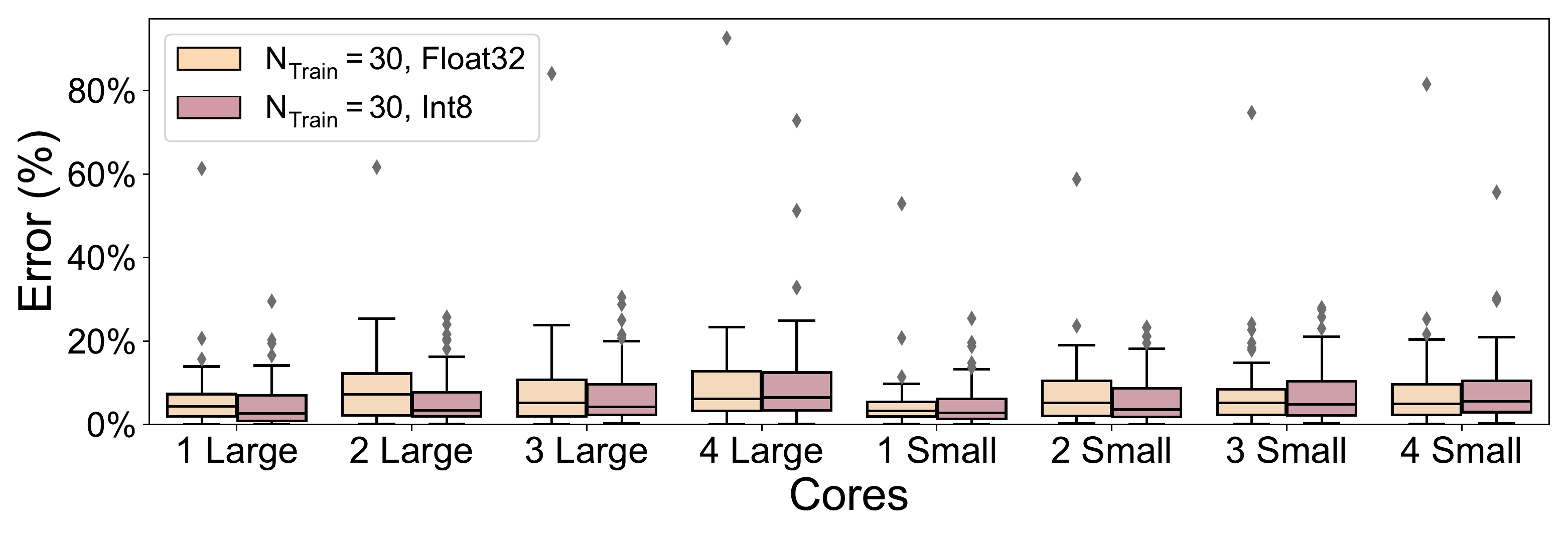}
		\caption{Helio P35}\label{fig:result_common_30_lasso_cpu_a03s_outliers}
	\end{subfigure}
	\begin{subfigure}[b]{.49\linewidth}
		\centering
		\includegraphics[width=\linewidth]{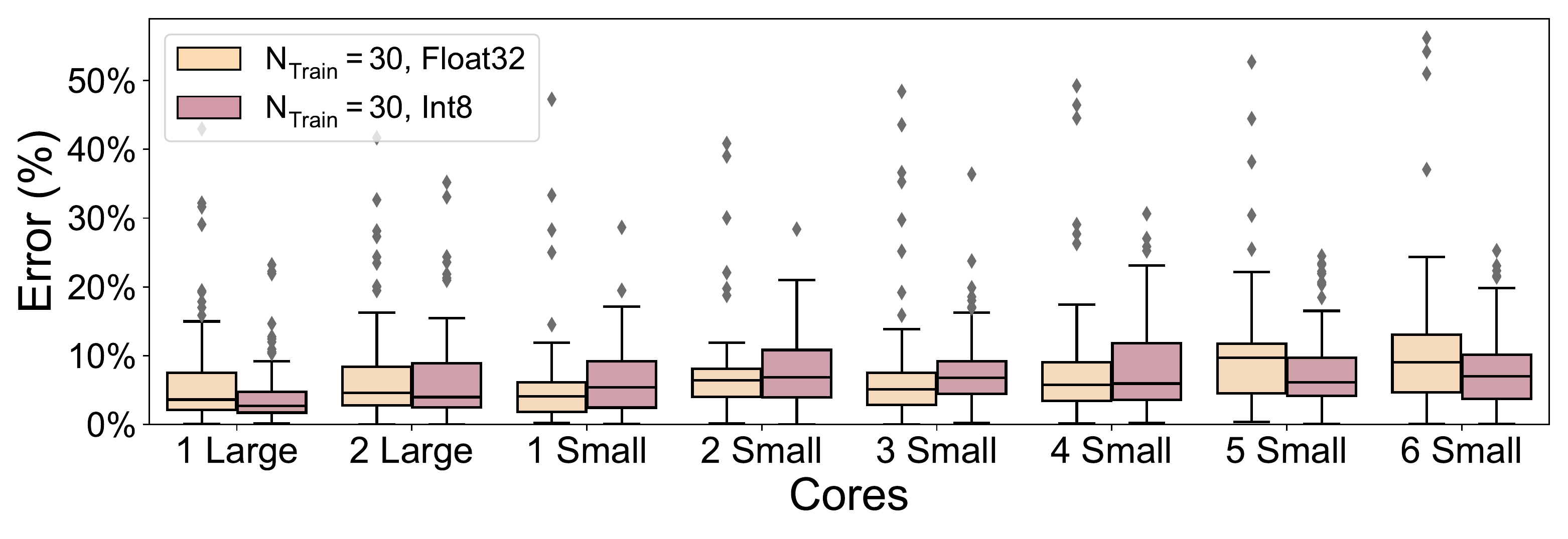}
		\caption{Snapdragon 710}\label{fig:result_common_30_lasso_cpu_mi8se_outliers}
	\end{subfigure}

	\caption{Predictions of Lasso on End-to-end Latency with Multiple CPU Cores (Real-world Neural Architectures)}
	\label{fig:appendix_common_cpu_end_to_end_mape}
\end{figure}

\begin{figure}[t]
	\centering
	\begin{subfigure}[b]{.49\linewidth}
		\centering
		\includegraphics[width=\linewidth]{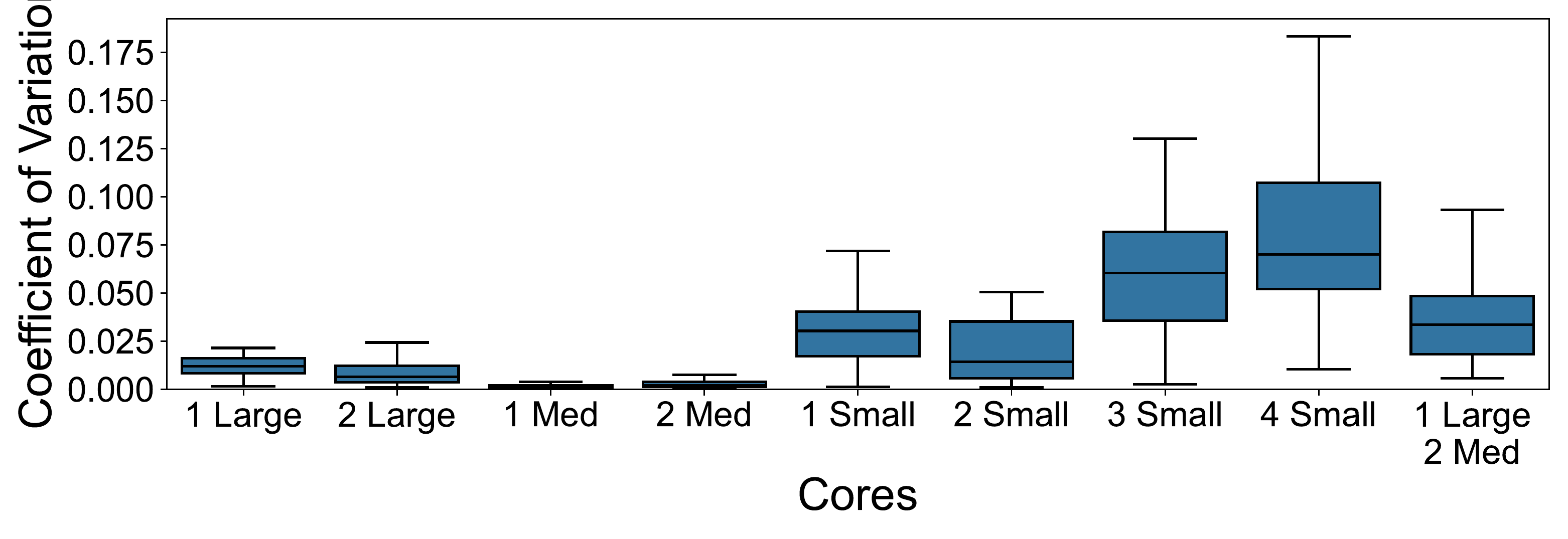}
		\caption{Exynos 9820}\label{fig:result_cv_cpu_s10}
	\end{subfigure}
	\begin{subfigure}[b]{.49\linewidth}
		\centering
		\includegraphics[width=\linewidth]{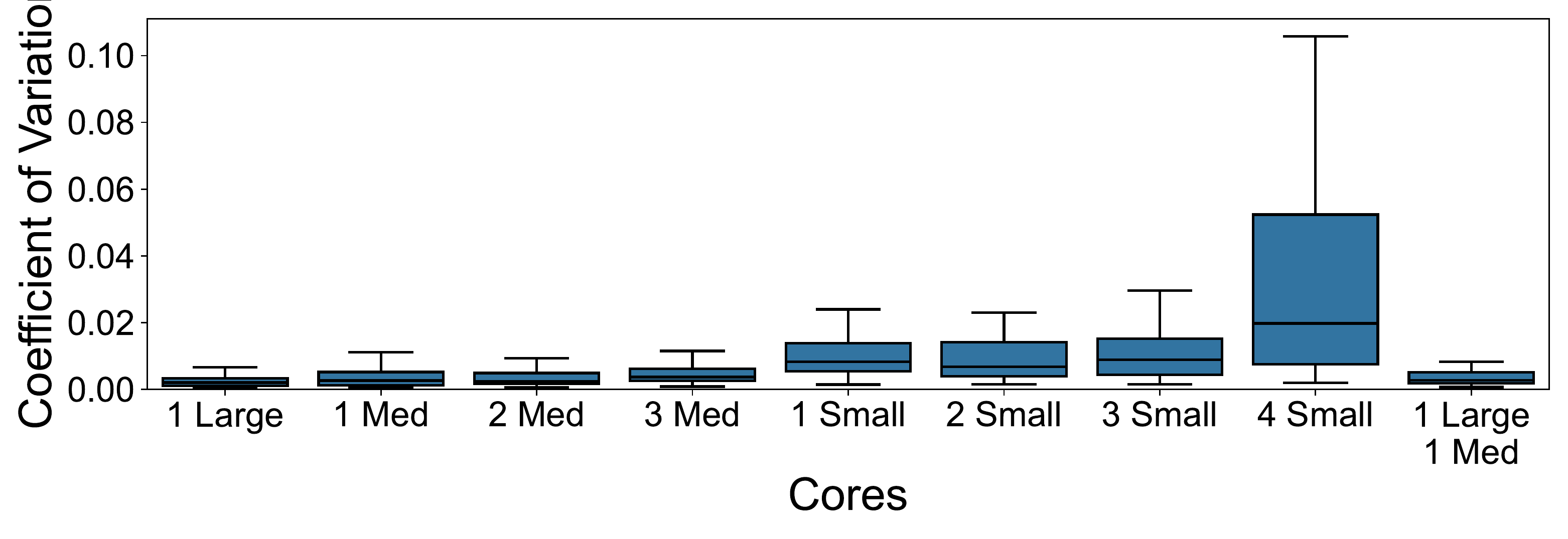}
		\caption{Snapdragon 855}\label{fig:result_cv_cpu_pixel4}
	\end{subfigure}
	\begin{subfigure}[b]{.49\linewidth}
		\centering
		\includegraphics[width=\linewidth]{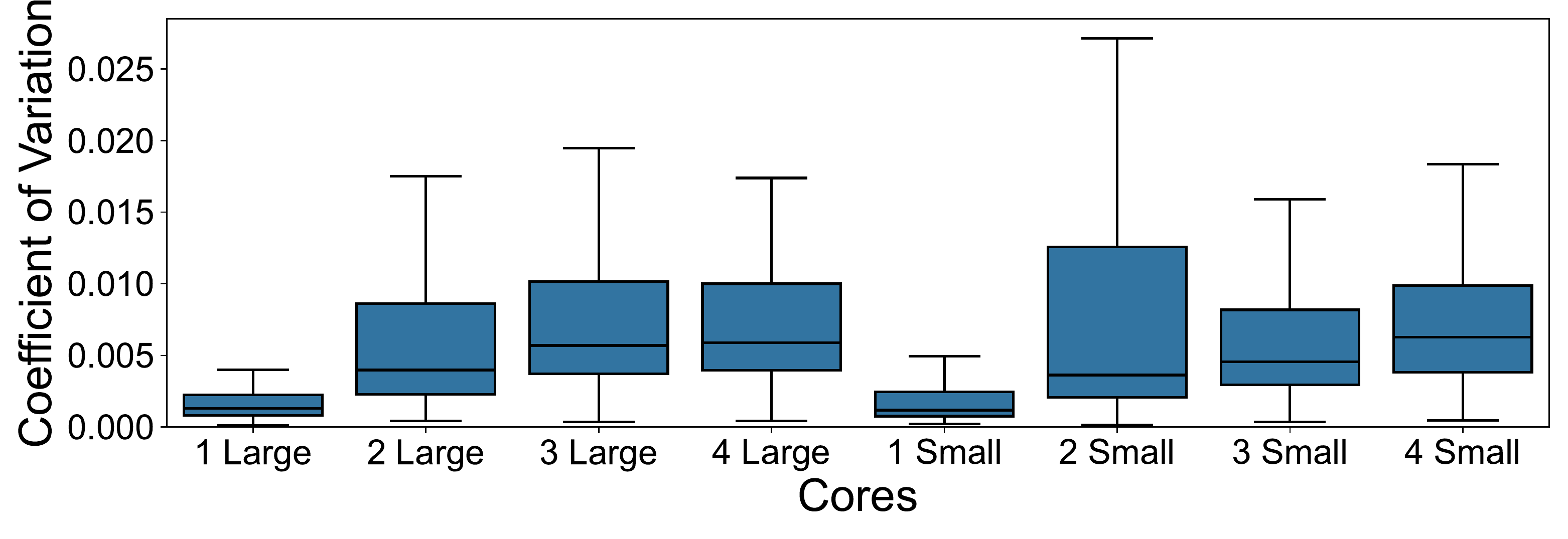}
		\caption{Helio P35}\label{fig:result_cv_cpu_a03s}
	\end{subfigure}
	\begin{subfigure}[b]{.49\linewidth}
		\centering
		\includegraphics[width=\linewidth]{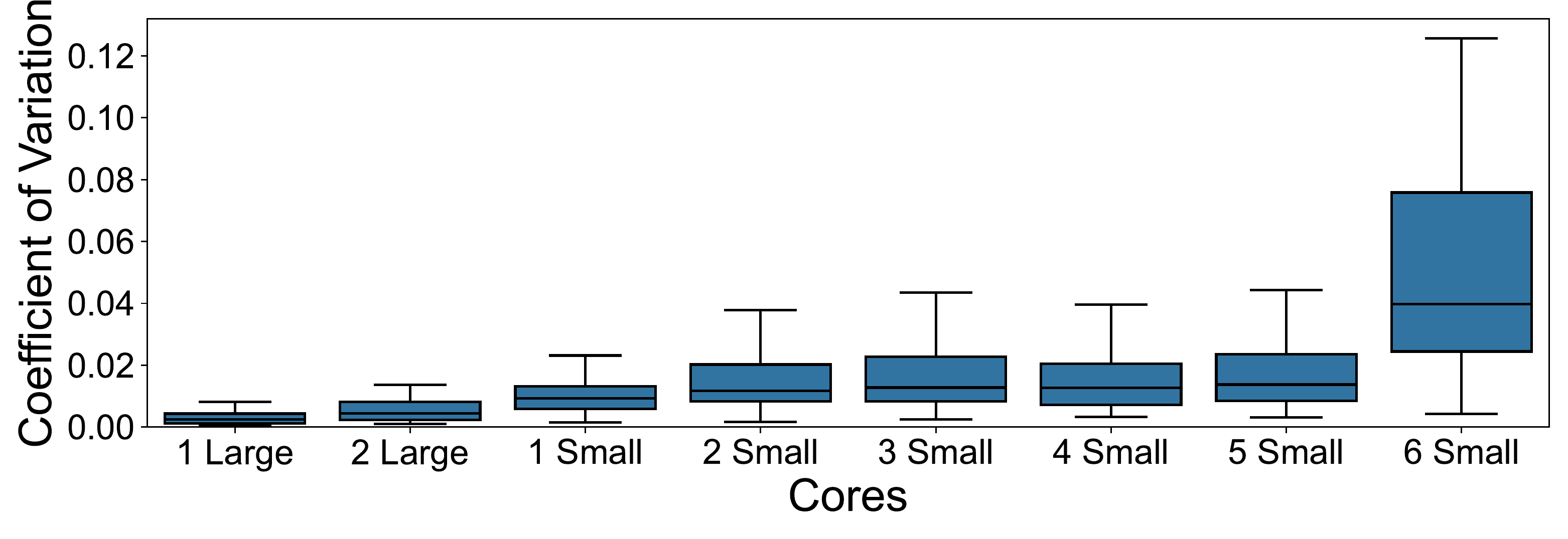}
		\caption{Snapdragon 710}\label{fig:result_cv_cpu_mi8se}
	\end{subfigure}

% 	\begin{subfigure}[b]{.49\linewidth}
% 		\centering
% 		\includegraphics[width=\linewidth]{figures/results_box_plot/result_cv_cpu_s10_outliers.pdf}
% 		\caption{Exynos 9820}\label{fig:result_cv_cpu_s10_outliers}
% 	\end{subfigure}
% 	\begin{subfigure}[b]{.49\linewidth}
% 		\centering
% 		\includegraphics[width=\linewidth]{figures/results_box_plot/result_cv_cpu_pixel4_outliers.pdf}
% 		\caption{Snapdragon 855}\label{fig:result_cv_cpu_pixel4_outliers}
% 	\end{subfigure}
% 	\begin{subfigure}[b]{.49\linewidth}
% 		\centering
% 		\includegraphics[width=\linewidth]{figures/results_box_plot/result_cv_cpu_a03s_outliers.pdf}
% 		\caption{Helio P35}\label{fig:result_cv_cpu_a03s_outliers}
% 	\end{subfigure}
% 	\begin{subfigure}[b]{.49\linewidth}
% 		\centering
% 		\includegraphics[width=\linewidth]{figures/results_box_plot/result_cv_cpu_mi8se_outliers.pdf}
% 		\caption{Snapdragon 710}\label{fig:result_cv_cpu_mi8se_outliers}
% 	\end{subfigure}
	\caption{Coefficient of Variance for 100 Test Synthetic Neural Architectures}
	\label{fig:appendix_nas_cv_cpu}
\end{figure}

\begin{figure}[t]
	\centering
	\includegraphics[width=.4\linewidth]{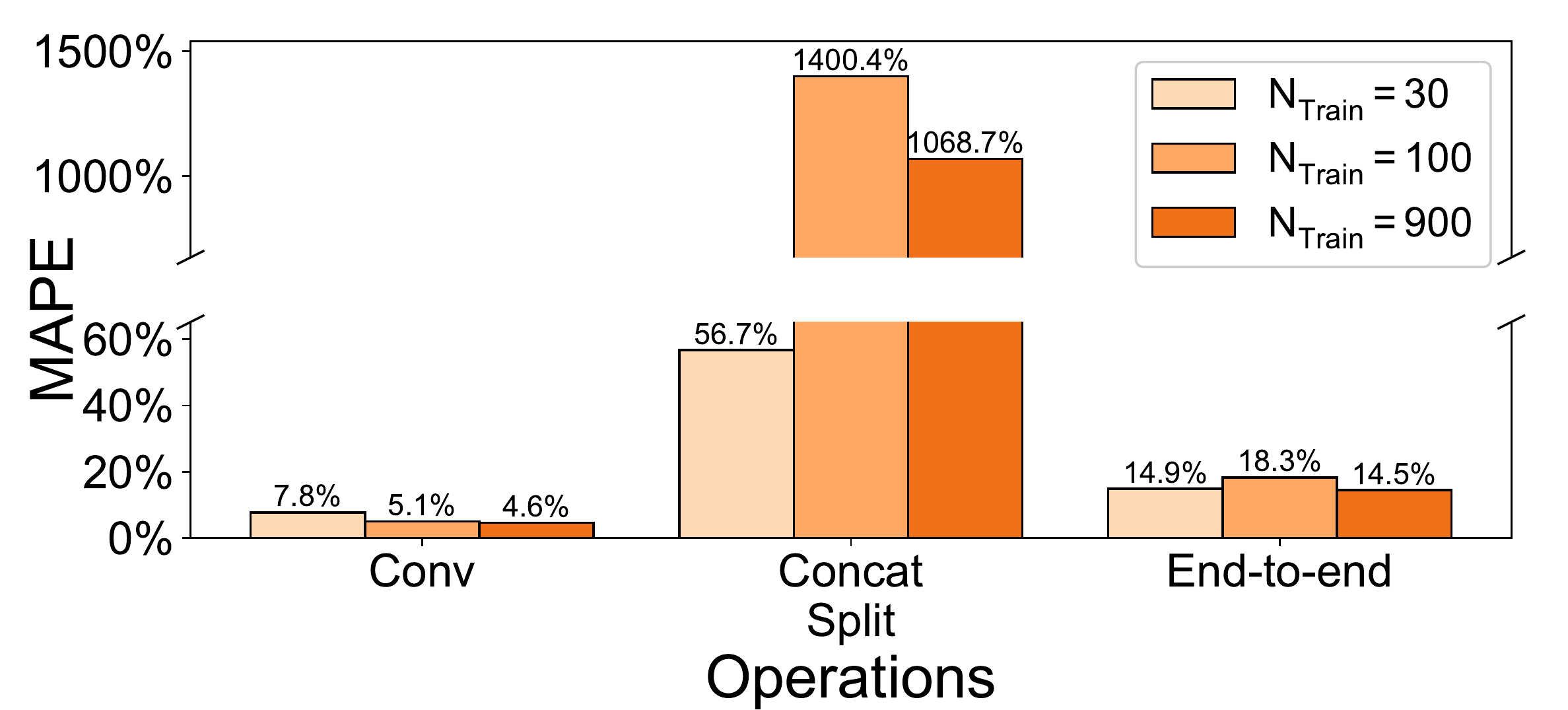}
	\caption{Predictions of MLP on Snapdragon 855 CPU (One Large Core)}
	\label{fig:appendix_comparison_common_cpu_mlp_pixel4}
\end{figure}

\cref{table:end_to_end_predictions_synthetic,table:end_to_end_predictions_real_world} report the complete MAPEs of end-to-end latency predictions on each hardware platform, for synthetic and real-world neural architectures, respectively, across different ML approaches, with varying training set sizes. This detailed data corresponds to the results in \cref{fig:comparison_nas,fig:comparison_common} in the main text where these errors were averaged across hardware platforms.
For predictions on different CPU core combinations and with both floating-point and integer representations, \cref{fig:appendix_nas_cpu_end_to_end_mape} shows the end-to-end latency predictions of GBDT for synthetic neural architectures on various core combinations, including the small set of outliers omitted in \cref{fig:nas_cpu_end_to_end_mape} (\cref{sec:result_hardware_heterogeneity}); 
\cref{fig:appendix_common_cpu_end_to_end_mape} presents the end-to-end latency predictions of Lasso for real-world neural architectures on various core combinations, including the small set of outliers omitted in \cref{fig:common_cpu_end_to_end_mape} (\cref{sec:result_limited_training_data_lasso_predictions}).
\cref{fig:appendix_nas_cv_cpu} depicts the coefficient of variation with multi-core on different platforms, to illustrate larger measurement variance when using multiple cores (\cref{sec:result_hardware_heterogeneity}).
\cref{fig:appendix_comparison_common_cpu_mlp_pixel4} shows the prediction errors of MLP with different training set sizes, to support the explanation of the lower prediction errors of MLP with a smaller training set of size 30 (\cref{sec:result_limited_training_data_comparison_ML_approaches}).

\begin{table}[t]\centering
\renewcommand\arraystretch{1.3}
\begin{tabular}{c c c c c c c c c c}\toprule
\multirow{2}{*}{Approach} & \multirow{2}{*}{Training Size} & \multicolumn{2}{c}{Snapdragon 855} & \multicolumn{2}{c}{Exynos 9820} & \multicolumn{2}{c}{Snapdragon 710} & \multicolumn{2}{c}{Helio P35} \\
& & CPU & GPU & CPU & GPU & CPU & GPU & CPU & GPU \\
\toprule
\multirow{3}{*}{Lasso} & 30 & 12.84\% & 17.95\% & 9.08\% & 10.29\% & 8.85\% & 14.15\% & 15.90\% & 6.05\%\tabularnewline
                      & 100 & 12.93\% & 18.71\% & 8.87\% & 10.23\% & 8.72\% & 14.46\% & 14.88\% & 5.59\% \tabularnewline
                      & 900 & 13.26\% & 16.36\% & 8.90\% & 9.63\% & 9.33\% & 12.67\% & 15.09\% & 5.31\% \tabularnewline
\midrule
\multirow{3}{*}{RF} & 30 & 10.71\% & 13.68\% & 13.52\% & 9.99\% & 11.83\% & 12.97\% & 9.98\% & 6.49\% \tabularnewline
                   & 100 & 6.20\% & 9.43\% & 4.90\% & 8.58\% & 6.13\% & 11.47\% & 7.79\% & 3.83\% \tabularnewline
                   & 900 & 2.83\% & 7.33\% & 2.82\% & 8.34\% & 2.29\% & 8.30\% & 3.09\% & 2.74\% \tabularnewline
\midrule
\multirow{3}{*}{GBDT} & 30 & 7.91\% & 12.52\% & 7.76\% & 9.59\% & 7.10\% & 15.97\% & 9.08\% & 4.93\% \tabularnewline
                     & 100 & 3.97\% & 9.77\% & 4.36\% & 8.59\% & 4.73\% & 12.29\% & 5.45\% & 3.43\% \tabularnewline
                     & 900 & 2.12\% & 7.60\% & 1.92\% & 8.41\% & 2.01\% & 6.56\% & 3.71\% & 2.77\% \tabularnewline
\midrule
\multirow{3}{*}{MLP} & 30 & 9.11\% & 10.02\% & 7.94\% & 8.55\% & 8.21\% & 10.12\% & 10.71\% & 4.84\% \tabularnewline
                    & 100 & 4.03\% & 9.17\% & 3.84\% & 9.01\% & 3.07\% & 9.28\% & 6.61\% & 4.35\% \tabularnewline
                    & 900 & 2.30\% & 6.37\% & 2.44\% & 8.19\% & 2.03\% & 6.35\% & 6.09\% & 3.35\% \tabularnewline
\bottomrule
\end{tabular}
\caption{End-to-end Predictions on Synthetic Neural Architectures (CPU Stands For a Large Core)}
\label{table:end_to_end_predictions_synthetic}
\end{table}

\begin{table}[t]\centering
\renewcommand\arraystretch{1.3}
\begin{tabular}{c c c c c c c c c c}\toprule
\multirow{2}{*}{Approach} & \multirow{2}{*}{Training Size} & \multicolumn{2}{c}{Snapdragon 855} & \multicolumn{2}{c}{Exynos 9820} & \multicolumn{2}{c}{Snapdragon 710} & \multicolumn{2}{c}{Helio P35} \\
& & CPU & GPU & CPU & GPU & CPU & GPU & CPU & GPU \\
\toprule
\multirow{3}{*}{Lasso} & 30 & 9.77\% & 12.04\% & 5.83\% & 12.68\% & 6.40\% & 4.78\% & 5.51\% & 6.79\% \tabularnewline
                      & 100 & 8.23\% & 14.41\% & 4.85\% & 11.77\% & 7.08\% & 5.21\% & 4.87\% & 6.51\% \tabularnewline
                      & 900 & 7.29\% & 12.10\% & 5.24\% & 12.28\% & 5.27\% & 4.59\% & 4.65\% & 6.06\% \tabularnewline
\midrule
\multirow{3}{*}{RF} & 30 & 14.79\% & 14.77\% & 20.15\% & 13.23\% & 14.37\% & 7.99\% & 18.86\% & 6.81\% \tabularnewline
                   & 100 & 11.67\% & 9.94\% & 10.85\% & 11.24\% & 9.10\% & 5.72\% & 10.26\% & 7.19\% \tabularnewline
                   & 900 & 7.43\% & 7.24\% & 8.01\% & 11.39\% & 5.02\% & 5.60\% & 5.71\% & 6.01\% \tabularnewline
\midrule
\multirow{3}{*}{GBDT} & 30 & 12.20\% & 12.13\% & 16.57\% & 12.50\% & 11.92\% & 9.03\% & 16.11\% & 6.92\% \tabularnewline
                     & 100 & 12.32\% & 7.83\% & 10.28\% & 12.32\% & 7.38\% & 5.24\% & 10.19\% & 6.44\% \tabularnewline
                     & 900 & 6.38\% & 6.68\% & 7.86\% & 11.87\% & 4.79\% & 4.15\% & 4.80\% & 5.86\% \tabularnewline
\midrule
\multirow{3}{*}{MLP} & 30 & 14.87\% & 7.79\% & 13.18\% & 9.94\% & 11.35\% & 8.52\% & 13.01\% & 7.03\% \tabularnewline
                    & 100 & 18.31\% & 9.05\% & 16.61\% & 10.51\% & 12.35\% & 10.37\% & 12.25\% & 7.91\% \tabularnewline
                    & 900 & 14.48\% & 7.59\% & 14.23\% & 11.06\% & 16.59\% & 11.06\% & 10.22\% & 7.08\% \tabularnewline
\bottomrule
\end{tabular}
\caption{End-to-end Predictions on Real-world Neural Architectures (CPU Stands For a Large Core)}
\label{table:end_to_end_predictions_real_world}
\end{table}

\section{Details of Kernel Fusion and Kernel Selection in TFLite}

% \cref{alg:kernel_fusion,alg:select_conv2d_kernel} present the implementation of kernel fusion and kernel selection in TFLite.

\cref{alg:kernel_fusion} presents the implementation details of kernel fusion in TFLite: two operations of the computational graph are fused when (1)~the first operation has only one output tensor (Line~\ref{line:check_dst_tensor}), (2) the second operation is the only operation in the graph using this output tensor (Line~\ref{line:check_candidate_nodes}), (3) the second operation uses this output tensor as its first input and produces a single output (Line~\ref{line:check_dst_tensor_output}), and (4) the next operation has a compatible type (Line~\ref{line:linkable_operation}).

\cref{alg:select_conv2d_kernel} summarizes the criteria used by TFlite to enable the use of the Winograd algorithm for convolution operations: when the input tensor and kernel size of a convolution operation both satisfy certain \emph{hardware-dependent} criteria (i.e., CheckWinograd), the kernel of Winograd is selected for the operation.

\begin{algorithm}
% \begin{verbatim}
% MergeNodes(N)
%     ready = []
%     for n in N:
%         ready = ready + n.outputs
%         if len(n.outputs) == 1:
%             consumers = []
%             for m in N:
%                 for k in m.inputs:
%                     if m.inputs[k] == n.outputs[0]:
%                         consumers.append(m)
%         if len(consumers) == 1 and
%             consumers[0].inputs[0] == n.outputs[0] and
%             consumers[0].inputs[0] in ready and  # always true
%             len(consumers[0].outputs) == 1 and
%             consumers[0].type in MERGEABLE_TYPES:
%             merge(n, consumers[0])
%             N.remove(consumers[0])
%     return N
% \end{verbatim}
\begin{pseudo}[fullwidth,indent-length=1.1em,font=\small]*
\toprule
\hd{MergeNodes}(\text{nodes})\\
% [bol=\midrule]
\text{ready\_tensors} = [] \\
\kw{for} \text{cur\_node} \kw{in} \text{nodes} \\+
    \kw{for} \text{dst\_tensor} \kw{in} \text{cur\_node.dst\_tensors} \\+
        \text{ready\_tensors.insert}(\text{dst\_tensor}) \\-
    \kw{if} \text{cur\_node.dst\_tensors.size()} $\ne$ 1  \label{line:check_dst_tensor}\\+
        \kw{continue} \\-
    \text{candidate\_nodes} = [] \\
    \text{candidate\_tensor\_index} = 0 \\
    \kw{for} \text{next\_node} \kw{in} \text{nodes} \\+
        \kw{for} k = 0 to next\_node.src\_tensors.size() - 1 \\+
            \kw{if} \text{next\_node.src\_tensors[k]} == \text{cur\_node.dst\_tensors[0]} \\+ 
                \text{candidate\_tensor\_index} = k\\
                \text{candidate\_nodes.insert(next\_node)} \\---
    \kw{if} \text{candidate\_nodes.size()} $\ne$ 1 or \text{candidate\_tensor\_index} $\ne$ 0 \label{line:check_candidate_nodes}\\+
        \kw{continue} \\-
    \text{next\_node} = \text{candidate\_nodes[0]} \\
    \kw{if} \text{next\_node.src\_tensors[0]} $\in$ \text{ready\_tensors} and \textsc{IsLinkable}(\text{next\_node}) \\+
        \textsc{Merge}(\text{cur\_node}, \text{next\_node}) \\
        \text{nodes.remove}(cur\_node) \\--
    \kw{return} \text{nodes} \\*
% \bottomrule
% \end{pseudo}
\\*
% \begin{pseudo}[fullwidth,indent-length=1.1em,font=\small]*
% \toprule
\hd{IsLinkable}(\text{node})\\
% [bol=\midrule]

\kw{if} \text{node.output\_tensors.size()} $\ne$ 1 \label{line:check_dst_tensor_output}\\+
    \kw{return} \cn{False} \\-
\kw{if} \text{node.type} $\in$ [\cn{ACTIVATION, COPY, ADD, SUB, MUL, DIV, EXP, LOG, SQRT, SQUARE, ABS, NEG, POW, EQUAL, GREATER, LESS, MAXIMUM, MINIMUM}]\label{line:linkable_operation}\\+
    \kw{return} \cn{True} \\-
\kw{return} \cn{False}\\*
\bottomrule
\end{pseudo}

\caption{Kernel Fusion in TFLite GPU Delegate} \label{alg:kernel_fusion}
\end{algorithm}

\begin{algorithm}

\begin{pseudo}[fullwidth,indent-length=1.1em,font=\small]*
\toprule
\hd{SelectConv2DKernel}(\text{gpu\_info}, \text{op\_info})\\
% [bol=\midrule]
\kw{If} \textsc{CheckGroupedConv2D}(\text{gpu\_info}, \text{op\_info})\\+
    \kw{return} \textsc{Kernel}(\cn{GroupedConv2D}, \text{gpu\_info}, \text{op\_info})\\-
\kw{Else} \kw{if} \textsc{CheckWinograd}(\text{gpu\_info}, \text{op\_info})\\+
    \kw{return} \textsc{Kernel}(\cn{Winograd}, \text{gpu\_info}, \text{op\_info})\\-
\kw{Else}
    \kw{return} \textsc{Kernel}(\cn{Conv2D}, \text{gpu\_info}, \text{op\_info})\\*
% \bottomrule
% \end{pseudo}
\\*
% \begin{pseudo}[fullwidth,indent-length=1.1em,font=\small]*
% \toprule
\hd{CheckGroupedConv2D}(\text{gpu\_info}, \text{op\_info})\\
% [bol=\midrule]

\text{src\_group\_size} = \text{op\_info.input\_channel} \\
\text{dst\_group\_size} = \text{op\_info.output\_channel} / \text{op\_info.group} \\
\kw{If} \text{op\_info.group} $\ne$ 1 and \text{src\_group\_size} $\%$ 4 == 0 and \text{dst\_group\_size} $\%$ 4 == 0 \\+
\kw{return} \cn{True}\\-
\kw{return} \cn{False}\\*
% \bottomrule
% \end{pseudo}
\\*
% \begin{pseudo}[fullwidth,indent-length=1.1em,font=\small]*
% \toprule
\hd{CheckWinograd}(\text{gpu\_info}, \text{op\_info})\\
% [bol=\midrule]
\kw{If} \text{op\_info.group} $\ne$ 1 or \text{op\_info.kernel\_shape} $\ne$ 3x3 or \text{op\_info.stride} $\ne$ 1 \\+
\kw{return} \cn{False}\\-

\text{src\_depth} $= \ceil{\text{op\_info.input\_channel} / 4}$ \\
\text{dst\_depth} $= \ceil{\text{op\_info.output\_channel} / 4}$ \\

\kw{If} \text{gpu\_info.type} == \cn{Adreno} and (\text{src\_depth} < 32 or \text{dst\_depth} < 32) \label{line:adreno_depth_check}\\+
    \kw{return} \cn{False}\\-
\kw{Else if} \text{gpu\_info.type} == \cn{AMD} and (\text{src\_depth} < 16 or \text{dst\_depth} < 8) \\+
    \kw{return} \cn{False}\\-
\kw{Else if} \text{src\_depth} < 16 or \text{dst\_depth} < 16 \label{line:others_depth_check}\\+
    \kw{return} \cn{False}\\-

\text{total\_tiles} $= \ceil{\text{op\_info.output\_height} / 4} * \ceil{\text{op\_info.output\_width} / 4}$\\

\kw{If} \text{gpu\_info.type} == \cn{Adreno6xx} and \text{total\_tiles} < 128 \label{line:adreno600_tiles_check} \\+
    \kw{return} \cn{False}\\-
\kw{Else if} \text{gpu\_info.type} == \cn{Adreno} and \text{total\_tiles} < 64 \\+
    \kw{return} \cn{False}\\-
\kw{Else if} \text{total\_tiles} < 32 \label{line:others_tiles_check} \\+
    \kw{return} \cn{False} \\-
\kw{return} \cn{True}\\*
\bottomrule
\end{pseudo}

\caption{Kernel Selection for Convolution Operations in TFLite GPU Delegate} \label{alg:select_conv2d_kernel}
\end{algorithm}

% \appendix

\end{document}